\documentclass[a4paper,fleqn,usenatbib,twocolumn]{aastex631}
\let\tablenum\relax
\usepackage{siunitx}

\usepackage{amsmath}
\received{\today}
\revised{XXX}
\accepted{XXX}
\shorttitle{Photometrically determined $M_{BH}$ to $M_{\ast}$ relation across cosmic time}
\shortauthors{A. R. Young et al.}
\submitjournal{ApJ}

\begin{document}

\title{Glimmers in the Cosmic Dawn. III. On the Photometrically Determined Black Hole Mass to Stellar Mass Relation Across Cosmic Time
\footnote{This research is based on observations made with the NASA/ESA Hubble Space Telescope obtained from the Space Telescope Science Institute, which is operated by the Association of Universities for Research in Astronomy, Inc., under NASA contract NAS 5–26555. These observations are associated with programs 1563,12498,17073.}}

\correspondingauthor{Alice R. Young}
\email{alice.young@astro.su.se}

\author[0000-0001-9136-3701]{Alice R. Young}
\affiliation{Stockholm University, Department of Astronomy and Oskar Klein Centre for Cosmoparticle Physics, AlbaNova University Centre, SE-10691 Stockholm, Sweden}

\author[0000-0001-8587-218X]{Matthew J. Hayes}
\affiliation{Stockholm University, Department of Astronomy and Oskar Klein Centre for Cosmoparticle Physics, AlbaNova University Centre, SE-10691 Stockholm, Sweden}

\author[0000-0001-8419-3062]{Alberto Saldana-Lopez}
\affiliation{Stockholm University, Department of Astronomy and Oskar Klein Centre for Cosmoparticle Physics, AlbaNova University Centre, SE-10691 Stockholm, Sweden}

\author[0000-0002-1025-7569]{Axel Runnholm}
\affiliation{Stockholm University, Department of Astronomy and Oskar Klein Centre for Cosmoparticle Physics, AlbaNova University Centre, SE-10691 Stockholm, Sweden}

\author[0000-0002-2070-9047]{Vieri Cammelli}
\affiliation{Department of Physics, Informatics \& Mathematics, University of Modena \& Reggio Emilia, via G. Campi 213/A, 41125, Modena, Italy}

\author[0000-0002-3389-9142]{Jonathan C. Tan}
\affiliation{Department of Space, Earth \& Environment, Chalmers University of Technology, SE-412 96 Gothenburg, Sweden}
\affiliation{Department of Astronomy, University of Virginia, Charlottesville, VA 22904, USA}

\author[0000-0001-7782-7071]{Richard S. Ellis}
\affiliation{Department of Physics and Astronomy, University College London, Gower Street, London WC1E 6BT, UK}

\author[0000-0002-9642-7193]{Benjamin W. Keller}
\affiliation{Department of Physics and Materials Science, University of Memphis, 3720 Alumni Avenue, Memphis, TN 38152, USA}

\author[0000-0003-0470-8754]{Jens Melinder}
\affiliation{Stockholm University, Department of Astronomy and Oskar Klein Centre for Cosmoparticle Physics, AlbaNova University Centre, SE-10691 Stockholm, Sweden}

\author[0000-0002-6260-1165]{Jasbir Singh}
\affiliation{INAF – Astronomical Observatory of Brera, via Brera 28, I-20121 Milan, Italy}

\begin{abstract}
We present the results from performing spectral energy distribution (SED) fitting on 121 variable active galactic nuclei (AGN) candidates in the Hubble Ultra Deep Field (HUDF) using photometry from both the Hubble Space Telescope (HST) and the James Webb Space Telescope (JWST) covering $0.2 - 4.8$ microns. We designed a bespoke SED fitting code which decomposes the total SED into its stellar and AGN contributions. Our SED fitting retrieves a significant contribution to the total SED from an AGN template for 26 of our variable sources with $0 < z < 7$. We leverage the model AGN spectrum to estimate black hole masses ($M_{BH}$) using the measured luminosity at 5100 \AA\ and local empirical calibrations. Common with recently discovered JWST broad line AGN (BL-AGN), we observe a trend in the $M_{BH} - M_{\ast}$ plane where low redshift sources have $M_{BH}$ which agree with local relations while high redshift sources have increasingly overmassive black holes with respect to the stellar mass ($M_{\ast}$) of their host galaxies. Within our sample, we identify two IMBH candidates hosted by dwarf galaxies at $z<1$ featuring overmassive BHs in the $M_{BH}-M_{\ast}$ plane, similarly to our high redshift sources. Finally, our SED fitter successfully retrieves the AGN nature of one source at $z >6$. This object has $z_{phot} = 6.74$ and we estimate a lower limit on its black hole mass of $\log_{10}(M_{BH}/M_{\odot}) > 7.36$. 
\end{abstract}



\section{Introduction} \label{sec:intro}
Despite evidence that supermassive black holes (SMBHs) can be found in the nuclei of most massive galaxies in the local universe, the physical mechanisms which lead to the tight correlations between SMBH mass and the bulge properties of their host galaxies remain uncertain \citep{KormendyHo_2013, Greene_IMBHreview_2020}. Theories backed by cosmological simulations suggest that the coevolutionary growth may be the result of galaxy and black hole (BH) mergers as well as the feedback of accreting BHs called active galactic nuclei (AGNs) which can regulate the accumulation of stellar mass within a galaxy \citep{ Granato_2004,hopkins_2006,Li_2007,Sijacki_2009,volonteri_2010,Valiante_2016,InayoshiReview_2020,Greene_IMBHreview_2020,Trinca_2022,fan2022, Volonteri_2023,bennett2023growthgargantuanblackholes,Koudmani_2022}. In order to better constrain theoretical predictions for the coevolutionary nature of SMBH and galaxy relations, it is essential to probe the SMBH population beyond the local volume and into the high redshift (high-z) universe. 

Before the advent of the James Webb Space Telescope (JWST), observations of SMBHs beyond $z \sim 4$ were difficult to achieve. This was due to the fact that relevant diagnostic lines were redshifted beyond the wavelength constraints of the available detectors as well as the lack of sensitivity to faint, less massive galaxies and SMBHs at higher redshift. As such, it was primarily only possible to detect the most luminous AGN, namely quasars at $z \gtrsim 4$ (e.g., \citealt{Warren_1987qsoz>4,Schneider_qsoz>41994,hook_1998,Fan_1999,schneider2000,Fan_2001}). With the discovery of quasars at $z > 7$ it was found that these objects had masses exceeding several billions of solar masses already within the first few billions of years after the Big Bang \citep{mortlock_2011_highzqso,Banados_2017_highzqso,Matsuoka_2018a_highzqso,wang_2018_highzqso, Yang2019_highzqso,Matsuoka_2019b_highzqso,Yang_2020b_highzqso,fan2022}. This was in contention with many simulations which struggled to grow SMBHs sufficiently to achieve such large SMBH masses at these redshifts \citep{McCaffrey_2025_heavyseeds}. This result has been exacerbated by recent discoveries with JWST which have identified overmassive black holes at high-z, necessitating either extreme growth rates or favouring a heavy seeding scenario in the early universe to explain the observed masses at these early cosmological times \citep{larson_ceers_2023, kokorev_uncover_2023,Ubler_23MBH, tripodi_2024, Maiolino_2024, furtak_high_2024,Harikane_23MBH, zhang2025_SCJCMBH, akins_2025,naidu_2025,taylor2025_capersz9LRD}. 

A variety of supermassive black hole seeding scenarios have been invoked to attempt to resolve this observational disparity with theoretical predictions \citep{InayoshiReview_2020,Greene_IMBHreview_2020,Regan_2024}, including direct collapse black holes (DCBH) \citep{haehneltrees_1993DCBH, Bromm_2003dcbh, begelman_DCBH_2006,Chon_2016DCBH,Wise_2019DCBH}, stellar mergers in dense star clusters \citep{Freitag_2006mergsc,Schleicher_2023mergsc}, super Eddington growth of population III (Pop III) stars \citep{Madau_Rees_2001,Trinca_2022} and unique processes of the very first generation of Pop III stars (designated Pop III.1; \citep{McKeetan_2008popiii.1}) which allow these stars to leave behind supermassive remnants that act as SMBH seeds with masses $\sim 10^5 M_{\odot}$ \citep{McKeetan_2008popiii.1,banik_2019popIII.1,Singh_2023,cammelli2024popiii.1,sanati2025}. Degeneracies that exist in the predictions of these seeding theories can be constrained by probing the population of SMBHs at high redshift ($z > 6$) highlighting the importance of observations which target the detection of faint, lower mass and more distant AGNs. 

A multitude of methods exist for the detection of these objects including X-ray detections with luminosities exceeding those produced by stellar populations, the detection of emission lines with ionization energies greater than the typical output of stars as well as the previously mentioned characteristic broadened emission lines that designate a broad line AGN (BL-AGN). An additional observational tool comes from using photometric variability to detect variations in brightness within the nucleus of a host galaxy which can be explained by changes in the accretion rate of a central SMBH. By using multiple images of approximately equal depth, it is possible to test an entire field for variability without any need for observational pre-selection unlike slit-based spectroscopic methods. Photometric variability also allows for a more complete sample of luminosities to be tested for the presence of AGN in comparison to other methods where typically only the brightest candidates are flagged for spectroscopic follow-up. The ability photometric variability offers to probe fainter luminosities also allows for lower mass BHs to be probed at each redshift. Leveraging photometric variability has been successful at identifying AGN at intermediate redshifts using deep images taken with the Hubble Space Telescope (HST) \citep{cohen_2006,Pouliasis_2019,obrien2024}. With longer baselines in time, variability with HST has even successfully probed AGN candidates out to $6 < z < 8$ 
\citep{Hayes_Glimmers_2024, cammelli_glimmers25}. Such a census of variable SMBHs can offer strong constraints for simulations predicting SMBH seeding in the early universe via the observed co-moving number density \citep{Rees_1978,volonteri_2010,Banik_2018,InayoshiReview_2020,Singh_2023}. In addition, the observational power of variability studies is fortified by the addition of legacy data across a wide range of wavelengths which can then be used to disentangle the AGN contribution to the total flux by modeling the spectral energy distributions (SEDs) of these sources through SED fitting.

In this paper, we present a bespoke SED fitting code we have developed utilising the semi-empirical AGN templates presented in \cite{Temple_2021} in combination with stellar population models from \cite{BruzualCharlot_2003}. We use this SED fitting tool to determine physical properties of the central SMBH and its host galaxy. We achieve this using spectral decomposition in order to understand the underlying contributions of both a stellar population and an AGN to the total SED of our variable sources. This decomposition method is inspired by recent works seeking to determine the nature of the mysterious population of compact objects at $z>3$ featuring characteristic `v-shaped' SEDs \citep{Kocevski_monsters_23MBH,barro2023LRD,labbe2024LRD} and often broad Balmer emission lines termed ``Little Red Dots" (LRDs) \citep{LRDFurtak_2023,Kocevski_monsters_23MBH,Greene_2024LRDs,LRDMatthee_2024,LRDKokorev_2024,Wang_2025_LRDs}. A popular method includes modeling the SEDs of these objects to understand if they are better represented by a pure stellar component or if an AGN is necessary (for examples, see: \cite{akins2024_LRDs,leung2024_LRDs,setton2024_LRDs,Ma_2025_LRDs,Wang_2025_LRDs}). However, we note that the goal of this work is not identical to those studying the nature of LRDs. Namely, since we have already identified our sources as variable AGN, we are not trying to ascertain whether an AGN is present in our sources (as in the LRD scenario), we are instead seeking to determine if our SED fitter can successfully recover the AGN nature of our sources.

Our sample consists of 121 sources in the Hubble UltraDeep Field (HUDF) with $z = 0 -8$ detected with variability measured at the $2.5\sigma$ level in \cite{cammelli_glimmers25}. From this sample of variable SMBH candidates, we aim to determine what fraction of our sources can be retrieved as AGN via SED fitting as well as physical properties of our variable sources. In particular, we seek to determine the black hole masses ($M_{BH}$) for a subset of variable BHs and identify how these evolve with respect to the stellar mass of the host galaxies for the range of redshifts represented by our sample. Our SED modeling is performed on PSF matched photometry from 26 HST+JWST filters covering a wavelength range from $0.2 - 4.8$ microns using only 7 free parameters. Our SED fits provide estimates for the stellar mass ($M_*$) as well as a model AGN spectrum which we use to determine the luminosity of the AGN at 5100 \AA\ ($L_{5100}$). This, in combination with the relation from \cite{kaspi_2000_MBH}, allows us to estimate $M_{BH}$. 

Additionally, this work offers a unique opportunity to compare the trends seen in the $M_{BH} - M_{*}$ plane using an alternative method to calculate $M_{BH}$ from those typically employed for the recently discovered high redshift BL-AGN with JWST.

This paper is structured as follows: Section \ref{sec:obs} details the HUDF imaging campaign and the data reduction processes used to achieve approximately equal depths in both epochs of each filter pair used for variability detection. Additionally, we detail the PSF matching executed to produce the photometry used in our SED fitting. Section \ref{sec:modeling} details the creation of our bespoke SED fitting tool and the methods we employ to select either a pure stellar population SED or a combined stellar and AGN model. Section \ref{sec:results} details how we retrieve AGN contributions for our variable sources using SED fitting as well as how we estimate the black hole masses for a subset of our variable sources. Finally, in Section \ref{sec:discussion} we discuss our estimated black hole masses and their evolution with redshift as well as the various factors which could effect our estimates. These include SED reddening and the impact of variability on our SED fits. We also present two dwarf galaxies hosting IMBH candidates and discuss the $M_{BH}$ calculated for our highest redshift source and what this estimate implies about black hole seeding. 

Throughout this work we use the AB magnitude system \citep{ABmag_OkeandGunn_1983} and assume a flat $\Lambda$CDM cosmology with $H_0=70\, \mathrm{km\,s^{-1}\,Mpc^{-1}}$, $\Omega_m = 0.3$ and $\Omega_{\Lambda}=0.7$.
\section{Observations and Data Reduction} \label{sec:obs}
\subsection{Processing WFC3/IR Epochs for Variable Detection}
In order to detect our variable sources, two sets of images corresponding to two epochs of observation were analysed for each of the F105W, F140W and F160W WFC3/IR filters. The data was originally presented in \cite{Hayes_Glimmers_2024}, however, a more detailed description of the reduction procedure is provided here. The HUDF was originally imaged with F105W, F125W and F160W in 2008-9 (GO 11563, UDF09; PI: Illingworth) followed by repeated imaging in 2012 with F105W, F140W and F160W (GO 12498, UDF12; PI: Ellis). Finally, in 2023 the HUDF was imaged a third time with F140W to match the depth of the UDF12 campaign (GO 17073; PI: Hayes). This longer baseline in time of $\sim 11$ years is particularly well suited to probing the variability of AGN at high-z in order to overcome the large cosmological time dilation factor.

We performed independent reductions of the 2009 and 2012 epochs for the F105W and F160W filters as well as a re-reduction of the F140W 2012 image. Additionally, we utilise the High Level Science Products (HLSP) from the Mikulski Archive for Space Telescopes (MAST) as reference images and compare the depth of our re-reductions to the original HLSPs. The reduction procedure was slightly altered between the F105W and F140W/F160W images as additional care is required to correct for the time-varying background resulting from Helium line emission being captured by the detector with the F105W filter when HST leaves Earth's shadow (see: \citet{Brammer_TVB_2016}). Therefore, for the F105W epochs we apply an additional correction to ``flatten the ramp" of the charge accumulation history before further processing. 

For F105W, the data is processed using the \texttt{calwf3} pipeline up to the cosmic ray identification step. The ramp is then flattened before being fed back into \texttt{calwf3} pipeline to complete the cosmic ray identification and result in the final reduced files as described in \cite{Brammer_TVB_2016}. For F140W and F160W images, this step was not applied, and all reduced files were retrieved from MAST having already been processed with the \texttt{calwf3} pipeline. 

The final calibrated images in each association output from the pipeline require additional processing, namely satellite trail identification and persistence correction. For the 2009 and 2012 epochs, persistence is corrected using persistence maps from the PERSIST Search STScI Archive to identify the affected pixels in the quality flags of each file so that these pixels are masked during the combination process. For the 2023 data, these persistence masks are created by-eye to flag areas of strong persistence effects. Further, for all epochs satellite trails are flagged in a similar manner, after being identified with the WfcWrapper for the \texttt{TrailFinder} tool within STScI's ACSTOOLS Python package \citep{acstools}. 

After these corrections, the images in each association are combined using \texttt{AstroDrizzle} \citep{DrizzlePac}. The relevant HLSP is used as the reference for the WCS, rotation and pixel scale, which is Drizzled to be $0.06\,''$/pixel. To be consistent with values used by \cite{Koekemoer_HUDF_2012} in the UDF12 data products, the pixfrac, $p= 0.8$. The scale, $s = 6/13$ arises from the WFCIR native pixel scale of $0.13\,''$/pixel and the Drizzled pixel scale of $0.06\,''$/pixel used in the HLSP data product. \texttt{TweakReg} is then used to refine the WCS, again, using the HLSP as a reference. Finally, for consistency with \cite{Koekemoer_HUDF_2012}, the images have a best fit 2D median background subtracted to remove residual structures seen in the exposures before being Drizzled together using \texttt{combine\_type = `median'} to create the final image at each epoch. 

\subsection{PSF-Matched Photometry} 
The images for which we perform the photometry used as input for our SED analysis come from several programs. In addition to the WFC3/IR filter images we reprocessed, the HLSP for the F125W filter (PI: Illingworth) is also used in the analysis. We also employ HLSPs from the Hubble Ultraviolet Ultra Deep Field (UVUDF; PI: Teplitz; \citealt{Teplitz_2013_UVUDF, Rafelski_2015_UVUDF}) in the  WFC3/UVIS F225W, F275W and F336W filters obtained from the MAST. Additionally, HST images in the ACS/WFC F435W, F606W, F775W, F814W and F850LP filters from the Hubble eXtreme Deep Field (XDF) which combines observations taken from July 2002 to December 2012 from 19 different HST programs \citep{Illingworth_XDF_2013}. We also include JWST HLSPs from the JWST Advanced Deep Extragalactic Survey (JADES; PI: Eisenstein \& Luetzgendorf; \cite{JADES_DR1, JADES_DR2}) and the JWST Extragalactic Medium-band Survey (JEMS; PI: Williams, Tacchella \& Maseda \cite{williams_2023_JEMS}) in 14 filters covering the GOODS-S field, which contains the HUDF footprint (F090W, F115W, F150W, F182M, F200W, F210M, F277W, F335M, F356W, F410M, F430M, F444W, F460M and F480M). 

Before performing PSF matching, we aligned the HST images to the GAIA coordinates using \texttt{TweakReg}. We also matched the pixel scale of the NIRCam images to the Drizzled HST pixel scale of 0.06"/pixel and cropped the NIRCam images to have the same pixel dimensions as the HUDF images using \texttt{SWarp} \citep{SWarp_Bertin_2002}. Some residual pixel grid/WCS alignment offsets were resolved using Astropy's \texttt{reproject\_exact} function \citep{astropyreproject}. 

With the images all aligned to the same pixel grid, we then performed the PSF matching. PSFs were obtained from STScI's instrumentation archive for the WFC3/IR and UVIS filters, while the ACS PSFs were generated from the STScI ACS/WFC Focus-Diverse ePSF generator. Finally, NIRCam PSFs were generated using \texttt{STPSF} \citep{STPSF}. All PSFs were matched to the F480M PSF using a matching kernel generated with PyPHER \citep{Boucaud_pypher_2016}. PyPHER applies Wiener filtering \citep{wiener_1949} to generate PSF matched kernels with an adjustable regularization parameter which we set to $r = 0.003$ to reduce high frequency noise in kernels \citep{Weaver_UNCOVERphot_2024}. As discussed in \cite{Weaver_UNCOVERphot_2024}, PyPHER is capable of producing reliable matching kernels without the need to select a window function as would be required by Photutils, thereby offering a more stable kernel solution over all of the filter PSFs used in this study. The matching kernels were then convolved to the relevant filter using the convolve tool from Astropy's convolution package \citep{astropy:2013,astropy:2018,astropy:2022}. An exception is applied to the UVIS filters for which the matching kernels produced significant convolution artifacts in the final convolved images. This is likely due to the fact the UVIS PSFs are significantly smaller than the F480M NIRCam PSF and, as such, for the UVIS filters the images were simply convolved using the F480M PSF directly to reduce the effects of convolution artefacts in the UVIS images. This is an appropriate approximation given that the UVIS PSFs are on the order of $\sim 20$ times narrower, meaning the true matching kernel will be very close to the F480M PSF.
\subsection{Defining a Photometric Catalogue} \label{subsec:photcat}
A deep WFC3/IR image was created using all six of the reprocessed F105W, F140W and F160W images as well as the F125W HLSP image. This ultra-deep image was used as the detection image in \texttt{Source Extractor} \citep{SExtractor} to produce a reference detection catalogue containing all of the sources in HUDF. We then iteratively performed aperture photometry on all 29 filter images using an aperture diameter of 4 pixels (corresponding to $\sim 0.24''$). 
This aperture was chosen to be about twice the full width at half maximum (FWHM) of the F480M PSF and its size is optimized to capture the bulk of the light from compact sources.  It is plausible that some stellar light may be missing from large extended galaxies at very low redshifts, but visual inspection of our variable sources invariably shows this to be negligible. 
The final photometry is then converted to universal physical units of mJy for all filters. The measured fluxes for our sources were cross-matched to sources in the JADES photometric catalogue \citep{JADES_DR2} and were found to be in good agreement for all of the available NIRCam filters. 

\begin{figure}[htb!]
    \centering
    \includegraphics[width=.49\textwidth]{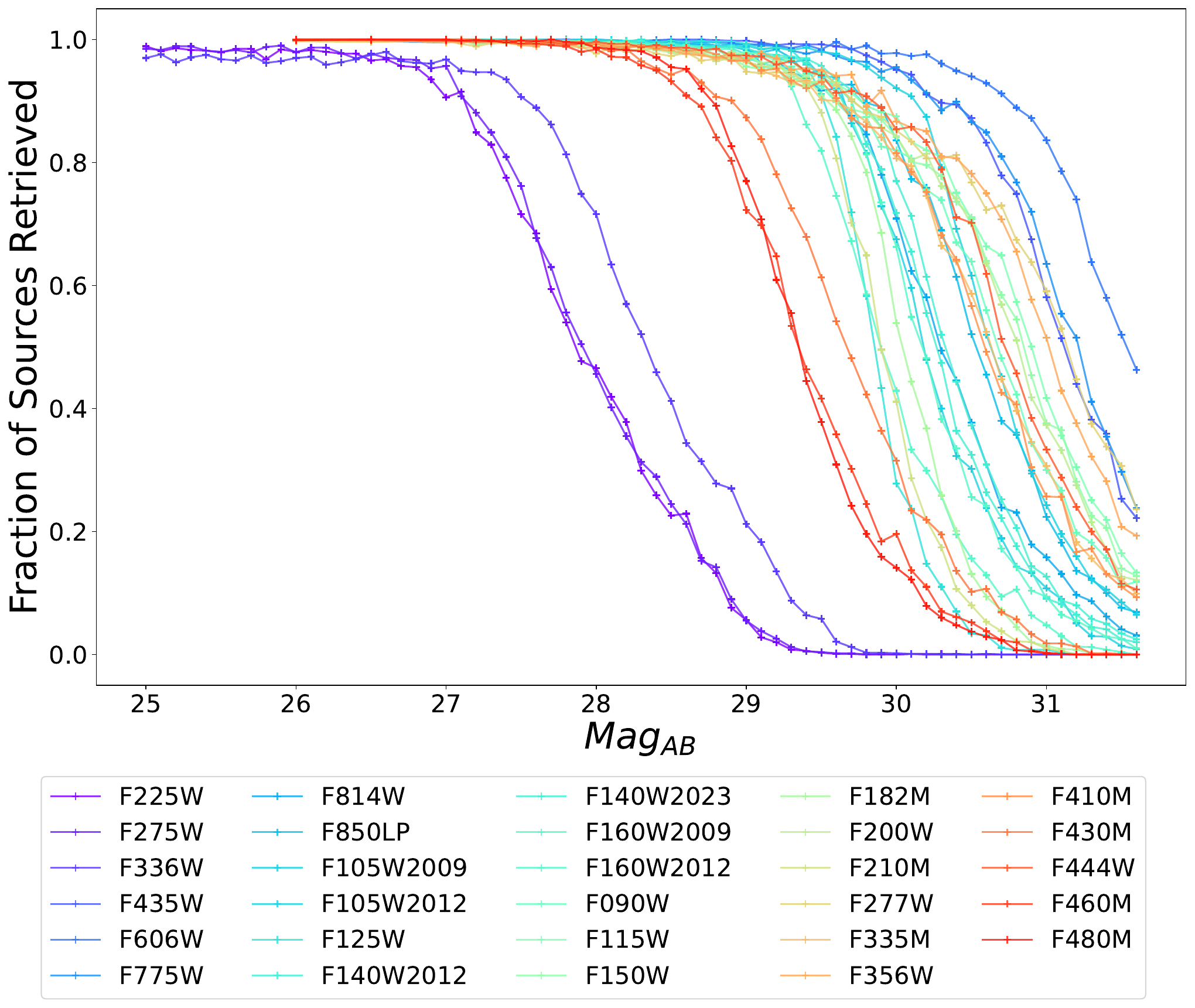}
    \caption{Completeness curves for each PSF matched filter. These are tabulated by injecting each image with point sources modeled by the F480M PSF iteratively normalised to a given injection magnitude and determining the number of injected sources retrieved by \texttt{Source Extractor}. The limiting magnitude of each filter is then set to the injected magnitude in the bin preceding the bin where the number retrieved is below 90\%.}
    \label{fig:filter90perccompleteness}
\end{figure}

\begin{table*}
    \centering
    \begin{tabular}{ccccc} \hline \hline
        Filter & $\mathrm{Mag_{AB,\ Lim}}$ & Program ID & PI Name & HLSP Program \\ \hline
        F225W  &  27.1  &   12534  &  Teplitz & UVUDF \\
F275W  &  27.1   &  12534  &  Teplitz & UVUDF \\
F336W  &  27.5   &  12534  &  Teplitz & UVUDF \\
F435W  &  30.2   & Numerous$^*$ & Numerous$^*$ & XDF \\
F606W  &  30.7  & Numerous$^*$ & Numerous$^*$ & XDF \\
F775W  &  30.2  & Numerous$^*$ & Numerous$^*$ & XDF \\
F814W  &  29.6  & Numerous$^*$ & Numerous$^*$ & XDF \\
F850LP  &  29.7  & Numerous$^*$ & Numerous$^*$ & XDF \\
$\mathrm{F105W_{2009}}$   &  29.6  &  11563  &  Illingworth  &  HUDF09  \\
$\mathrm{F105W_{2012}}$   &  30.1  &  12498  &  Ellis  &  HUDF12  \\
F125W  &  29.5  &  11563  &  Illingworth  &  HUDF09  \\
$\mathrm{F140W_{2012}}$  &  29.7  &  12498  &  Ellis  &  HUDF12  \\
$\mathrm{F140W_{2023}}$  &  29.7  &  17073  &  Hayes  &  HUDF23  \\
$\mathrm{F160W_{2009}}$  &  29.6  &  11563  &  Illingworth  &  HUDF09  \\
$\mathrm{F160W_{2012}}$  &  29.3  &  12498  &  Ellis  &  HUDF12  \\
F090W  &  29.5  &  1180  &  Eisenstein & JADES \\
F115W  &  29.7  &  1180  &  Eisenstein & JADES \\
F150W  &  29.5  &  1180  &  Eisenstein & JADES \\
F182M  &  29.5  &  1895  &  Oesch & JADES \\
F200W  &  29.7  &  1180  &  Eisenstein & JADES \\
F210M  &  29.4  &  1895  &  Oesch & JADES \\
F277W  &  29.7  &  1180  &  Eisenstein & JADES \\
F335M  &  29.6  &  1180  &  Eisenstein & JADES \\
F356W  &  29.7  &  1180  &  Eisenstein & JADES \\
F410M  &  29.6  &  1180  &  Eisenstein & JADES \\
F430M  &  28.9  &  1963  &  Williams & JADES \\
F444W  &  29.8  &  1180  &  Eisenstein & JADES \\
F460M  &  28.6  &  1963  &  Williams & JADES \\
F480M  &  28.7  &  1963  &  Williams & JADES \\
          \hline  
    \end{tabular}
    \caption{Summary of filter images used for PSF-matched photometry. This includes the limiting magnitudes of each image ($\mathrm{Mag_{AB,\ Lim}}$) calculated from the source injections described in Section \ref{subsec:photcat}, the original Program ID and PI name for the relevant observations and the program which developed the HLSP used in this analysis. * Numerous observing programs were used for the ACS HLSPs, see \cite{Illingworth_XDF_2013} for details. Finally, we list the relevant PIs and programs for the WFC3/IR filters, however, we note that the HLSPs were only used as references because the F105W, F140W and F160W images were all reprocessed to create two seperate epoch images for each of the aforementioned filters in this work.}
    \label{tab:photsummary}
\end{table*}
Finally, we corrected for the underestimated errors reported by \texttt{Source Extractor}. For this we normalised the F480M PSF to an injection magnitude and iteratively inserted ten of these PSFs into each PSF-matched filter image 100 times to reach a total of 1000 injected sources in each image at each injected magnitude. For most filters this range of magnitudes was between 26.0 and 31.5, however UVIS filters required a wider range from 25\,-\,31.5 magnitudes to determine the 90\% completeness limit for each image. The completeness curves for each filter are shown in Figure \ref{fig:filter90perccompleteness} and a summary of the images and their limiting magnitudes is presented in Table \ref{tab:photsummary}. The standard deviation for the distribution of retrieved source magnitudes were compared to \texttt{Source Extractor} magnitude errors determined for the HUDF sources in our reference detection catalogue. This comparison was then used to determine a correction factor which could be applied to the photometric errors of our sources as a function of source magnitude for each filter. A minimum error of 5\% was defined after the error correction was applied to account for the absolute calibration uncertainties of the telescopes and minor uncertainties resulting from the PSF matching. For a more detailed discussion of the error correction applied, see Appendix \ref{appendix:errcorr}.

Our final sample of sources for which we perform SED fitting consists of the 121 sources with variability detected at the 2.5 $\sigma$ level reported in \cite{cammelli_glimmers25}. For each source in the sample we also check that the photometry does not contain any pixels which are masked in a given filter (i.e., due to gaps in mosaic images for JADES and JEMS images) using the segmentation maps generated by \texttt{Source Extractor} while defining our reference detection catalogue. Finally, fluxes which are measured to be below the 90\% completeness limit of a given filter are designated as upper limits and the value is set to half the limiting magnitude converted to mJy with corresponding errors of the same value as suggested as one of the recommended options for \texttt{CIGALE} \citep{Boquien_cigale_2019}. 
\section{Modeling}\label{sec:modeling}
In order to learn more about the physical properties of our variable sources we designed a bespoke SED fitting code which decomposes the contribution of a stellar population from the host galaxy along with an AGN component to model our 26 band photometry covering ultraviolet to infrared wavelengths. We apply a cut to the sample from \cite{cammelli_glimmers25} which stipulates that we only perform SED fitting on sources which have a signal-to-noise, $S/N > 5$ in at least eight filters. This equates to the maximal number of free parameters in our SED fitting code (when we use the combined AGN + stellar model) plus one in order to be confident that our photometry is sufficient to properly constrain our SED model.

Our motivation to develop a new SED fitting code derived from numerous tests which found that using the \texttt{SKIRTOR} model implemented in other SED fitting codes (i.e., \texttt{CIGALE}) did not provide good fits for many of our variable sources. In particular, it was not possible to achieve fluxes representative of the photometry observed in our redder wavelength NIRCam filters using the \texttt{SKIRTOR} model without significantly under-predicting the fluxes at bluer wavelengths. Even when tuned to be as blue as possible, the \texttt{SKIRTOR} model was unsuccessful at replicating the flux at wavelengths which include the WFC3/IR filters. Since these are the filters where the variability of our objects is detected, we would, therefore, expect to see a significant contribution to the flux from an AGN template at those wavelengths. We observed this effect in a number of sources for which the AGN classification is well supported. One such case includes a variable source at $z=1.95$ which was originally presented in \cite{Hayes_Glimmers_2024} as source 1051264. For this source, the difference images show a point-like variable component in the nucleus of an elongated disk, strongly indicating the presence of an AGN. Despite this evidence, when fit with \texttt{CIGALE}, this source consistently came back with a 0\% AGN fraction. A desire to use an AGN model that provided better fits to our photometry prompted our development of a new SED fitting algorithm which we detail below. 

\subsection{Stellar Population Model}
The stellar spectra for our SED fitter are generated using the Bayesian Analysis of Galaxies for Physical Inference and Parameter EStimation (\textsc{bagpipes}) code \citep{BagpipesCarnall_2018}. \textsc{bagpipes} uses the updated \cite{BruzualCharlot_2003} stellar population synthesis templates with a \cite{Kroupa2001} stellar initial mass function, including the MILES stellar spectral library \citep{MILES2011} and the stellar evolutionary tracks of \citet{Bressan2012} and \citet{Marigo2013}. Nebular continuum and emission lines were added by processing the stellar emission through the \textsc{cloudy} v17.00 photoionization code \citep{Ferland2017}. We created a multi-dimensional grid of SED models by adopting a parametric star formation history following a delayed-$\tau$ prescription, where $SFR(t) = t e^{-t/\tau}$. We sample a hundred logarithmically-spaced bins for the stellar age, $t$ (between 1~Myr and 10~Gyr), as well as ten bins for the $e$-folding time, $\tau$, designed to have increased sampling for low values of $\tau$ starting with a minimum timescale of 2~Myr. All these parameters are summarized in Table \ref{tab:modelparams}. 

The metallicity is assumed to be the same for the stars and the gas, with possible values of 0.05, 0.1, 0.2, 0.5, 1 and $2~Z_{\odot}$, where $Z_{\odot} = 0.02$. All models formed a total mass of $10^6M_{\odot}$, and assume a fixed ionization parameter for the nebular emission of $\log U = -2.5$. The resulting SEDs were then linearly interpolated within the grid using the \texttt{RegularGridInterpolator} function, from the SciPy \citep{2020SciPy} package. 

\begin{table*}
    \centering
    \begin{tabular}{cc} \hline \hline
         Parameter  & Range \\ \hline
         IMF & \cite{Kroupa2001}  \\
         $t_i~$[Myr], $i=1, 2..., 100$ & $t_i = t_{\rm min} \cdot (t_{\rm max}/t_{\rm min})^{(i-1)/(100-1)}; t_{\rm min} = 1, t_{\rm max} = 10^4$ \\
          $\tau$ [Myr] & 2, 5, 10, 20, 50, 100, 200, 500, 1000, 2000, 5000, 10000 \\
          Metallicity [$Z_{\odot}$] & 0.1, 0.2, 0.5, 1, 2 \\
          Ionization Parameter, $\log ~U$ & $-2.5$ \\ \hline
    \end{tabular}
    \caption{Parameter space defined to generate the set of stellar spectra with \textsc{bagpipes} which were used to make the model grid for our bespoke SED fitting code.}
    \label{tab:modelparams}
\end{table*}
\subsection{AGN Model}
This work employs the AGN semi-empirical templates from \cite{Temple_2021}. The templates are constructed to reproduce the average SDSS-UKIDSS-WISE colours of quasars with $0 < z < 5$. Compared to previous quasar catalogues, the \cite{Temple_2021} templates are generated from a more complete sample since the SDSS DR 16 catalogue of quasars represents an increased range of luminosities per redshift bin. Additionally, these templates include a diverse population of the strongest rest-frame ultraviolet (UV) and optical emission lines while maintaining a small number of free parameters. These qualities, in conjunction with the fact that great care has been taken to account for host galaxy contamination to the quasar continuum, make these templates a valuable tool for SED modeling \citep{Temple_2021, Marshall_2022}. 

The template incorporates a broken power law to model the UV-optical emission from the accretion disk, blackbody emission of hot dust dominating the NIR emission as well as broad and narrow emission lines. Our code generates an AGN model with the publicly available \texttt{qsogen} code by setting the template redshift to the known galaxy redshift from cross-matching performed by \cite{cammelli_glimmers25}. Redshift classifications in \cite{cammelli_glimmers25} prioritized spectroscopic redshifts from JADES \citep{bunkerspecz2023} and VLT/MUSE \citep{Bacon_2023}, using photometric redshifts from JADES \citep{JADES_DR1} and the UVUDF \citep{Rafelski_2015_UVUDF} when spectroscopic redshifts were unavailable. Additionally, the `lyForest' parameter which implements IGM absorption, and the `gflag' parameter which includes a host galaxy flux contribution are all set to False and the `lylim' is forced to zero. These choices are made such that we can independently calculate a host-galaxy contribution to the flux and uniformly apply IGM transmission and dust reddening to both components of our combined AGN + stellar population models. 
\subsection{Minimization and Model Selection}
We simultaneously fit the parameters which govern our stellar models, namely the duration of the star formation, also known as the e-folding time ($\log_{10}(\tau)$), the age of the stellar population ($\log_{10}(t_{age})$), a normalization factor for the stellar templates ($norm_{stellar}$), the metallicity ($Z/Z_{\odot}$), a diffuse reddening parameter which applies to both the stellar population and the AGN following the Calzetti Law \citep{calzetti_2000} ($E(B-V)_{diffuse}$) as well as those which define our AGN model, including a normalisation factor for the AGN template ($norm_{AGN}$) and a nuclear reddening term following the SMC law \citep{Prevot_SMC_1984} which applies only to the AGN template ($E(B-V)_{AGN}$). For our pure stellar population model (denoted as the $SP$ model) this corresponds to 5 free parameters, while for our combined stellar and AGN model (denoted as the $AGN + SP$ model) this gives us a total of 7 free parameters. Additionally, our model includes IGM absorption following \cite{Madau_IGM_1995}. 

In order to determine our best fit model parameters we use three steps in sequence. First the parameter space is explored by an optimizer, namely Non-Linear Least-Squares Minimization and Curve-Fitting for Python, or \texttt{lmfit} \citep{lmfit_2016}. The results of this fit are then used to initialize a Markov chain Monte Carlo Ensemble sampler, namely \texttt{emcee} \citep{emcee_Foreman_Mackey_2013} to estimate the posterior distributions of our model parameters. Finally, since we expect there to be multiple modes in the posterior distribution and to prevent the scenario where the walkers miss the actual location of the maximum likelihood (corresponding to the ``best fit'' model) we use the results from the MCMC fit to inform a final run of \texttt{lmfit} in order to allocate the best fit model parameters to those which minimize the residual of our model. Numerous tests showed that the final fits were significantly improved (i.e., the $\chi^2$ was significantly reduced) when this additional run of \texttt{lmfit} was implemented. This final model output is used as our best fit model for each run. 

Operating under the baseline assumption that all of the sources in our sample vary in brightness between epochs due to the presence of an AGN, we test what fraction of these AGNs can be recovered from SED fitting alone. We therefore model the same set of photometry for each source with a pure stellar component and for a combine stellar + AGN model. We then compare the pure stellar population (SP) model and the combined (AGN + SP) model for each source to identify those sources which have photometry that is better reproduced by including the AGN templates into our SED fitting. To test this, we define two criteria. Firstly, we calculate the Akaike information criterion (AIC) from:
\begin{equation}
    AIC = 2k - 2log(L),
\end{equation}
where $k$ is the number of free parameters and $L$ is the maximum likelihood. The AIC penalizes models with additional free parameters, meaning we can use it to compare our combined model and pure stellar model to determine which set of parameters provides an improved fit to the data. We consider a fit to be statistically improved by including an AGN component if the AIC decreases by at least 6 for the combined model compared to the pure stellar model. This corresponds to a relative uncertainty of 
\begin{equation}
\begin{split}
\exp\left(\frac{AIC_{AGN+SP}-AIC_{SP}}{2}\right) \\
= \exp(-6/2) = 0.0497, 
\end{split}
\end{equation}

\noindent suggesting the combined model is $\sim 0.95$ times more likely to minimize the information loss from the ``true'' model compared to the pure stellar population model \citep{statsAICbook}.  

Additionally, we employ a second criterion denoted by $AGN_{frac}$ which is also determined from our best fit AGN + SP model. $AGN_{frac}$ is used to estimate the maximal contribution of the best fit AGN template to the total flux of the SED in each filter. We convolve the best fit AGN spectrum with each filter transmission function to determine an ``AGN SED" which we divide by the best fit model SED for the total flux (i.e., AGN + SP) in each filter and take the maximal result from all 26 filters to be $AGN_{frac}$. We then apply the condition that, in order to be successfully retrieved as an AGN via SED fitting, the source must have an $AGN_{frac} > 20\%$ to complement the fact that we demand our sources have S/N $>5$ in at least 8 filters.

\begin{figure*}[htb!]
    \begin{center}
    \includegraphics[width=.45\linewidth]{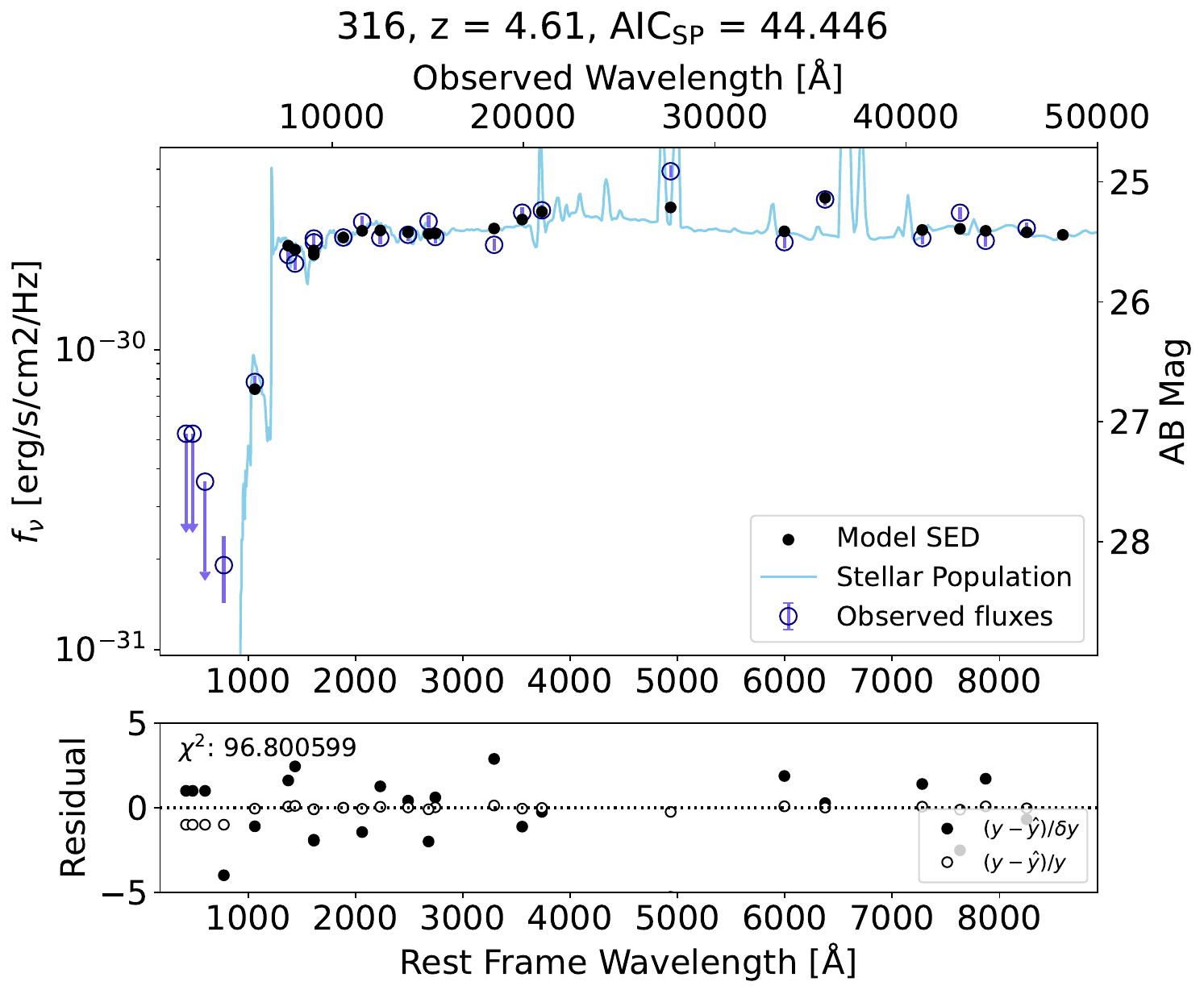}  
    \includegraphics[width=.45\linewidth]{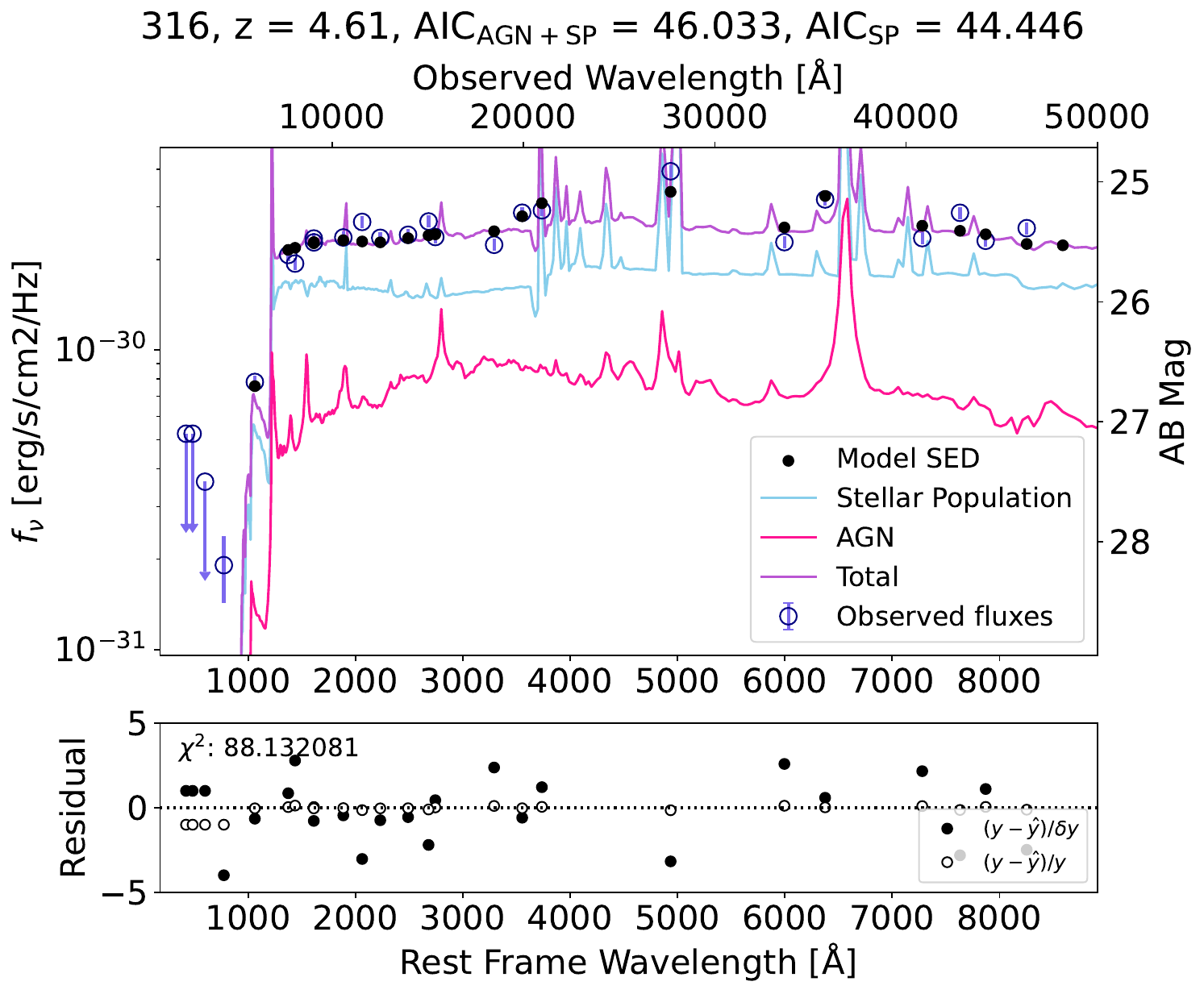}  
    \caption{SED fitting results for source 316 at z= 4.61 for the pure stellar population (SP) model (\textit{left}) and the combined (AGN + SP) model (\textit{right}). These fits are performed on the same set of photometry for this source, namely in both cases, the WFC3/IR epochs are selected where the variable AGN is in the bright state. The stellar population is shown in blue, the AGN contribution in pink and the combined stellar + AGN in purple. The observed data points are plotted in navy and the corresponding best fit SED data are represented by the black points. Below we show the residuals between the best fit SED points and data for each model in two ways, the black points are the residuals divided by the error in the data, whereas the white points are the residuals divided by the data values. Here we can see that the AIC resulting from the pure stellar fit ($AIC_{SP}$) is less than the AIC for the combined fit $AIC_{AGN + SP}$ meaning the SED fitting was not successful in retrieving the AGN nature of this source.}
    \label{fig:exampleSEDs_bad}
    \end{center}
\end{figure*}

\begin{figure*}[htb!]
    \begin{center}
    \includegraphics[width=.45\linewidth]{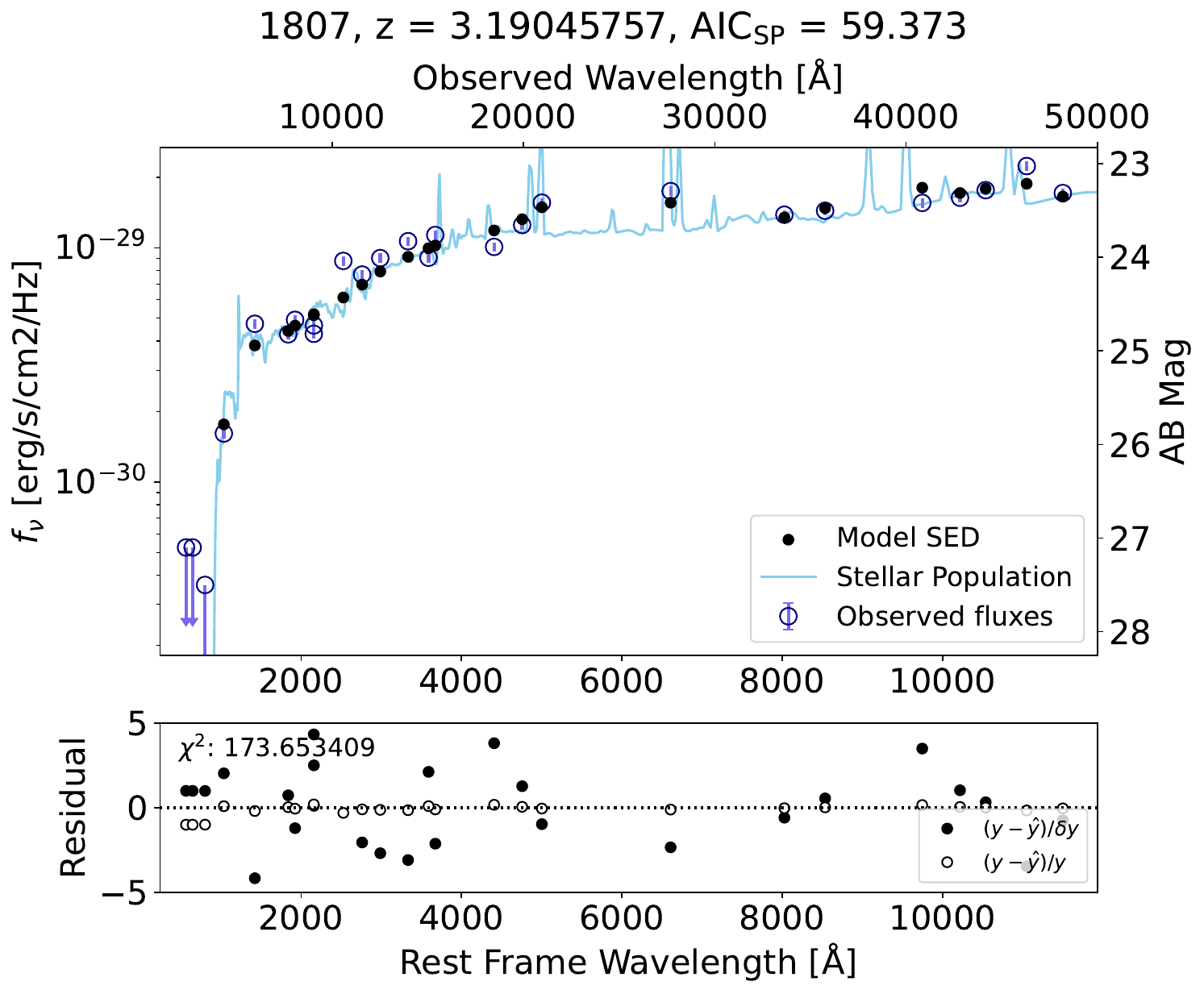}  
    \includegraphics[width=.45\linewidth]{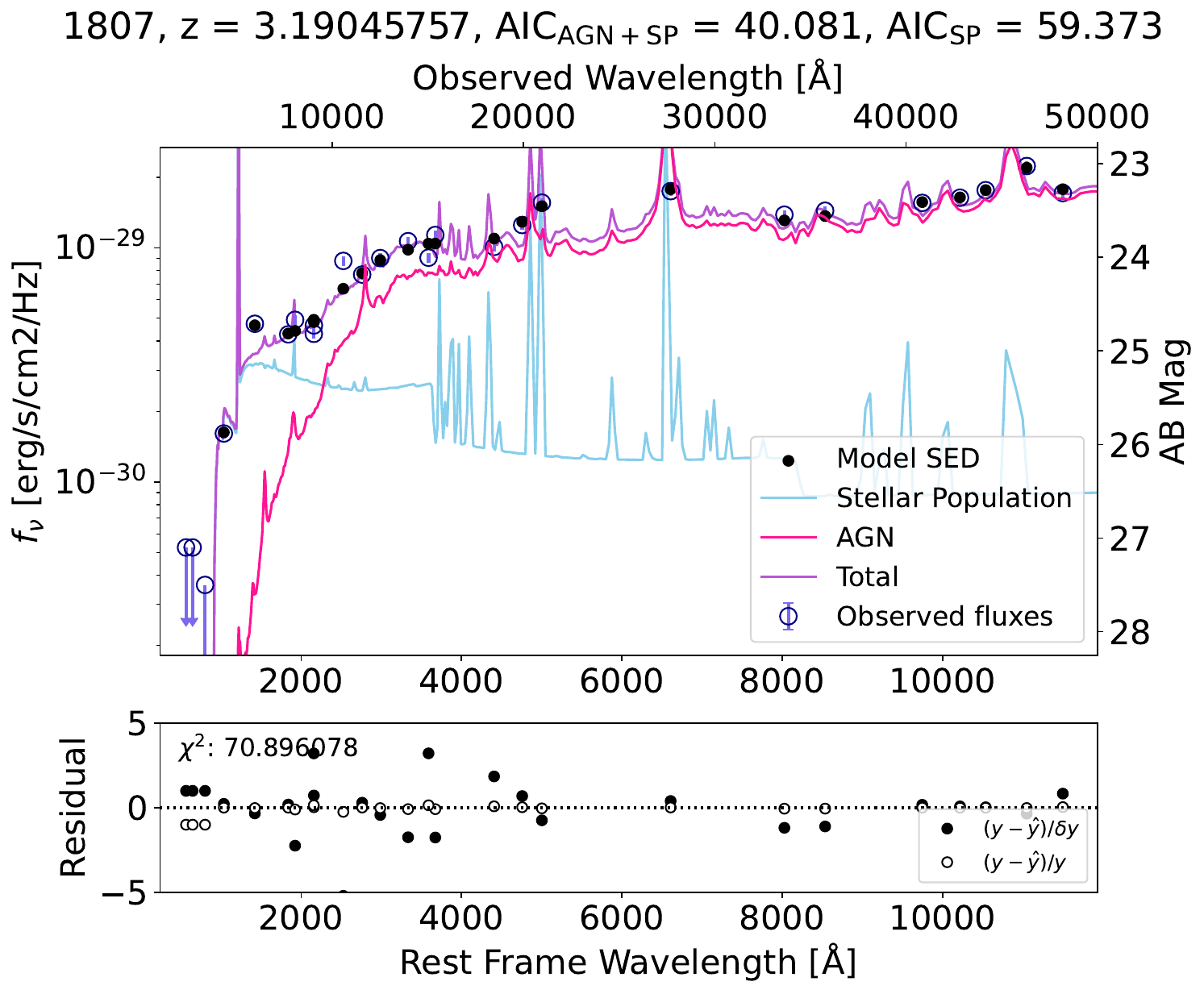}  
    \caption{The same as Figure \ref{fig:exampleSEDs_bad} except for source 1807 at $z\simeq3.19$. This source, however, has an AIC for the combined model which is less than the AIC for the pure stellar population model and the difference between these AIC values is $> 6$, meaning that it is possible to retrieve the AGN nature of this source using our SED fitting. Additionally, this source is a confirmed broad line emitter from NIRISS spectroscopy presented in \cite{Hayes_Glimmers_2024} (see also \cite{pirzkal+2023_NGDEEP_paper}). This provides additional evidence that the SED fitter is able to accurately retrieve AGN contributions to the flux when spectroscopic AGN features are observed. It is also clear that his source has a significant contribution to the flux from the AGN template, reaching 95\% in at least one filter. Therefore, this source also meets our $AGN_{frac}$ condition.  }
    \label{fig:exampleSEDs_good}
    \end{center}
\end{figure*}

Figure \ref{fig:exampleSEDs_bad} and Figure \ref{fig:exampleSEDs_good} show the fitting results for two sources. In particular, Figure \ref{fig:exampleSEDs_bad} presents both the best fit pure stellar population model and the best fit combined model. We can see that while the total $\chi^2$ is smaller when the AGN template is included, the AIC condition is not met by this source. We would therefore not consider that the AGN nature of this source is retrieved by the SED fitter in this case. However, in Figure \ref{fig:exampleSEDs_good} the AIC is drastically reduced for the combined model.  In this instance, a reasonable fit is recovered using a pure stellar model, but when an AGN is included in the fit it almost entirely replaces the stellar SED, and the source shows only an excess of starlight at the blue end. In this case, we consider the AGN nature of the source successfully retrieved by SED fitting.

\subsection{Parameter Estimation}
In the previous subsection we discuss determining the best fit model and using statistics from this fit to identify whether our pure stellar model or our combined model best reproduces the observed photometry for our sources. We note that since our posteriors are expected to be quite complex in seven dimensions, the use of the final \texttt{lmfit} result is essential as optimizing the posterior (as is done with the minimizer) is not equivalent to sampling the posterior (as is the case for the MCMC method) when the posterior distributions are not narrow \citep{Hogg_formanMackey_2018MCMCrecipes}. However, the shape of the posterior becomes important to properly determine the confidence intervals (i.e., the upper and lower limits) on the model parameters of our fits. Therefore, we use the MCMC results for parameter estimation. This means for each source we have both a best fit model which we use to determine the AIC and $AGN_{frac}$ conditions and a most probable model which we use to estimate the values of our free parameters. Namely, for each of our (maximally) seven free parameters, we take the median of the 
posterior distribution to be the most probable solution and determine the errors from the 16th and 84th percentiles. These values are used to determine the stellar masses ($M_{\ast}$) and reddening parameters for each of our sources. We also use the resulting parameters from the MCMC fits to generate an AGN spectrum and measure $L_{5100}$ to estimate $M_{BH}$ for each source. 

We note that the model parameters may vary between the results from the minimizer and those from the MCMC sampler. However, the final values calculated using both sets of parameters (i.e., $M_{BH}$ and $M_{\ast}$) typically agree within errors between both fit results for a given source.

\section{Results}\label{sec:results}

\subsection{Determining AGN contribution from SED fitting}
\begin{figure*}[htb!]
    \begin{center}
    \includegraphics[width=.99\linewidth]{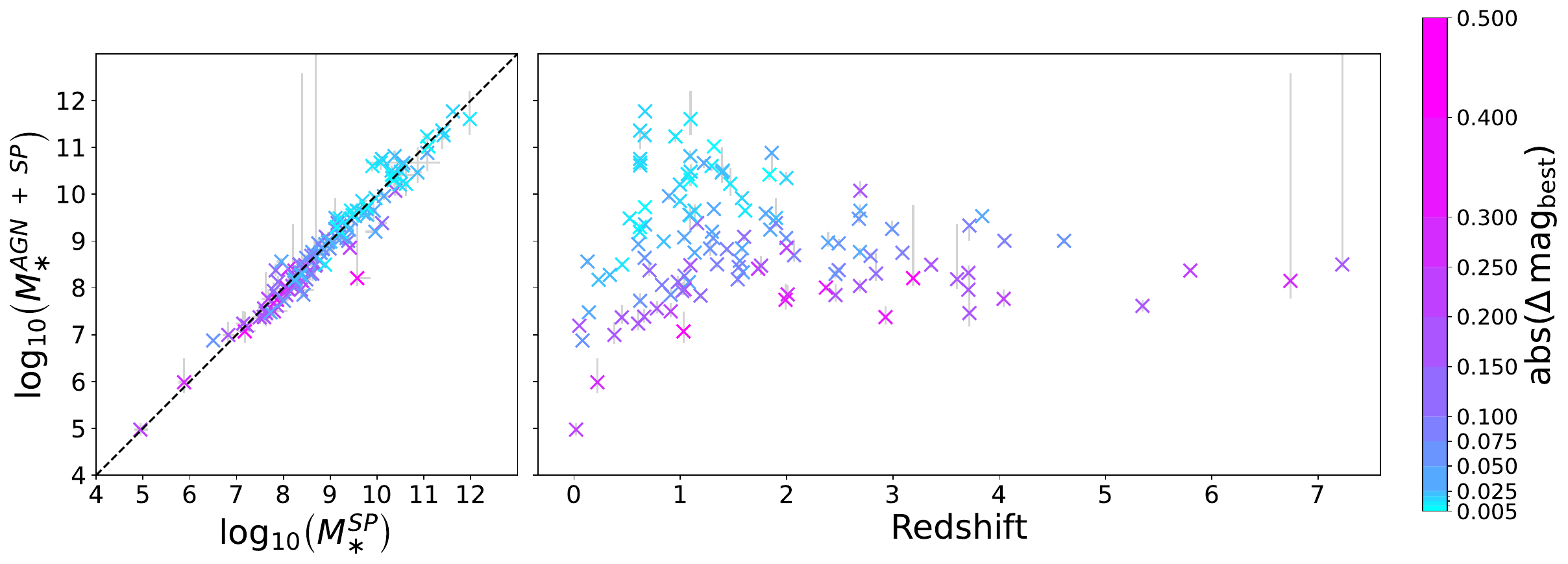}  
    \caption{The output stellar mass from the SED fitter when using our combined model, $M_{*}^{AGN + SP}$, vs. the stellar mass when a pure stellar population is used, $M_{*}^{SP}$, (\textit{left}) and redshift (\textit{right}). The points are coloured by the largest change in AB magnitude between the WFC3/IR epochs in either F105W, F140W or F160W, $\mathrm{abs(\Delta\,mag_{best})}$. $M_{*}^{AGN + SP} = M_{*}^{SP} $ is given by the black dashed line and shows that the stellar masses typically agree well between runs of the SED fitter.}
    \label{fig:massvsredshift}
    \end{center}
\end{figure*}

A summary of the stellar masses which result from fitting our full set of 121 variable sources with both a purely stellar population ($M_{\ast}^{SP}$) as well as our combined model featuring both a best AGN template and a stellar population model ($M_{\ast}^{AGN + SP}$) is depicted in Figure \ref{fig:massvsredshift}. The right panel of Figure \ref{fig:massvsredshift} also shows the stellar mass from our combined model as a function of redshift. Generally, the two sets of models result in stellar masses that agree well between runs of the SED fitter.

The colourbar in Figure \ref{fig:massvsredshift} is a proxy for the variability of the source and is determined by taking the largest change in AB magnitude between two epochs in any of the WFC3/IR filters. The evolution of this variability measurement with stellar mass and redshift is a reflection of the source selection procedure. Here, bright sources which result in larger stellar masses at lower redshifts required a smaller change in flux between epochs to be considered a significant variable in \cite{cammelli_glimmers25}. For sources approaching the limiting magnitudes of the images (resulting in correspondingly lower stellar masses) the difference in flux between epochs needed to be larger to account for the additional uncertainty in the measured flux. The shape of this distribution reflects the completeness of our source selection and represents our ability to probe lower masses near z = 0 than at high redshift for a given magnitude limit. 

\begin{figure}[htb!]
    \begin{center}
    \includegraphics[width=.99\linewidth]{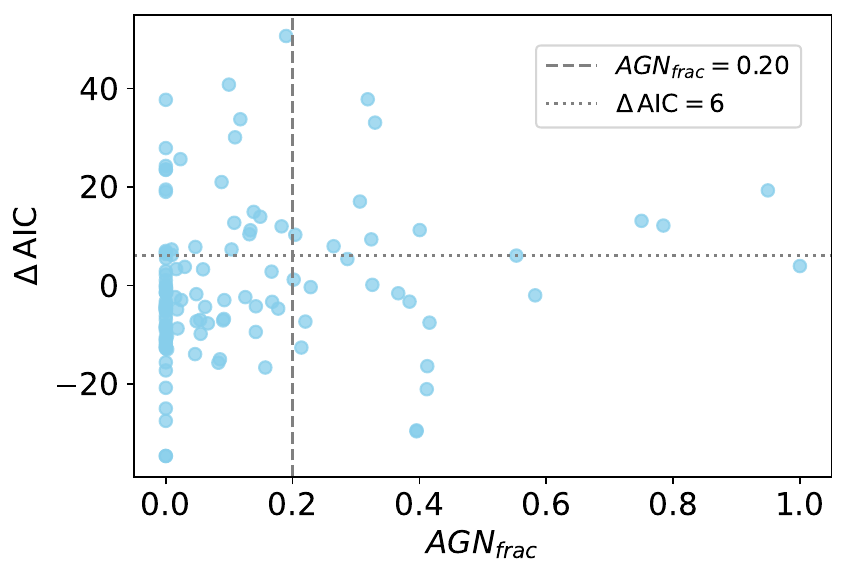}  

    \caption{The difference between the Akaike information criterion (AIC) calculated for the best fit SED modeled with a pure stellar population vs. our combined model with a stellar and AGN component ($\mathrm{\Delta\,AIC = AIC_{SP} - AIC_{AGN+SP}}$) as a function of $AGN_{frac}$. The horizontal grey dotted line indicates the $\mathrm{\Delta\,AIC} = 6$ criterion used to define which sources have SEDs which are statistically improved by including an AGN template in the model. The vertical grey dashed line indicates the $AGN_{frac} = 0.2$ line, the sources to the right of which meet our $AGN_{frac}$ criterion. We calculate $M_{BH}$ for the sources with $AGN_{frac}> 0.2$ since these are the sources which have a contribution to the total flux from the best fit AGN template exceeding 20\% in at least one filter. In total, 37/121 variables meet our AIC criterion, 26/121 meet the $AGN_{frac}$ criterion and 11/121 meet both criteria.}
    \label{fig:aiccomparison}
    \end{center}
\end{figure}

  We present $\Delta\,AIC$ as a function of $AGN_{frac}$ in Figure \ref{fig:aiccomparison}. The horizontal dotted line indicates where $\Delta\,AIC = 6$ meaning sources above this dashed line have a statistically improved SED fit when using the AGN + stellar population model. All sources with an $AGN_{frac} > 20\%$ are located rightward of the vertical grey dashed line in Figure \ref{fig:aiccomparison}. We present the sources meeting our $AGN_{frac}$ criterion in Table \ref{tab:summaryagnfracgt20}.

\begin{table}\label{tab:summaryagnfracgt20}
\centering
    \begin{tabular}{ccccc} \hline \hline
    
 ID & RA & Dec. & z & $z_{type}$ \\ \hline 
2199 & 53.179405 & -27.781833 & 0.22 & p \\
116 & 53.163439 & -27.799619 & 0.6649 & s \\
1059 & 53.157257 & -27.785336 & 0.6676 & s \\
458 & 53.172081 & -27.797478 & 0.71 & p \\
3902 & 53.165592 & -27.763133 & 0.8274 & s \\
3976 & 53.165344 & -27.762849 & 1.03 & p \\
222 & 53.156769 & -27.795656 & 1.0966 & s \\
1441 & 53.160980 & -27.786578 & 1.1348 & s \\
3404 & 53.166354 & -27.768660 & 1.2954 & s \\
1975 & 53.184353 & -27.783443 & 1.3458 & s \\
1013 & 53.160445 & -27.790442 & 1.6098 & s \\
3752 & 53.164102 & -27.765497 & 1.76 & p \\
2156 & 53.155676 & -27.779348 & 1.8391 & s \\
3005 & 53.159608 & -27.774646 & 1.9 & p \\
1775 & 53.174344 & -27.782600 & 1.9982 & s \\
446 & 53.166077 & -27.798129 & 2.01 & p \\
2460 & 53.144954 & -27.779554 & 2.68 & p \\
1540 & 53.180808 & -27.786333 & 2.6900 & s \\
2810 & 53.183466 & -27.776658 & 2.6933 & s \\
2774 & 53.183386 & -27.776454 & 2.6933 & s \\
2810 & 53.183466 & -27.776658 & 2.6933 & s \\
2774 & 53.183386 & -27.776454 & 2.6933 & s \\
45 & 53.161165 & -27.806155 & 2.9928 & s \\
1807 & 53.178511 & -27.784108 & 3.1905 & s \\
3384 & 53.166094 & -27.771261 & 3.6040 & s \\
258 & 53.152758 & -27.800964 & 3.72 & p \\
316 & 53.151454 & -27.799838 & 4.61 & p \\
1511 & 53.161944 & -27.787065 & 6.74 & p \\
\hline     
    \end{tabular}
    \caption{A summary of the variable sources meeting our $AGN_{frac}$ condition. Here, $z_{type}$ indicates whether the source has an available spectrscopic redshift (s) or if only a photometric redshift (p) is available. }
    
\end{table}

In total, 37/121 variables meet our AIC criterion, 26/121 meet the $AGN_{frac}$ criterion and 11/121 meet both criteria. Sources which do not meet our AIC criterion are not necessarily misidentified as AGN in our sample selection. This instead reflects the well-known challenge of identifying AGN purely from their SEDs: in these instances we simply cannot demonstrate statistical evidence for the presence of an AGN from the SED data alone. 

In the following analysis we determine BH properties for the sources which meet the $AGN_{frac}$ criterion. Due to the fact that we are assuming all of our 2.5 $\sigma$ variables are, indeed, varying due to the presence of an AGN, we selected the sample which meet the $AGN_{frac}$ criterion as the SED fitter was able to successfully retrieve a significant AGN contribution to the total flux for these sources.
\subsection{Black hole mass estimation}
Due to the decomposition of our total SED into separate stellar and AGN contributions, we are able to use the MCMC results to reconstruct our model AGN spectrum to determine $L_{5100}$ by taking the median flux between 4850 and 5350 \AA. The calculated $L_{5100}$ can then be used with the relation defined in \cite{kaspi_2000_MBH} for black hole mass as a function of 5100\AA \  luminosity. To our knowledge this is the first time that photometric decomposition of stellar and AGN contributions has been leveraged to estimate SMBH masses, and is facilitated in this case by the exquisite wide bandwidth, 26-filter photometry in the HUDF.  
\cite{kaspi_2000_MBH} define this relation by fitting the mass estimates from reverberation mapping (RM) analysis for the mean spectra of a sample of 34 AGN. This gives:
\begin{equation}
    \label{eq:mbh}
    M_{BH} = 5.71\,\times\, 10^7 \left(\frac{L_{5100}}{10^{44}\,ergs\,s^{-1}}\right)^{0.545} M_{\odot}.
\end{equation}
 To estimate the errors on $M_{BH}$ we use the 16th and 84th percentiles from the MCMC results for the $norm_{agn}$ parameter since the luminosity we measure from the model spectrum will be proportional to its normalization. We do not include the intrinsic scatter of the relation in our error propagation. The tabulated results for the various AGN criteria, $L_{5100}$, $M_{BH}$ and $M_{*}$ are presented in Table \ref{tab:resultssedfitter}. 

\begin{figure*}[htb!]
    \begin{center}
    \includegraphics[width=.99\linewidth]{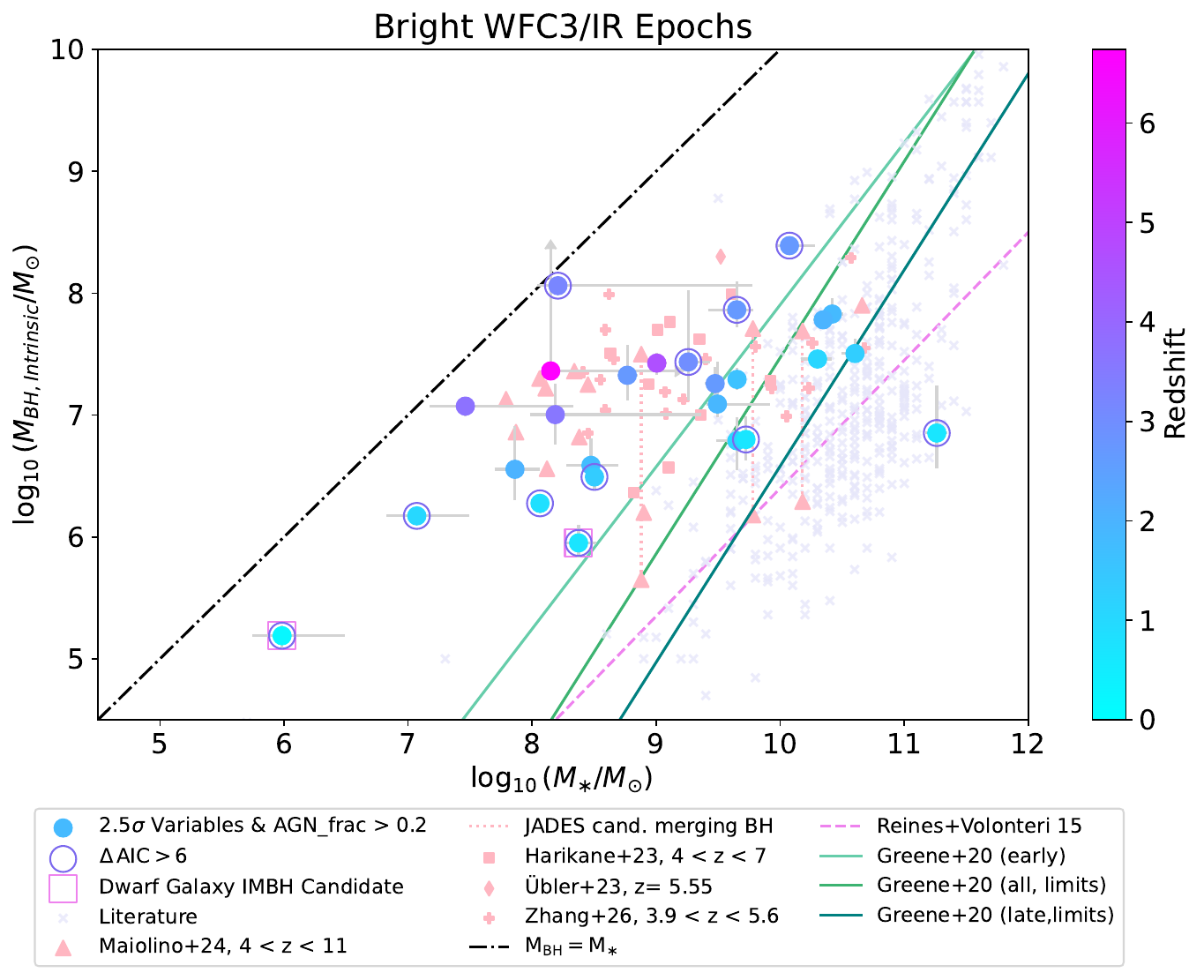}  
    \caption{Black hole masses calculated from the \cite{kaspi_2000_MBH} relation utilising the luminosity at 5100\AA\ in the AGN spectrum resulting from the parameters determined from the MCMC SED fit as a function of stellar mass for each variable meeting the $AGN_{frac}$ condition. Sources which also meet our $\Delta\, AIC$ condition are outlined by purple circles. Sources are coloured by redshift and the $L_{5100}$ values used to calculate $M_{BH}$ are determined from spectra which have undergone a reddening correction calculated from the MCMC results for $E(B-V)_{diffuse}$ and $E(B-V)_{AGN}$ as well as IGM absorption along the line of sight. The scaling relation between black hole mass and stellar mass determined from a sample of 262 $z < 0.055$ BL-AGN from \cite{Reines_Volonteri_2015} is shown as the pink dashed line as well as similar scaling relations from \cite{Greene_IMBHreview_2020} in shades of green. The 1:1 line for $M_{BH}$ and $M_{\ast}$ is plotted as the dashed-dotted black line. Low-redshift objects with values for black hole mass and stellar mass from the literature are included from the table compiled by \cite{Zaw_lowzMbhMstellar_catalogue_2020}. These are marked by the lilac ``x'' symbols. Finally, we offer a direct comparison to black hole masses determined from fitting broad H$\alpha$ emission lines from \cite{Maiolino_2024} (\textit{pink triangles}), \cite{Harikane_23MBH} (\textit{pink squares}), \cite{Ubler_23MBH} (\textit{pink diamonds}) and \cite{zhang2025_SCJCMBH} (\textit{pink crosses}). Our two IMBH candidates identified in dwarf galaxies are marked by the pink boxes.}
    \label{fig:bhmassvsstellarmass_single}
    \end{center}
\end{figure*}
\begin{figure*}[htb!]
    \begin{center}
    \includegraphics[width=.99\linewidth]{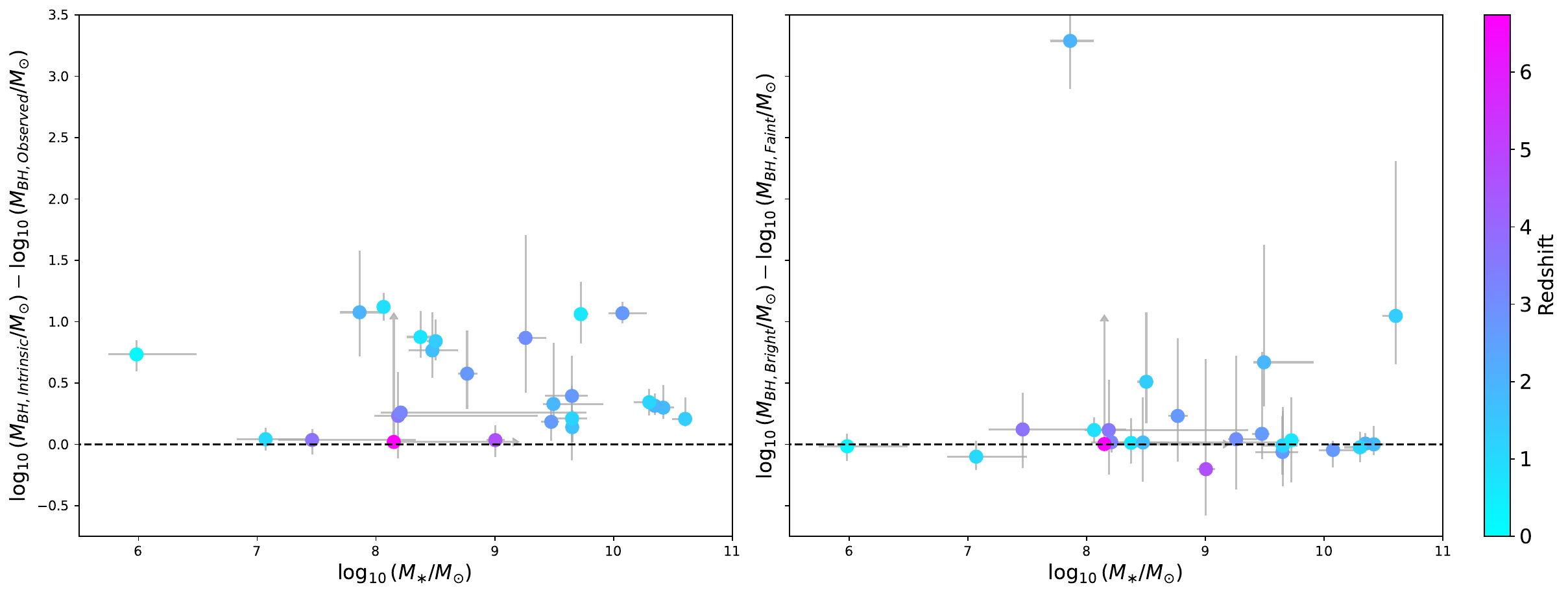}  
    \caption{\textit{Left:} The difference in $M_{BH}$ resulting from the SED model AGN spectra when corrected for reddening and IGM absorption ($M_{BH,\ Intrinsic}$) and the $M_{BH}$ resulting from the uncorrected spectra ($M_{BH,\ Observed}$) as a function of $M_{\ast}$. The sources which have large deviations from the black $M_{BH,\ Intrinsic} = M_{BH,\ Observed}$ line demonstrate the importance of accurately correcting for reddening when determining $M_{BH}$. \textit{Right:} The difference in the $M_{BH}$ calculated when we run our SED fitter on the set of photometry containing WFC3/IR fluxes from the epochs where the AGN is bright ($M_{BH,\ Bright}$) and the $M_{BH}$ resulting from the SED run when the faint WFC3/IR epochs are used ($M_{BH,\ Faint}$) as a function of $M_{\ast}$. Large deviations from the $M_{BH,\ Bright} = M_{BH,\ Faint}$ line show that the intrinsic variability of the AGN can have a significant impact on the estimated $M_{BH}$.}
    \label{fig:bhmassvsstellarmass_subcompare}
    \end{center}
\end{figure*}
\subsection{Source 1511}
Our highest redshift variable source meeting our $AGN_{frac}$ condition has been previously studied in multiple works and in this section we will discuss our results in the context of previous literature.

Source 1511 was originally presented in \cite{Hayes_Glimmers_2024} as source 1052123. It has a photometric redshift of 6.74 and interestingly, our SED fitter finds that the photometry from this source is consistent with a contribution almost entirely dominated by a best fit \cite{Temple_2021} AGN template. \cite{decoursey2025_1511stellar} report a spectrum from DDT Program 6541 (PI: E. Egami) for this source that they claim is not well matched by any supernova, galaxy or AGN template at $5 < z < 7$, and is instead fit best by a late-type star at $z \sim 0$. Our SED fitting results and photometry from 26 filters does appear to be consistent with an AGN template, so we do not remove this source from our sample. Using this best fit results, we retrieve an $AGN_{frac}=1$. This is slightly lower when we instead consider our MCMC results, from which we derive a lower limit on the black hole mass of $\log_{10}(M_{BH}/M_{\odot}) > 7.36$. The stellar mass is not well constrained by the SED fitter and we again provide a lower limit of $\log_{10}(M_{*}/M_{\odot}) > 8.15$. This source is likely best described by a bright AGN with very little contribution to its total flux from any stellar population. Given that we do not include individual stellar templates in our suite of SED models and that the errors on the resulting MCMC parameters remain large, we cannot rule out the stellar classification from \cite{decoursey2025_1511stellar}, however, we treat 1511 as a $z_{phot} \simeq 6.74$ AGN candidate in the remainder of the analysis in this work.

\begin{table*}
\centering
  \setlength{\tabcolsep}{2pt}
    \begin{rotatetable}
    \begin{tabular}{ccccccccc} \hline \hline
    
 ID &$AGN_{frac}$ & $AIC_{AGN + SP}$ & $AIC_{SP}$ & $L_{5100}\ [\mathrm{erg\, s^{-1}}]$ & $\log(M_{BH, int}/M_{\odot})$  & $\log(M_{*}/M_{\odot})$& $E({B-V})_{diffuse}$ & $E({B-V})_{agn}$ \\ \hline 
2199 & \num{0.75} & \num{13.39} & \num{26.50} & \num{1.95E+39}  &$ 5.19 _{-0.10}^{+0.08}$  & $5.98 _{-0.24}^{+0.51}$  &$ 0.07 _{-0.05}^{+0.07}$  &$ 0.90 _{-0.63}^{+0.64}$ \\
116 & \num{0.32} & \num{-8.06} & \num{29.74} & \num{2.19E+42}  &$ 6.85 _{-0.29}^{+0.39}$  & $11.26 _{-0.09}^{+0.09}$  &$ 0.06 _{-0.03}^{+0.03}$  &$ 0.61 _{-0.41}^{+0.94}$ \\
1059 & \num{0.32} & \num{12.57} & \num{21.93} & \num{1.75E+42}  &$ 6.80 _{-0.17}^{+0.19}$  & $9.72 _{-0.06}^{+0.07}$  &$ 0.43 _{-0.03}^{+0.04}$  &$ 0.88 _{-0.62}^{+0.73}$ \\
458 & \num{0.26} & \num{46.35} & \num{54.31} & \num{4.87E+40}  &$ 5.95 _{-0.12}^{+0.15}$  & $8.37 _{-0.11}^{+0.14}$  &$ 0.08 _{-0.05}^{+0.05}$  &$ 1.08 _{-0.77}^{+0.65}$ \\
3902 & \num{0.40} & \num{46.47} & \num{57.71} & \num{1.92E+41}  &$ 6.28 _{-0.08}^{+0.08}$  & $8.06 _{-0.04}^{+0.04}$  &$ 0.04 _{-0.02}^{+0.02}$  &$ 1.46 _{-0.67}^{+0.40}$ \\
3976 & \num{0.78} & \num{27.27} & \num{39.43} & \num{1.24E+41}  &$ 6.17 _{-0.06}^{+0.07}$  & $7.07 _{-0.24}^{+0.43}$  &$ 0.05 _{-0.03}^{+0.06}$  &$ 0.00 _{-0.00}^{+0.00}$ \\
222 & \num{0.40} & \num{35.30} & \num{5.88} & \num{2.85E+43}  &$ 7.46 _{-0.08}^{+0.08}$  & $10.30 _{-0.13}^{+0.12}$  &$ 0.37 _{-0.02}^{+0.02}$  &$ 0.00 _{-0.00}^{+0.00}$ \\
1441 & \num{0.20} & \num{39.55} & \num{40.73} & \num{1.68E+42}  &$ 6.79 _{-0.24}^{+0.19}$  & $9.65 _{-0.17}^{+0.13}$  &$ 0.21 _{-0.03}^{+0.02}$  &$ 0.02 _{-0.02}^{+0.08}$ \\
3404 & \num{0.41} & \num{63.98} & \num{47.59} & \num{3.45E+43}  &$ 7.50 _{-0.05}^{+0.12}$  & $10.60 _{-0.11}^{+0.05}$  &$ 0.20 _{-0.02}^{+0.02}$  &$ 0.02 _{-0.02}^{+0.96}$ \\
1975 & \num{0.33} & \num{10.47} & \num{43.53} & \num{4.79E+41}  &$ 6.49 _{-0.11}^{+0.12}$  & $8.50 _{-0.07}^{+0.06}$  &$ 0.07 _{-0.07}^{+0.03}$  &$ 1.04 _{-0.48}^{+0.57}$ \\
1013 & \num{0.40} & \num{47.95} & \num{18.35} & \num{1.41E+43}  &$ 7.29 _{-0.09}^{+0.10}$  & $9.65 _{-0.06}^{+0.04}$  &$ 0.10 _{-0.02}^{+0.03}$  &$ 0.06 _{-0.05}^{+0.95}$ \\
3752 & \num{0.33} & \num{35.67} & \num{35.80} & \num{7.15E+41}  &$ 6.59 _{-0.16}^{+0.22}$  & $8.47 _{-0.20}^{+0.22}$  &$ 0.06 _{-0.04}^{+0.07}$  &$ 0.96 _{-0.85}^{+0.70}$ \\
2156 & \num{0.38} & \num{23.83} & \num{20.55} & \num{1.36E+44}  &$ 7.83 _{-0.06}^{+0.13}$  & $10.42 _{-0.10}^{+0.09}$  &$ 0.29 _{-0.02}^{+0.02}$  &$ 0.03 _{-0.03}^{+0.41}$ \\
3005 & \num{0.42} & \num{28.97} & \num{21.42} & \num{5.99E+42}  &$ 7.09 _{-0.11}^{+0.35}$  & $9.49 _{-0.09}^{+0.42}$  &$ 0.14 _{-0.09}^{+0.04}$  &$ 0.27 _{-0.25}^{+1.36}$ \\
1775 & \num{0.41} & \num{32.65} & \num{11.58} & \num{1.11E+44}  &$ 7.78 _{-0.05}^{+0.07}$  & $10.35 _{-0.08}^{+0.12}$  &$ 0.34 _{-0.02}^{+0.03}$  &$ 0.00 _{-0.00}^{+0.01}$ \\
446 & \num{0.22} & \num{-13.76} & \num{-21.09} & \num{6.24E+41}  &$ 6.55 _{-0.25}^{+0.36}$  & $7.86 _{-0.17}^{+0.20}$  &$ 0.16 _{-0.08}^{+0.08}$  &$ 1.25 _{-0.75}^{+0.55}$ \\
2460 & \num{0.29} & \num{35.23} & \num{40.59} & \num{1.22E+43}  &$ 7.26 _{-0.11}^{+0.14}$  & $9.48 _{-0.08}^{+0.07}$  &$ 0.14 _{-0.02}^{+0.02}$  &$ 0.07 _{-0.05}^{+0.11}$ \\
1540 & \num{0.21} & \num{76.53} & \num{63.92} & \num{1.62E+43}  &$ 7.33 _{-0.20}^{+0.25}$  & $8.77 _{-0.08}^{+0.09}$  &$ 0.04 _{-0.03}^{+0.04}$  &$ 0.72 _{-0.35}^{+0.38}$ \\
2810 & \num{0.55} & \num{42.22} & \num{48.29} & \num{1.45E+45}  &$ 8.39 _{-0.06}^{+0.06}$  & $10.07 _{-0.12}^{+0.20}$  &$ 0.73 _{-0.04}^{+0.03}$  &$ 0.52 _{-0.11}^{+0.20}$ \\
2774 & \num{0.31} & \num{-2.67} & \num{14.37} & \num{1.57E+44}  &$ 7.86 _{-0.14}^{+0.23}$  & $9.65 _{-0.23}^{+0.13}$  &$ 0.24 _{-0.07}^{+0.03}$  &$ 0.23 _{-0.12}^{+0.10}$ \\
2810 & \num{0.55} & \num{42.22} & \num{48.29} & \num{1.45E+45}  &$ 8.39 _{-0.06}^{+0.06}$  & $10.07 _{-0.12}^{+0.20}$  &$ 0.73 _{-0.04}^{+0.03}$  &$ 0.52 _{-0.11}^{+0.20}$ \\
2774 & \num{0.31} & \num{-2.67} & \num{14.37} & \num{1.57E+44}  &$ 7.86 _{-0.14}^{+0.23}$  & $9.65 _{-0.23}^{+0.13}$  &$ 0.24 _{-0.07}^{+0.03}$  &$ 0.23 _{-0.12}^{+0.10}$ \\
45 & \num{0.20} & \num{46.59} & \num{56.88} & \num{2.57E+43}  &$ 7.44 _{-0.32}^{+0.59}$  & $9.26 _{-0.07}^{+0.18}$  &$ 0.04 _{-0.02}^{+0.06}$  &$ 1.12 _{-0.95}^{+0.73}$ \\
1807 & \num{0.95} & \num{40.08} & \num{59.37} & \num{3.62E+44}  &$ 8.06 _{-0.03}^{+0.02}$  & $8.21 _{-0.17}^{+1.56}$  &$ 0.03 _{-0.02}^{+0.02}$  &$ 0.31 _{-0.02}^{+0.02}$ \\
3384 & \num{0.23} & \num{2.93} & \num{2.58} & \num{4.15E+42}  &$ 7.00 _{-0.24}^{+0.25}$  & $8.19 _{-0.20}^{+1.18}$  &$ 0.04 _{-0.03}^{+0.06}$  &$ 0.26 _{-0.25}^{+1.11}$ \\
258 & \num{0.58} & \num{32.71} & \num{30.71} & \num{5.57E+42}  &$ 7.07 _{-0.08}^{+0.06}$  & $7.46 _{-0.29}^{+0.87}$  &$ 0.02 _{-0.02}^{+0.02}$  &$ 0.02 _{-0.01}^{+0.03}$ \\
316 & \num{0.37} & \num{46.03} & \num{44.45} & \num{2.47E+43}  &$ 7.43 _{-0.10}^{+0.08}$  & $9.00 _{-0.07}^{+0.08}$  &$ 0.02 _{-0.01}^{+0.02}$  &$ 0.02 _{-0.01}^{+0.02}$ \\
1511 & \num{1.00} & \num{26.97} & \num{30.91} & \num{1.89E+43}  &$ >7.36$  & $>8.15$  &$ 0.01 _{-0.01}^{+0.73}$  &$ 0.01 _{-0.01}^{+0.56}$ \\
\hline     

    \end{tabular}
    \caption{A summary of the variable sources meeting our $AGN_{frac}$ condition and the relevant quantities determined from our SED fitting code for the BH and host galaxy. Here $L_{5100}$ refers to the luminosity at 5100\,\AA\ measured from the corrected spectra which is used to determine the reported $M_{BH, int}$ values. All values reported except $AIC_{SP}$ are determined from the results of the SED model corresponding to our combined (AGN + SP) model using the set of photometry which includes the measurements taken at the bright epochs in the WFC3/IR filters. $AIC_{SP}$ is reported from the SED run which modeled the same photometry, however, using only the stellar population model.}
    \end{rotatetable}
    \label{tab:resultssedfitter}
\end{table*}

\section{Discussion}\label{sec:discussion}
\subsection{Mass calculations with $L_{5100}$}
As shown in Figure \ref{fig:bhmassvsstellarmass_single} and summarized in Table \ref{tab:resultssedfitter}, our modeling recovers black hole masses which range from the intermediate mass regime at $\sim 10^5\ M_{\odot}$ through to the supermassive at $\gtrsim 10^8\ M_{\odot}$.

\cite{kaspi_2000_MBH} detail many caveats that enter into the determination of $M_{BH}$ using the scaling relations presented therein, including the assumptions that are necessary for deriving physical parameters from the properties of Balmer emission lines. However, additional considerations are also relevant when applying this relation to our sample specifically. Firstly, the population studied here is not representative of the sample for which the relation was derived. In particular, we probe sources with systematically fainter $L_{5100}$ than \cite{kaspi_2000_MBH}. Our median value is $\sim 10^{42.9}\ \mathrm{erg\ s^{-1}}$, and we probe down to $L_{5100}\sim 10^{39}\ \mathrm{erg\ s^{-1}}$ versus a median of $\sim 10^{44.2}\ \mathrm{erg\ s^{-1}}$ from \cite{kaspi_2000_MBH} where the source with the lowest $L_{5100}\sim10^{41}\ \mathrm{erg\ s^{-1}}$. \cite{kaspi_2000_MBH} does warn that using the relations dependent on luminosity determinations for fainter sources would lead to increased contributions from the host galaxy light. However, to first order this effect should be mitigated by our approach of decomposing the stellar and AGN light in the SED modeling.  
Additionally, many sources in our sample are also at significantly higher redshift than the z $<1$ sample from \cite{kaspi_2000_MBH}. It is not yet known whether the relations that can be used to determine black hole masses at low redshift are accurate also at high redshift \citep{Ubler_23MBH,Maiolino_2024}. Further, although it was found that $M_{BH}$ and $L_{5100}$ had a Pearson linear correlation coefficient of 0.646 at a significance level of $3.7 \times 10^{-5}$, Eq. \ref{eq:mbh} is determined from a low-number sample featuring significant scatter. \cite{KormendyHo_2013} also emphasise that all mass determinaitons based on virialised motion in the BLR depend on the scaling of masses such that they obey the robust $M_{BH}-\sigma_{\ast}$ relation. This is a shortcoming of this method as the $M_{BH}-\sigma_{\ast}$ relations used were derived incorporating sources featuring both classical and pseudobulges, despite $M_{BH}$ not correlating with pseudobulges \citep{KormendyHo_2013}. Depending on the galaxy sample used to define the $M_{BH}-\sigma_{\ast}$ sample, this can introduce an uncertainty up to a factor of two \citep{Ho_2016,Greene_IMBHreview_2020}. Finally, we are applying this function to a model spectrum resulting from photometric SED analysis, rather than an observed spectrum. 

With this in mind, the resulting black hole mass estimates for our variable sources are consistent with those determined in other works at similar redshifts which utilise the luminosity and FWHM of the broad component of $\mathrm{H\alpha}$ (see comparison to JWST high-z BL-AGN masses in Figure \ref{fig:bhmassvsstellarmass_single}). 

The black hole masses estimated here are also consistent with the trend towards overmassive BHs with increasing redshift frequently observed with JWST detected high-z SMBHs. This is despite the fact we use a different method to determine the black hole mass which, in this case, is independent of Balmer emission line profiles.

Additionally, we are able to directly compare the black hole mass calculated using $L_{5100}$ to that using the luminosity and FWHM of H$\beta$ for one source, namely source 1807. The best fit SED for this source is shown in the right panel of Figure \ref{fig:exampleSEDs_good} and it is identified as a BL-AGN in a NIRISS spectrum \cite[see also \citealt{pirzkal+2023_NGDEEP_paper}]{Hayes_Glimmers_2024}. We find an H$\beta$ luminosity ($L_{H\beta}$) of 9.5$\times 10^{41}\ \mathrm{erg\ s^{-1}}$ with a $S/N \sim 8$ and a FWHM ($FWHM_{H\beta}$) of $\simeq 3600\pm 500\ \mathrm{km\ s^{-1}}$. With these measurements, we apply the relation from \cite{Greene_2005} which is given as:
\begin{equation}
\begin{split}
M_{BH} & = 2.4 \times 10^{6}\left(\frac{L_{H\beta}}{10^{42}\, \mathrm{erg\ s}^{-1}}\right)^{0.59} \\
& \times \left(\frac{FWHM_{H\beta}}{10^3\mathrm{km/s}}\right)^2\ M_{\odot} .
\end{split}
\end{equation}
Using this relation we find $\log_{10}(M_{BH}/M_{\odot}) = 7.48 \pm 0.13$ if we consider the uncertainties in our measured $H\beta$ line properties, but excluding the intrinsic scatter from the scaling relation. This is compared to the value we retrieve using $L_{5100}$ which gives $\log_{10}(M_{BH}/M_{\odot}) = 8.06^{+0.02}_{-0.03}$. This means that the $M_{BH}$ calculated for this source using these two methods agree by a factor of $\sim3.5$. Differences in calculated $M_{BH}$ are expected between alternate methods \citep{Korista_2004, AbuterGRAVITY_2024, Dalla_Bont_2025compGRAVtoSE}. \cite{Dalla_Bont_2025compGRAVtoSE} examined the comparison between single epoch methods to determine $M_{BH}$ and compared them to the results using interferometry to spatially resolved the rotation of broad line regions using the GRAVITY instrument on the ESO VLTI (e.g., \cite{gravityquasarmass_2018}). The level of agreement between our method using $L_{5100}$ and the line properties of H$\beta$ are at a similar level to the agreement in $M_{BH}$ calculated for the quasar 3C 273 using the luminosity and FWHM of H$\beta$ which gave $\log_{10}(M_{BH}/M_{\odot}) = 9.383 \pm 0.372$ \citep{kaspi_2000_MBH} and that determined from GRAVITY, $\log_{10}(M_{BH}/M_{\odot}) = 8.41 \pm 0.18$ \citep{gravityquasarmass_2018} when the errors on the measurements are considered. The two masses calculated for 1807 also show a similar level of agreement as was found for 3C 273 when comparing the masses retrieved using H$\beta$ and H$\alpha$ ($\log_{10}(M_{BH,H\alpha}/M_{\odot}) = 9.403 \pm 0.328$, from \cite{kaspi_2000_MBH}). 

Although we can only make this comparison for a single source, the agreement found between the $M_{BH}$ determined using these separate methods for source 1807 suggests that masses calculated from the $L_{5100}$ relation in \cite{kaspi_2000_MBH} can be used to provide reliable estimates for the black hole masses of our variable sources.
\subsection{The $M_{BH}-M_*$ relation}
We plot the 26 sources meeting our $AGN_{frac}$ condition in the $M_{BH}-M_*$ plane in Figure \ref{fig:bhmassvsstellarmass_single}. Sources from this work are compared to the corresponding BH and stellar masses determined for low-z AGN from the catalogue compiled by \cite{Zaw_lowzMbhMstellar_catalogue_2020} as well as the masses calculated for the BL-AGN at $4 < z < 7$ from recent studies utilising data from JWST \citep{Maiolino_2024,Harikane_23MBH,Ubler_23MBH,zhang2025_SCJCMBH}. The scaling relation derived from local BL-AGN at $z < 0.055$ from \cite{Reines_Volonteri_2015} as well as those from \cite{Greene_IMBHreview_2020} are also included. It is clear that our sources agree with the trends observed by other recent studies probing high redshift SMBH masses showing that black holes appear to become overmassive with respect to the stellar mass of their host galaxies at $z \gtrsim 4$ (e.g., \cite{Ubler_23MBH, Maiolino_2024, Harikane_23MBH, zhang2025_SCJCMBH}). The agreement of trends in SMBH masses determined in this work and those from recent works utilising JWST observations is additionally interesting due to the fact that we have used a method independent of those used in other studies. The other works all utilise single epoch spectroscopy to derive black hole masses based on the width of broad lines, whereas our method uniquely relies on the decomposition of our SEDs into their component parts, allowing us to use only the luminosity at 5100\,\AA\  determined from our SED model AGN spectra. The works utilising the properties of broad lines typically utilise the scaling relations originally derived in \cite{Greene_2005} which translates the virial mass system from \cite{kaspi_2000_MBH} derived using reverberation mapping into a system that depends only on observations of broad $H\alpha$. These results are often compared to the scaling relations derived from \cite{Reines_Volonteri_2015} which follows the approach outlined in \cite{Greene_2005} to define a relation for their sample of SDSS broad $H\alpha$ emitters. All of which, therefore, assume the BLR is in virial equilibrium and that the relations for $M_{BH}$ which appear to hold at low redshift are also applicable at high redshift \citep{Ubler_23MBH, Maiolino_2024}. These are also assumptions intrinsic to the scaling relation with $L_{5100}$ from \cite{kaspi_2000_MBH} since the masses originally used to produce this scaling relation were calculated from broad emission lines. Previous studies have reported systematic uncertainties in the relations used to determine $M_{BH}$ from single epoch measurements leading to mass disparities up to factors of $\sim 1.5 - 2.25$ \citep{SDSS_signleepoch_2020,AbuterGRAVITY_2024,Dalla_Bont_2025compGRAVtoSE}.

As mentioned by previous works, black hole masses at higher redshifts being more overmassive than those at low redshift could imply more rapid growth of central SMBHs at early cosmological times. We examine the implication this could have on seeding mechanisms for black holes in the early universe further in Section \ref{subsec:seeds}. 
\subsubsection{The effects of reddening on black hole mass}
Due to the nature of our SED fitting code which works on the premise of spectral decomposition, we are able to fit two extinction parameters. One applies globally to the stellar population and the central AGN, $E(B-V)_{diffuse}$ and the other which is a nuclear reddening term that is only applied to the AGN spectrum,  $E(B-V)_{AGN}$. This means we are able to accurately correct for the reddening at each wavelength of our AGN spectrum using these reddening values and the Calzetti \citep{calzetti_2000} and SMC \citep{Prevot_SMC_1984} extinction laws, respectively.

In the left panel of Figure \ref{fig:bhmassvsstellarmass_subcompare} we plot the difference between the logarithm of the black hole masses we estimate when the AGN spectra have been corrected for reddening and IGM absorption ($M_{BH,\ Intrinsic}$) and the masses calculated from the un-corrected spectra ($M_{BH,\ Observed}$). Our results show that when estimating the mass from the uncorrected luminosity at 5100\AA\ the masses can be underestimated by factors of up to 10, with a standard deviation on the difference between estimates of $\sim0.36$. This indicates the potential importance of applying accurate reddening corrections in order to estimate black hole masses. An additional consideration is that the relations derived by \cite{kaspi_2000_MBH} use spectrophotometry that was not corrected for reddening, and it is unlikely that all of their RM sources are dust-free.  This obscuration is a possible contributor to the large scatter that is seen on single-epoch luminosity--$M_\mathrm{BH}$ calibrations, which we here estimate to be on the order of a factor of $\simeq 3$, albeit in a sample much fainter than those used to derive the calibrations.  

\subsubsection{The effects of variability on black hole mass} \label{subsec:variabilityeffects}
As well as running two different sets of models for each source we also test two sets of photometry. Since we have three WFC3/IR filters each with two epochs we construct two sets of photometric catalogues. The first contains F105W/F140W/F160W fluxes from the epochs where each source is detected to be faint, while the second includes the fluxes from the epochs where the sources are bright. 

By comparing the estimated black hole masses for our sample using the SED fits performed on the two sets of photometry we can examine the effects of variability on our SED fitting results. Of course, the black hole masses between epochs must, in reality, be almost entirely unchanged due to the small time lapse between observations. Thus, changes observed in the calculated masses reflect the ability of SED modeling to retrieve an AGN contribution to the total flux when the AGN is observed in its faint state. We compare the difference in $M_{BH}$ calculated from SED results performed on these two sets of photometry in the right panel of Figure \ref{fig:bhmassvsstellarmass_subcompare}. In many cases, it is observed that, without the bright epochs from WFC3/IR, the recovered black hole masses are significantly smaller, and the large error bars resulting from adding the two $M_{BH}$ errors in quadrature indicate that the black hole mass is not well constrained by the SED fitting when the faint epoch photometry is used. This suggests that the bright phase of the variability is necessary for these sources to be recovered as AGN using our AIC and $AGN_{frac}$ conditions. Therefore, it can be understood that it may be difficult to recover AGN via SED fitting without considering variability and that if, by chance, the AGN is observed only when it is in its faint state the SED may falsely underestimate the AGN contribution to the total flux. It also highlights the fact that variability is an important tool for identifying AGN as a significant fraction of the AGN identified via variability did not meet the requirements of our AIC and $AGN_{frac}$ conditions to be correctly identified as AGN from SED fitting alone.

\subsection{Dwarf galaxies with overmassive black holes}
Dwarf galaxies hosting AGN offer a unique opportunity to gain further information about the population of BHs at low redshift overall. In particular, these lower mass BHs are important for probing the overall number density of BHs, understanding the BH occupation fraction and can provide powerful constraints for the BH mass function. All of which can inform models which seek to predict the formation and co-evolution of BHs in galaxies accross cosmic time.

Interestingly, two of our sources at $z \lesssim 1$ occupy the dwarf galaxy regime with $M_{*} < 3\times10^9 M_{\odot}$ and have $M_{BH} < 10^6  M_{\odot}$. We classify these objects as intermediate mass black hole (IMBH) candidates and see that they are also overmassive BHs with respect to the calculated stellar masses of their hosts. This is consistent with literature which has found that black holes residing in dwarf galaxies occupy a similar region of the $M_{BH} -M_{*}$ plane as the high-z SMBH discovered by JWST \citep{Burke_dwarfgals_2022, Mezcua_dwarf_2023, Siudek_dwarf_2023, Ubler_23MBH}. 

Our selection of dwarf galaxies have black hole masses $10^5< M_{BH}/M_{\odot} < 10^{6}$. These black holes masses are systematically lower than the black hole masses computed for 25/155 dwarf galaxies exhibiting broad H$\alpha$ emission lines in \cite{reines_dwarfgal_2013} which had $10^{8.5}< M_{BH}/M_{\odot} < 10^{9.5}$ using the approach from \cite{Greene_2005} to determine $M_{BH}$ from the properties of the H$\alpha$ emission line. This suggests that variability is able to probe lower mass black holes in dwarf galaxies and could be a unique and important approach for identifying black holes in dwarf galaxies. 
This was also suggested in the study by \cite{Kimbro_dwarf_2025} which used bulge masses determined from a morphological study of photometrically variable dwarf galaxies to determine black hole masses. The masses determined by \cite{Kimbro_dwarf_2025} were systematically even lower than our dwarf galaxy black hole masses at $10^{3}< M_{BH}/M_{\odot} < 10^{5}$.

Additionally, \cite{Wasleske_Baldassare2024dwarfxray} found that for their sample of 23 UV variable AGN candidates in active low-mass galaxies, 11/23 featured X-ray excess typical of AGNs. They further identified a sample of sources which were both UV variable and X-ray bright which did, however, not feature any spectroscopic signatures typically found in the optical for active BHs. This further indicates that variability could be an important tool for probing the lower mass end of the black hole mass function since it has been shown that BH indicators which work for more massive galaxies often fail to identify BHs in dwarf galaxies.
\subsection{Implications for black hole seeding}\label{subsec:seeds}
\begin{figure*}[htb!]
    \begin{center}
    \includegraphics[width=.99\linewidth]{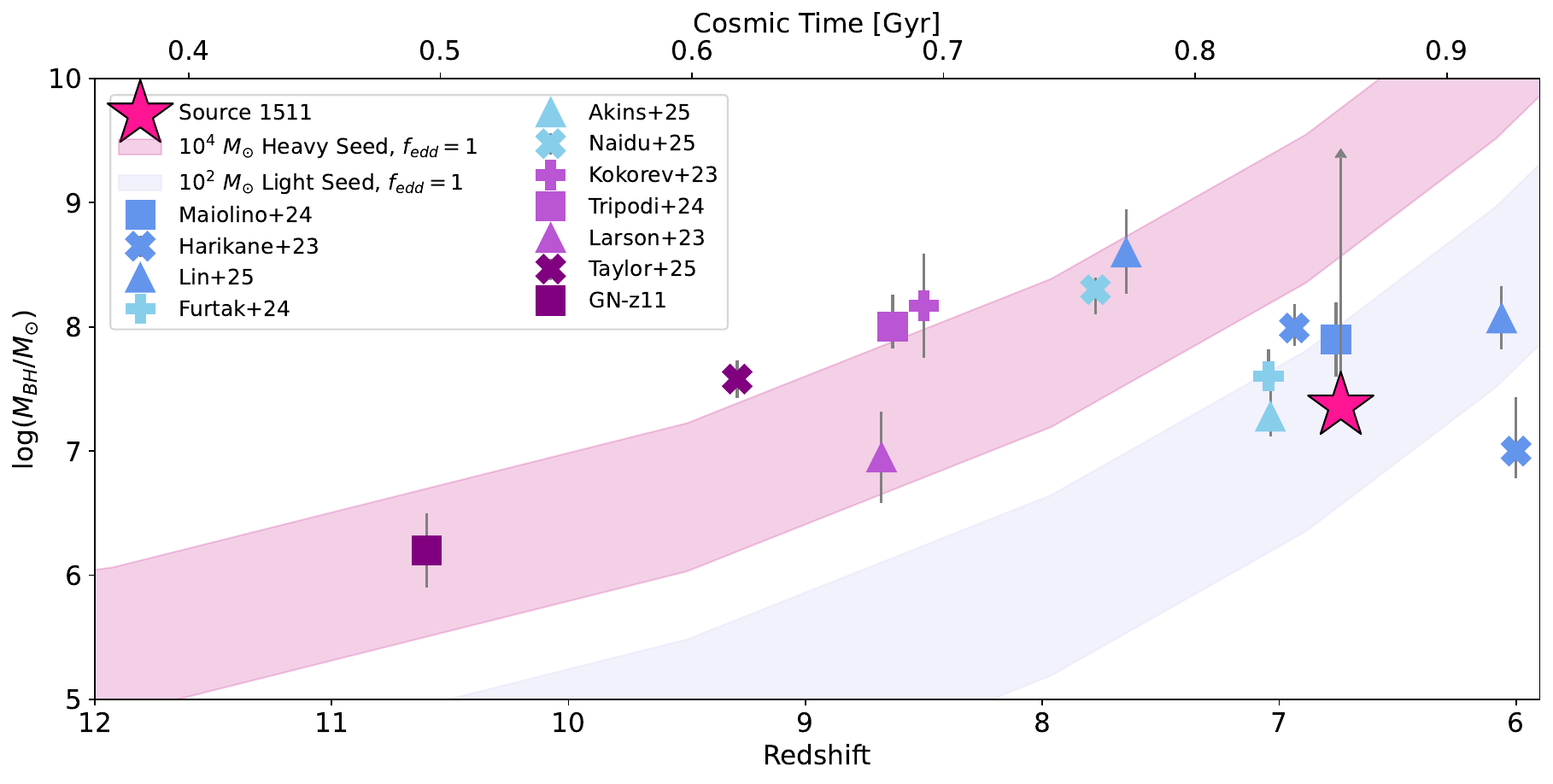}  
    \caption{Adapted version of Figure 6 from \cite{taylor2025_capersz9LRD} showing black hole mass as a function of redshift plotted for our highest redshift variable source, 1511, marked here as the pink star. Spectroscopically confirmed BL-AGN from \cite{akins_2025,furtak_high_2024,Harikane_23MBH,kokorev_uncover_2023,larson_ceers_2023,lin_2025, Maiolino_2024, maiolino_gnz112024a,naidu_2025,tacchella_2023gnz11,taylor2025_capersz9LRD,tripodi_2024} are shown by the filled markers. We note that GN-z11 is included in the analysis in \cite{Maiolino_2024}, however, the parameters are determined in \cite{tacchella_2023gnz11} and \cite{maiolino_gnz112024a}. We overplot the growth of light and heavy seeds following a simple Eddington growth model as presented in \cite{taylor2025_capersz9LRD}. The bounds of the pink shaded region correspond to the growth of massive seeds ($10^4M_{\odot}$) formed at redshifts of 25 and 15 which then immediately begin to accrete mass at the Eddington limit. The bounds of the blue shaded region show the mass evolution of light seeds ($10^2M_{\odot}$) formed at redshifts of 30 and 15 which do not build up mass for the first 100 Myr, at which point they start accreting at the Eddington limit. The lower limit on the black hole mass for Source 1511 is consistent with sources at similar redshift from the literature and can be explained by both a heavy seeding scenario and our simple model for the Eddington growth of a light seed. }
    \label{fig:bhmassvsz_seedinggrowth}
    \end{center}
\end{figure*}

From our sample of variables meeting the $AGN_{frac}$ condition a single source, namely source 1511, has $z_{phot}>6$. We calculate a lower limit on $log_{10}(M_{BH}/M_{\odot}) >7.36$ and plot this mass as a function of redshift along with masses from various $z > 6 $ AGN from the literature in Figure \ref{fig:bhmassvsz_seedinggrowth} \citep{akins_2025,furtak_high_2024,Harikane_23MBH,kokorev_uncover_2023,larson_ceers_2023,lin_2025,maiolino_gnz112024a,Maiolino_2024,naidu_2025,tacchella_2023gnz11,taylor2025_capersz9LRD,tripodi_2024}. We reproduce Figure 6 from \cite{taylor2025_capersz9LRD} by overplotting the evolution of a light mass seed and heavy mass seed following a simplified Eddington-limited black hole growth model. Light seeds are formed at $10^2 M_{\odot}$ and the bounds of the lilac shaded region correspond to the evolution of a light seed formed at $z = 30$ and $z = 15$. As in \cite{taylor2025_capersz9LRD}, for both light seeds, formation is followed by 100 Myr of no growth due to heating of the surrounding gas by the stellar progenitor of the seed \citep{Yoshida_2006_100Myr,johnsonbromm_100Myr_2007} before immediately accreting with $f_{edd} = 1$. The bounds of the magenta shaded region correspond to a heavy seed formed at $10^4 M_{\odot}$ between $z = 25$ and $z = 15$ which undergoes accretion with $f_{edd} = 1$ immediately. We note that differences in the shaded regions from \cite{taylor2025_capersz9LRD} are due the slightly different cosmology used in this work. As shown in Figure \ref{fig:bhmassvsz_seedinggrowth}, the lower limit of the mass for source 1511 is consistent with a light seed growing  at constant Eddington limited accretion until $z = 6$ \citep{haiman_loeb_2001, Madau_Rees_2001,Volonteri_2003, Li_2007, InayoshiReview_2020}. Additionally, our source has a black hole mass consistent with the other $6 < z < 7$ SMBH from \cite{Maiolino_2024, Harikane_23MBH} and \cite{lin_2025} as well as those with $z < 7.5$ from \cite{furtak_high_2024} and \cite{akins_2025}. This result would suggest that at $6 \lesssim z \lesssim 7.5$ SMBH which appear to be overmassive in the $M_{BH} - M_{*}$ plane do not necessarily require heavy seeds to reach the masses observed. This is consistent with our understanding of Eddington limited accretion, which predicts growth timescales of light SMBH seeds (100 $M_{\odot}$) up to $\sim 10^9$ to be approximately the age of the universe at $z \sim 6$ (with $f_{duty} \sim 1$) \citep{InayoshiReview_2020}. However, this is a simplistic model and a more realistic approach would invoke long periods of sub-Eddington accretion followed by rapid bursts of super Eddington accretion to better test if the masses observed can be reached via the growth of light BH seeds formed in the early Universe \citep{sanati2025}. 
\section{Conclusion}\label{sec:concl}
In this work, we conduct SED fitting of 121 variable AGN identified in the HUDF by \cite{cammelli_glimmers25} using 26 filters of PSF matched photometry from HST and JWST. We designed a bespoke SED fitting code to test whether each source has an SED which is better modeled by either a pure stellar population using a delayed-$\tau$ SFH or if the fit is statistically improved by including an AGN template from \cite{Temple_2021}. We designed two conditions which would indicate if our SED fitting retrieved the AGN nature of the source, namely the AIC condition, $\Delta\,AIC \geq 6$, and the $AGN_{frac}$ condition, where the AGN template contributes at least 20\% to the total flux in at least one filter. In total, we identified 37 sources which meet the first condition, 26 sources which meet the latter and 11 sources which meet both conditions. For the 26 sources meeting the $AGN_{frac}$ condition, we use the model spectrum resulting from the MCMC SED fitting to calculate $L_{5100}$. We use this luminosity in conjunction with the scaling relation from \cite{kaspi_2000_MBH} to determine estimates for the black hole masses of these objects. These mass estimates allow us to conclude the following:
\begin{itemize} 
    \item Using the stellar masses from the SED fitting procedure we plot our sources on the $M_{BH} - M_{*}$ plane and see that our sources become increasingly overmassive as a function of host redshift. This provides independent support for a systematic offset from the local relations as observed for other high-z BL-AGN detected with JWST. These overmassive BHs are possibly explained by either heavy seeding in the early universe or episodes of super Eddington accretion, especially for those sources in our sample with $M_{BH} \sim M_{*}$.
    \item Our SED fitting results allow us to correct both our model stellar template and AGN template for reddening. We find that neglecting this correction can reduce the calculated black hole mass by factors of up to $\simeq 10$. 
    \item We compare the black hole masses estimated for each source when photometry from the WFC3/IR epochs where the AGN is faint is used versus when the AGN is bright and find this can have considerable impact on the resulting black hole mass by up to $\sim 3\,$dex where bright epochs yield higher $M_{BH}$.  
    \item We identify two dwarf galaxies at $z < 1$ for which an AGN contribution to the SED is statistically preferred. The sources also exhibit black hole masses which are overmassive with respect to the stellar mass of their host galaxies which is consistent with observations of other IMBHs in dwarf galaxies \citep{Burke_dwarfgals_2022,Mezcua_dwarf_2023,Siudek_dwarf_2023}. Our dwarf galaxies also host black holes with masses that are systematically lower than those identified from spectroscopic features in \cite{reines_dwarfgal_2013}. This could indicate that variability offers a unique probe for identifying lower mass BHs in dwarf galaxies which is consistent with the results from \cite{Kimbro_dwarf_2025}.
    \item Finally, our highest redshift variable for which we calculate a black hole mass has an SED best explained by an $AGN_{frac} \sim 1$. When compared to simple models for the Eddington growth of heavy seeds and light seeds as a function of redshift, the black hole mass of 1511 is consistent with both seeding scenarios within errors. However, the value calculated as the lower limit for the black hole mass can be explained by a light seed growing at the Eddington rate starting from 100 Myr after its formation. 
\end{itemize}
Our work highlights the important role variability can play in identifying BHs and the effect this intrinsic phenomena can have on the calculated mass of black holes. It also offers confirmation of observed trends in the $M_{BH} - M_{*}$ plane which have been identified by JWST BL-AGN at high-z. Additionally, it does so by estimating black hole masses using a method which is does not require the measurement of broad line properties and instead leverages the decomposition of the SEDs of our sources into separate AGN and stellar components. 

Spectroscopic follow-up for the sample of $2.5\sigma$ variables presented in \cite{cammelli_glimmers25} and studied here would allow us to better understand the intersection of different detection methods in identifying AGN. This work would be similar to the studies conducted by \cite{Pouliasis_2019, Lyu_2022goods-sAGN}, however, spectra from JWST would allow us to probe the broad-line emission of AGN to much higher redshifts than was possible for the HUDF in these previous studies. This would be complimented by the relatively longer time baseline of our photometric variability study and the depth of the WFC3/IR images which allow us to probe faint, low-mass AGN as well as the broad photometric coverage leveraged in this work. Studies of this nature are essential for determining how the properties of host galaxies and their SMBHs are linked and how they evolve through cosmic time.
%

\vspace{5mm}
\facilities{HST (ACS/WFC, WFC3/UVIS, WFC3/IR), JWST (NIRCam, NIRISS)}


\software{Astropy \citep{astropy:2013, astropy:2018, astropy:2022}; DrizzlePac \citep{DrizzlePac}; SWarp \citep{SWarp_Bertin_2002}; PyPHER \citep{Boucaud_pypher_2016}; ACSTOOLS \citep{acstools}; STPSF \citep{STPSF}; Source Extractor \citep{SExtractor}; SciPy \citep{2020SciPy}; LMFIT \citep{lmfit_2016}, emcee \citep{emcee_Foreman_Mackey_2013}}

\begin{acknowledgements}
We thank Pierluigi Monaco for useful discussions and invaluable contributions to the project. 
A.R.Y. is funded by the Swedish National Space Agency, SNSA.
M.J.H. is supported by the Swedish Research Council
(Vetenskapsr\aa{}det) and is Fellow of the Knut \& Alice Wallenberg Foundation.
V.C. thanks the BlackHoleWeather project and PI Prof. Gaspari for salary support.
\end{acknowledgements}



\appendix
\section{Deriving error corrections for Source Extractor Photometry} \label{appendix:errcorr}
This appendix details the error correction applied to \texttt{Source Extractor} output photometry derived from the source injections introduced in Section \ref{sec:obs}. The F480M PSF was used as the injection PSF and renormalized to fluxes corresponding to an injection magnitude iteratively until 1000 sources had been injected at each magnitude in each filter image (however, only 10 sources were injected into an image a time to avoid crowding). A segmentation map of the HUDF was used as a mask and sources could only be injected into unmasked regions with a 50 pixel buffer to ensure the PSF wings were not overlapping with true sources. Sources were injected with AB magnitudes from 26 to 31.5 with $\delta_{mag} = 0.5$ for $26 \leq mag \leq 27$ and $\delta_{mag} = 0.1$ for $27 < mag  \leq 31.5$. The exceptions were the UVIS filters for which $\delta_{mag} = 0.1$ for the full magnitude range of $25 \leq mag \leq 31.5$. 

\begin{figure}
    \centering
    \includegraphics[width=.3\textwidth]{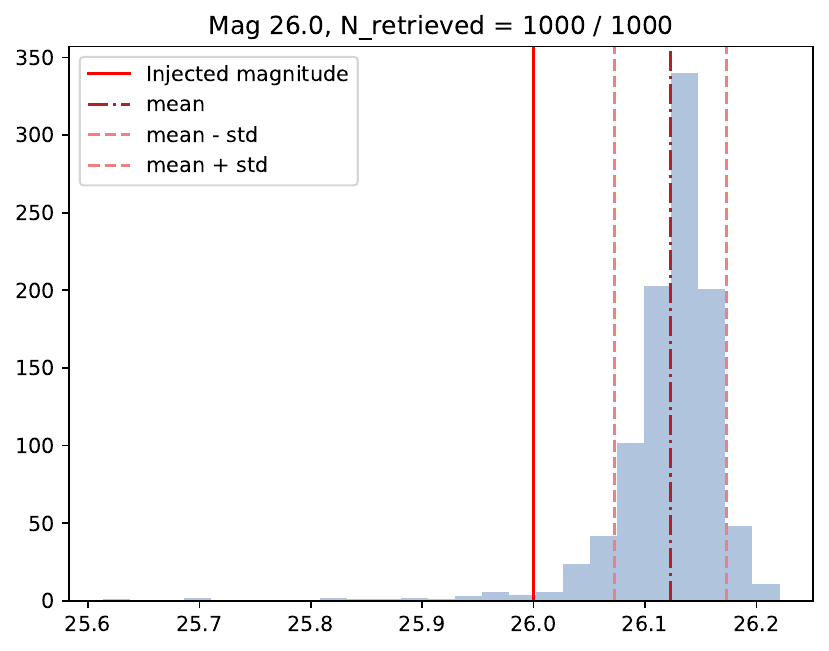}
    \includegraphics[width=.3\textwidth]{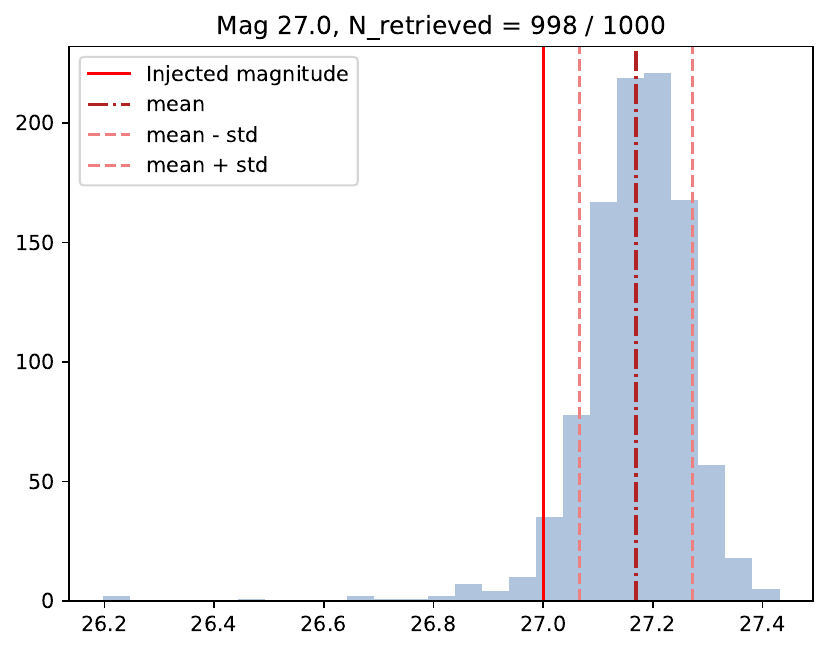}
    \includegraphics[width=.3\textwidth]{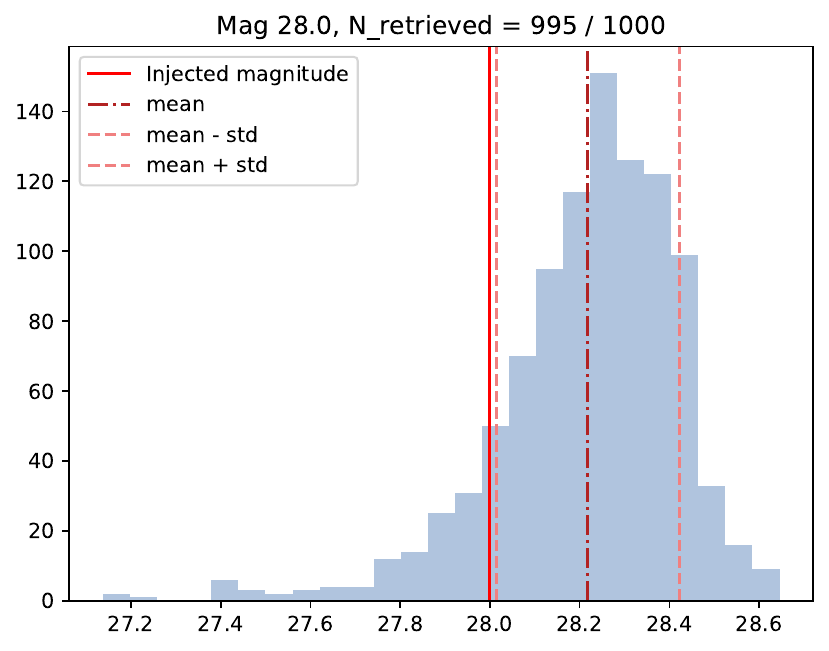}
    \includegraphics[width=.3\textwidth]{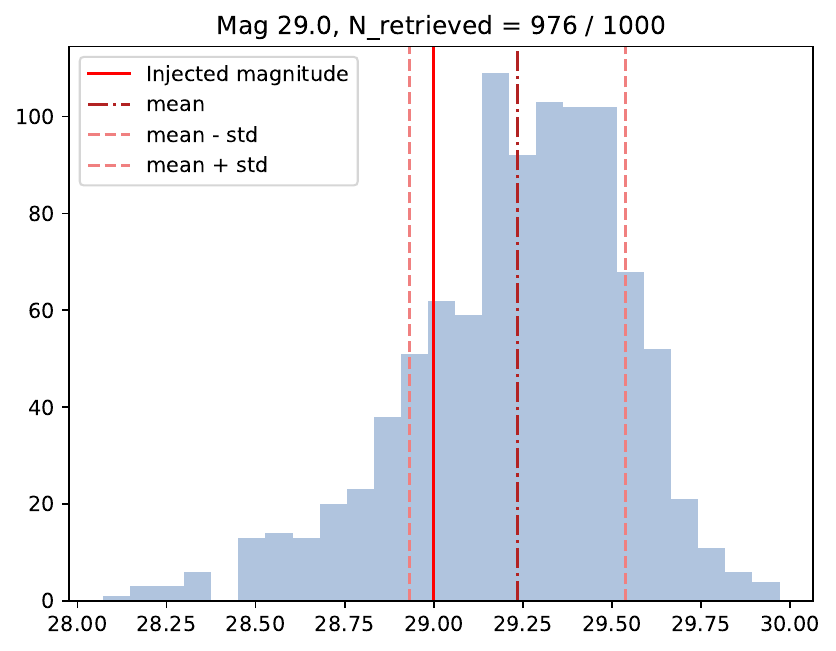}
    \includegraphics[width=.3\textwidth]{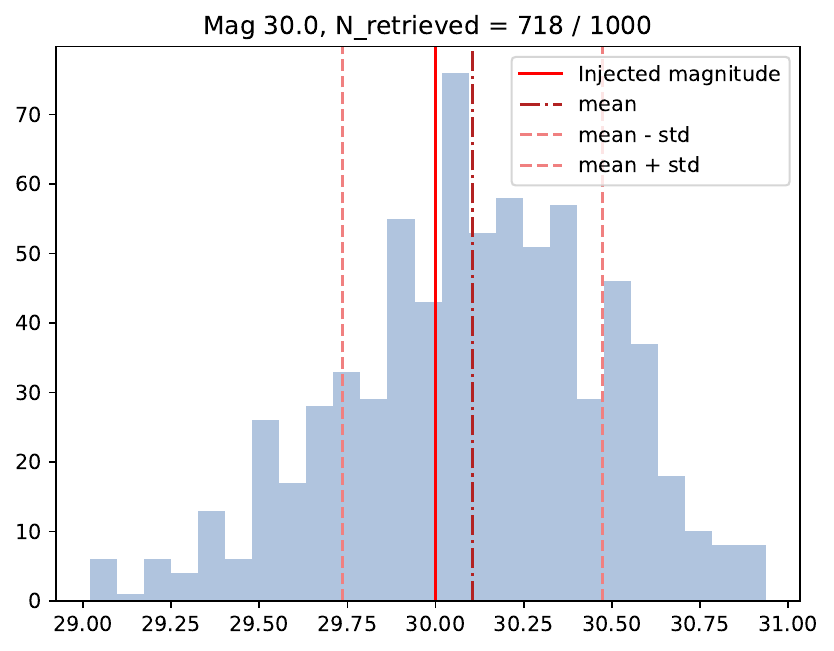}
    \includegraphics[width=.3\textwidth]{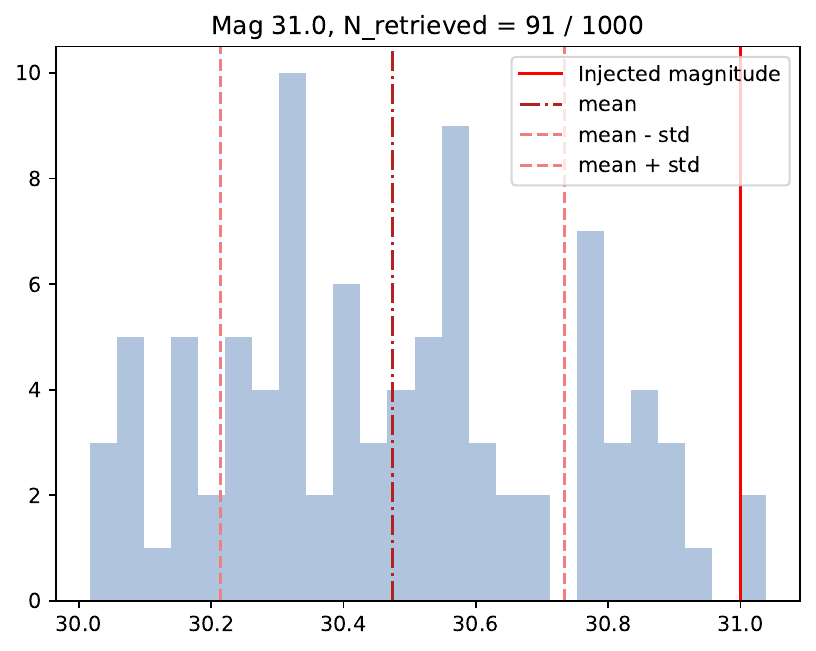}
    \caption{Histograms showing the magnitudes returned from \texttt{Source Extractor} for each retrieved source injected with a given magnitude which is recorded in the subtitle of each plot. The red solid line shows in the injected magnitude, the burgundy dash-dotted line indicates the mean of the retrieved magnitudes and the pink dashed lines show the one-sigma values of the retrieved magnitude distribution. This represents a subset of the injected magnitudes for the F140W filter. Here, we can also see that when the injected magnitude is brighter than the limiting magnitude the mean of the retrieved magnitudes is always fainter than the injected magnitude, even for bright source injections. This difference between the mean of the retrieved magnitude and the injected magnitude captures the aperture correction since a reduced fraction of the flux is captured for sources which are not point-like when aperture photometry is used. }
    \label{fig:errcorrhistos}
\end{figure}

\begin{figure}
    \centering
    \includegraphics[width=.95\textwidth]{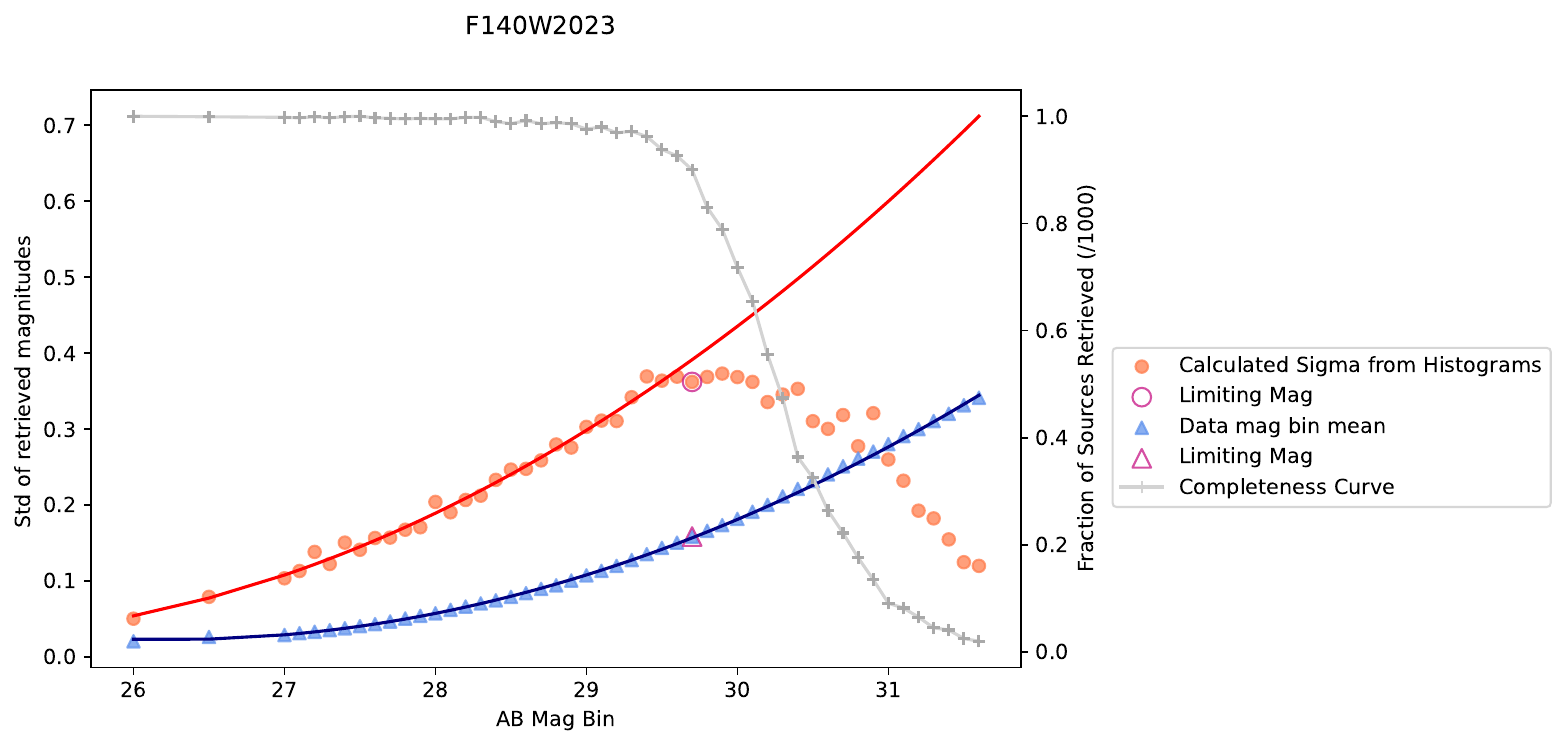}
    \caption{Demonstration of how the \texttt{Source Extractor} error correction factor is determined for each filter, here shown for the F140W filter. The grey curve represents the number of sources retrieved for each injected magnitude, this is our completeness curve. When this curve drops below the 90\% completeness limit, we consider the previous injected magnitude our limiting magnitude. This limiting magnitude is represented by the magenta outlined points. The standard deviation of the retrieved magnitudes for each injected magnitude are shown by the orange circular points. The blue triangles represent the mean \texttt{Source Extractor} error for sources in our photometric catalogue with \texttt{Source Extractor} magnitudes in each magnitude bin. The best fit second order polynomial for the orange points is shown as the red curve and the best fit curve for the blue points is given by the navy curve. The difference between the polynomials represented by the red and blue curves are used to calculate the error correction factor for each source in our catalogue. See text for details. }
    \label{fig:errcorr}
\end{figure}

Once sources were injected into the image with a given magnitude, \texttt{Source Extractor} was used to determine the recovered magnitudes of each source. The injected sources and the recovered magnitudes were extracted from the \texttt{Source Extractor} catalogues using the known positions assigned to the injected sources and a tolerance of 1.5 pixels in x and y. For a given set of 1000 injected sources at a given magnitude, the retrieved magnitudes are plotted as a histogram. Histograms of the retrieved magnitudes for a subsample of injected magnitudes for the F140W image in shown in Figure \ref{fig:errcorrhistos}. For each histogram, we calculate the standard deviation of the retrieved magnitudes to give the 1-sigma error for sources with a true magnitude corresponding to the injected magnitude. We then use these values to derive our \texttt{Source Extractor} error correction factor. The derivation of this correction factor is visualised in Figure \ref{fig:errcorr} and we will refer to this figure as an example in the text that follows.

The standard deviations of each injected magnitude for a given filter image is then plotted, as represented by the orange points in Figure \ref{fig:errcorr}. To compare to the errors provided by \texttt{Source Extractor}, the photometry from the full catalogue of sources in the HUDF is sorted into the same magnitude bins described above and the mean error reported from the \texttt{Source Extractor} output is determined for each magnitude bin. These values correspond to the blue triangle data points in Figure \ref{fig:errcorr}. The offset between the orange and blue points then corresponds to the error correction which must be applied to the photometric errors output from \texttt{Source Extractor}. To determine this offset, second order polynomials are fit using \texttt{np.polyfit} to both the standard deviation in the retrieved magnitudes in all magnitude bins (the best fit curve corresponding to the red line in Figure \ref{fig:errcorr}) as well as the mean error of the \texttt{Source Extractor} data (the best fit curve corresponding to the navy line in Figure \ref{fig:errcorr}). These were fit from the brightest magnitude bin until the magnitude bin corresponding to the last magnitude bin before the retrieved fraction of the injected sources reached 90\% (in Figure \ref{fig:errcorr} this magnitude bin in marked by magenta outlines around the relevant data point and corresponds to $\mathrm{mag}_{AB} = 29.7$ for the F140W 2023 epoch image). We use this magnitude bin as the limiting magnitude for each filter image. We do not fit the polynomials beyond this limiting magnitude due to the fact that the number of retrieved sources is reduced such that the distribution of retrieved magnitudes suffers from low number statistics. These low numbers result in artificially reduced standard deviations represented by the inflection point in the orange points in Figure \ref{fig:errcorr}. The coefficients of these polynomials are determined and used to calculate the correction for any source with a reported \texttt{Source Extractor} magnitude (\texttt{MAG\_SE}) and magnitude error (\texttt{MAGERR\_SE}). For this, we calculate the value of the polynomial best fit to the injected source standard deviations (using the polynomial coefficients corresponding to the red curve in Figure \ref{fig:errcorr}) for the \texttt{Source Extractor} magnitude of a source ($\mathrm{injected\_poly\left(\texttt{MAG\_SE}\right)}$) as well as the value of the polynomial best fit to the mean errors from the HUDF catalogue (using the polynomial coefficients corresponding to the navy curve in Figure \ref{fig:errcorr}) for the same source ($\mathrm{data\_poly\left(\texttt{MAG\_SE}\right)}$). The correction factor ($err\_corr$) then becomes:
\begin{equation}
     err\_corr = \frac{\mathrm{injected\_poly\left(\texttt{MAG\_SE}\right)}}{\mathrm{data\_poly\left(\texttt{MAG\_SE}\right)}}.
\end{equation}
This allows for the calculation of the corrected error as $\mathrm{\texttt{MAGERR\_corr}} = \mathrm{\texttt{MAGERR\_SE}} + err\_corr$. This correction must be applied to each source for the photometric measurement in each filter using the relevant polynomial coefficients determined from source injections in each filter.

We also note that Figure \ref{fig:errcorrhistos} demonstrates that our correction factor additionally contains an aperture correction. In these histograms, when the injected magnitude is brighter than the limiting magnitude the mean retrieved magnitude is always fainter than the injected magnitude, even for bright source injections. This difference between the mean retrieved magnitude and the injected magnitude captures the aperture correction since a reduced fraction of the flux is captured for sources which are not point-like when aperture photometry is used.

\section{ Artificial Star Injections: testing the surplus of detected variables getting brighter between epochs} \label{appendix:asymmetryinjection}
\begin{figure}[h!]
    \centering
    \includegraphics[width=.28\textwidth]{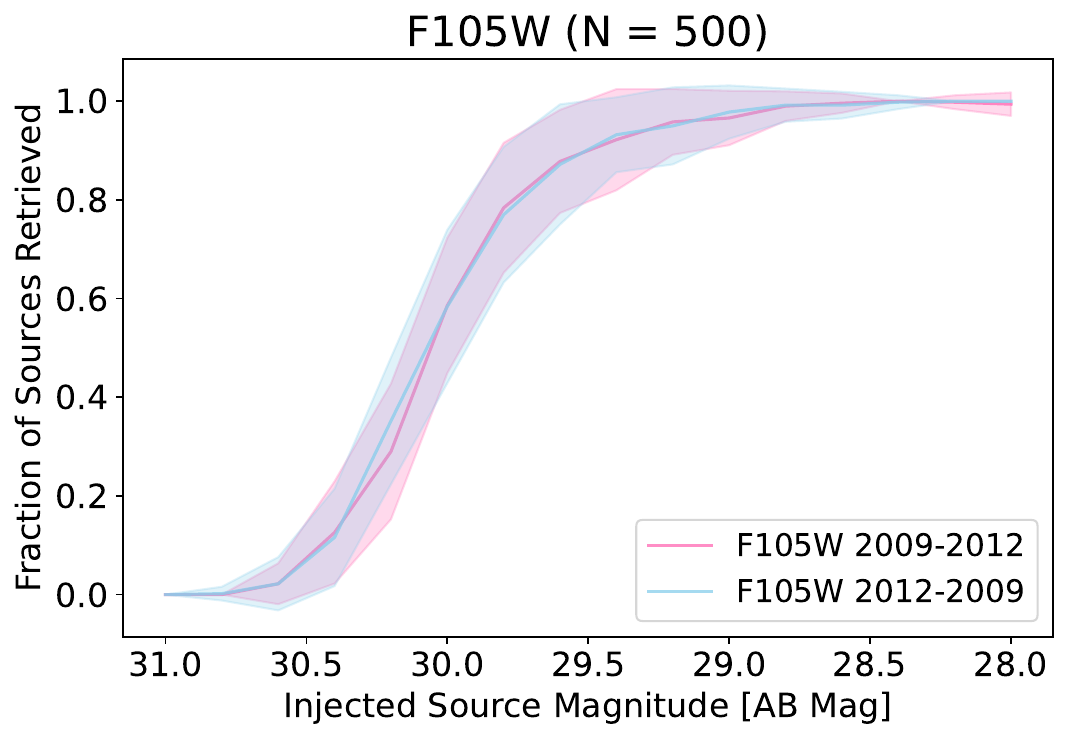}
    \includegraphics[width=.28\textwidth]{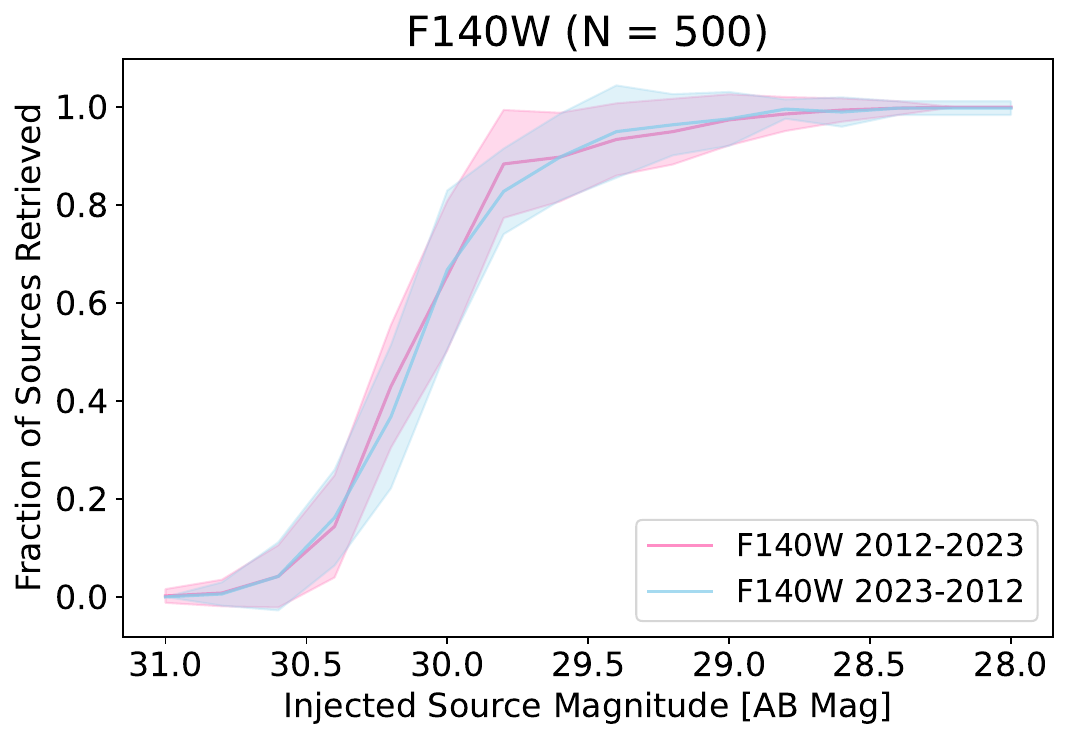}
    \includegraphics[width=.28\textwidth]{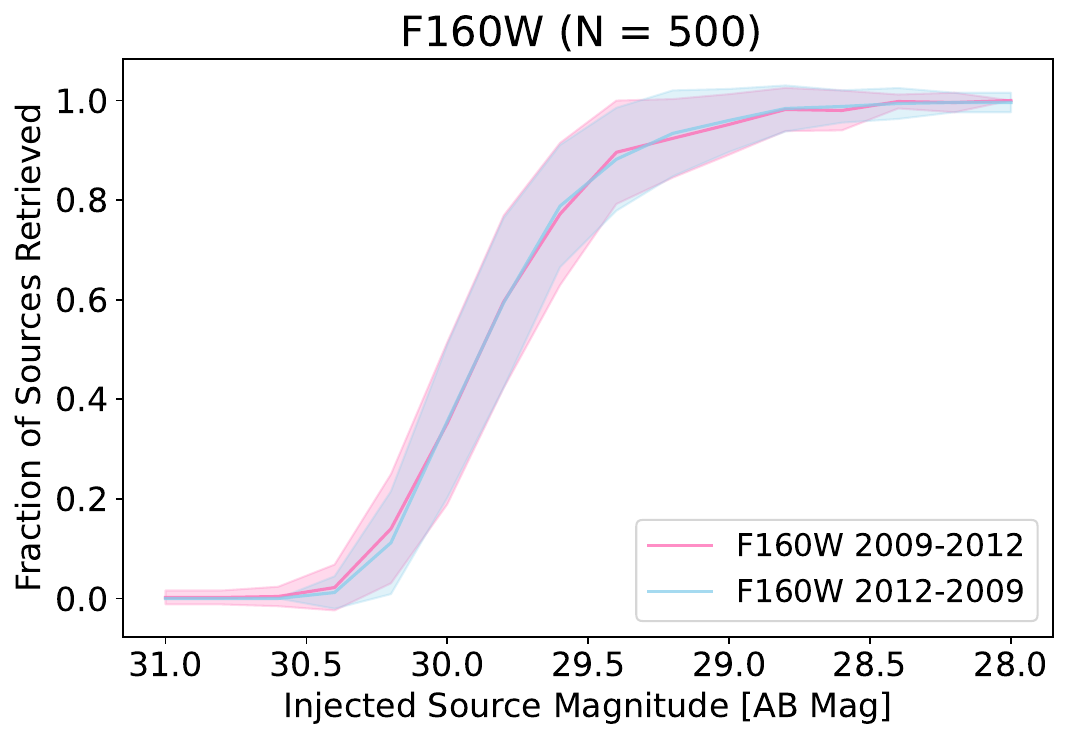}
    \caption{Results from the artificial stars tests seeking to determine if there is a detection bias for variables which get brighter between epochs. Plots show the fraction of injected sources retrieved from the total sample in each magnitude bin (N=500) for the F105W, F140W and F160W filters. The different curves show the different results when difference images are generated both by inserting the artificial sources into the early epoch and performing image subtraction in pink and by inserting the sources into the late epoch and performing image subtraction in blue. Within errors, the two scenarios are consistent meaning no significant bias exists in our photometry to select either preferentially dimming nor brightening sources.}
    \label{fig:artificialstars_diffim}
\end{figure}

We noticed that a surplus of variable sources in the HUDF appeared to get brighter between epochs when compared to the number which appear to get fainter. To test whether there was a bias in our detection methods, a series of tests were executed by inserting artificial sources into the difference images. This was done by iteratively inserting 10 artificial stars into the first epoch in the time series, subtracting the second epoch and performing photometry with \texttt{Source Extractor} on the resulting difference image to determine the fraction of inserted sources retrieved at a given magnitude. 50 images containing 10 artificial sources were generated in each magnitude bin. For each filter this is performed twice such that the fraction of sources retrieved when the difference image is created by inserting artificial stars into the first epoch can be tested against the number of sources retrieved when the difference image is instead generated by inserting stars into the second epoch before subtraction. Stars were generated at a given magnitude between 29.0 and 31.0 with a step size of 0.2 mag using the \texttt{make\_gaussian\_sources\_image} from the \textit{Photutils} affiliated package of \textit{Astropy}. These sources are assigned a random position in the masked field which excludes areas of the image which are less deep because of either pointing offsets (at the edge of the image) or persistence (and therefore have fewer images contributing to the total stack in the respective area of the image). \texttt{Source Extractor} is then used to perform source detection and photometry on the resulting difference image. The recovered source positions are then compared to the input positions of the artificial stars with a tolerance of 1.5 pixels in x and y as well as the recovered magnitude of the source which must be with 0.5 mag of the input magnitude. The fraction of retrieved sources from these tests for the F105W, F140W and F160W filters are presented in Figure \ref{fig:artificialstars_diffim}. Within errors, the results show that the artificial star tests are not affected by which epoch the artificial stars are inserted into. Physically, this would mean that our results are not biased towards either variable objects which get brighter, nor variables which get fainter. 


\bibliography{sample631}{}

\begin{thebibliography}{}
\expandafter\ifx\csname natexlab\endcsname\relax\def\natexlab#1{#1}\fi
\providecommand{\url}[1]{\href{#1}{#1}}
\providecommand{\dodoi}[1]{doi:~\href{http://doi.org/#1}{\nolinkurl{#1}}}
\providecommand{\doeprint}[1]{\href{http://ascl.net/#1}{\nolinkurl{http://ascl.net/#1}}}
\providecommand{\doarXiv}[1]{\href{https://arxiv.org/abs/#1}{\nolinkurl{https://arxiv.org/abs/#1}}}

\bibitem[{Abuter {et~al.}(2024)Abuter, Allouche, Amorim, Bailet, Berdeu, Berger, Berio, Bigioli, Boebion, Bolzer, Bonnet, Bourdarot, Bourget, Brandner, Cao, Conzelmann, Comin, Clénet, Courtney-Barrer, Davies, Defrère, Delboulbé, Delplancke-Ströbele, Dembet, Dexter, de~Zeeuw, Drescher, Eckart, Édouard, Eisenhauer, Fabricius, Feuchtgruber, Finger, Förster~Schreiber, Garcia, Garcia~Lopez, Gao, Gendron, Genzel, Gil, Gillessen, Gomes, Gonté, Gouvret, Guajardo, Guieu, Hackenberg, Haddad, Hartl, Haubois, Haußmann, Heißel, Henning, Hippler, Hönig, Horrobin, Hubin, Jacqmart, Jocou, Kaufer, Kervella, Kolb, Korhonen, Lacour, Lagarde, Lai, Lapeyrère, Laugier, Le~Bouquin, Leftley, Léna, Lewis, Liu, Lopez, Lutz, Magnard, Mang, Marcotto, Maurel, Mérand, Millour, More, Netzer, Nowacki, Nowak, Oberti, Ott, Pallanca, Paumard, Perraut, Perrin, Petrov, Pfuhl, Pourré, Rabien, Rau, Riquelme, Robbe-Dubois, Rochat, Salman, Sanchez-Bermudez, Santos, Scheithauer, Schöller, Schubert, Schuhler, Shangguan, Shchekaturov,
  Shimizu, Sevin, Soulez, Spang, Stadler, Sternberg, Straubmeier, Sturm, Sykes, Tacconi, Tristram, Vincent, von Fellenberg, Uysal, Widmann, Wieprecht, Wiezorrek, Woillez, \& Zins}]{AbuterGRAVITY_2024}
Abuter, R., Allouche, F., Amorim, A., {et~al.} 2024, Nature, 627, 281–285, \dodoi{10.1038/s41586-024-07053-4}

\bibitem[{Akins {et~al.}(2024)Akins, Casey, Lambrides, Allen, Andika, Brinch, Champagne, Cooper, Ding, Drakos, Faisst, Finkelstein, Franco, Fujimoto, Gentile, Gillman, Gozaliasl, Harish, Hayward, Hirschmann, Ilbert, Kartaltepe, Kocevski, Koekemoer, Kokorev, Liu, Long, McCracken, McKinney, Onoue, Paquereau, Renzini, Rhodes, Robertson, Shuntov, Silverman, Tanaka, Toft, Trakhtenbrot, Valentino, \& Zavala}]{akins2024_LRDs}
Akins, H.~B., Casey, C.~M., Lambrides, E., {et~al.} 2024, COSMOS-Web: The over-abundance and physical nature of "little red dots"--Implications for early galaxy and SMBH assembly.
\newblock \doarXiv{2406.10341}

\bibitem[{Akins {et~al.}(2025)Akins, Casey, Berg, Chisholm, Cloonan, Franco, Finkelstein, Fujimoto, Koekemoer, Kokorev, Lambrides, Robertson, Taylor, Coulter, Fox, \& Karmen}]{akins_2025}
Akins, H.~B., Casey, C.~M., Berg, D.~A., {et~al.} 2025, The Astrophysical Journal Letters, 980, L29, \dodoi{10.3847/2041-8213/adab76}

\bibitem[{{Astropy Collaboration} {et~al.}(2013){Astropy Collaboration}, {Robitaille}, {Tollerud}, {Greenfield}, {Droettboom}, {Bray}, {Aldcroft}, {Davis}, {Ginsburg}, {Price-Whelan}, {Kerzendorf}, {Conley}, {Crighton}, {Barbary}, {Muna}, {Ferguson}, {Grollier}, {Parikh}, {Nair}, {Unther}, {Deil}, {Woillez}, {Conseil}, {Kramer}, {Turner}, {Singer}, {Fox}, {Weaver}, {Zabalza}, {Edwards}, {Azalee Bostroem}, {Burke}, {Casey}, {Crawford}, {Dencheva}, {Ely}, {Jenness}, {Labrie}, {Lim}, {Pierfederici}, {Pontzen}, {Ptak}, {Refsdal}, {Servillat}, \& {Streicher}}]{astropy:2013}
{Astropy Collaboration}, {Robitaille}, T.~P., {Tollerud}, E.~J., {et~al.} 2013, \aap, 558, A33, \dodoi{10.1051/0004-6361/201322068}

\bibitem[{{Astropy Collaboration} {et~al.}(2018){Astropy Collaboration}, {Price-Whelan}, {Sip{\H{o}}cz}, {G{\"u}nther}, {Lim}, {Crawford}, {Conseil}, {Shupe}, {Craig}, {Dencheva}, {Ginsburg}, {Vand erPlas}, {Bradley}, {P{\'e}rez-Su{\'a}rez}, {de Val-Borro}, {Aldcroft}, {Cruz}, {Robitaille}, {Tollerud}, {Ardelean}, {Babej}, {Bach}, {Bachetti}, {Bakanov}, {Bamford}, {Barentsen}, {Barmby}, {Baumbach}, {Berry}, {Biscani}, {Boquien}, {Bostroem}, {Bouma}, {Brammer}, {Bray}, {Breytenbach}, {Buddelmeijer}, {Burke}, {Calderone}, {Cano Rodr{\'\i}guez}, {Cara}, {Cardoso}, {Cheedella}, {Copin}, {Corrales}, {Crichton}, {D'Avella}, {Deil}, {Depagne}, {Dietrich}, {Donath}, {Droettboom}, {Earl}, {Erben}, {Fabbro}, {Ferreira}, {Finethy}, {Fox}, {Garrison}, {Gibbons}, {Goldstein}, {Gommers}, {Greco}, {Greenfield}, {Groener}, {Grollier}, {Hagen}, {Hirst}, {Homeier}, {Horton}, {Hosseinzadeh}, {Hu}, {Hunkeler}, {Ivezi{\'c}}, {Jain}, {Jenness}, {Kanarek}, {Kendrew}, {Kern}, {Kerzendorf}, {Khvalko}, {King}, {Kirkby}, {Kulkarni},
  {Kumar}, {Lee}, {Lenz}, {Littlefair}, {Ma}, {Macleod}, {Mastropietro}, {McCully}, {Montagnac}, {Morris}, {Mueller}, {Mumford}, {Muna}, {Murphy}, {Nelson}, {Nguyen}, {Ninan}, {N{\"o}the}, {Ogaz}, {Oh}, {Parejko}, {Parley}, {Pascual}, {Patil}, {Patil}, {Plunkett}, {Prochaska}, {Rastogi}, {Reddy Janga}, {Sabater}, {Sakurikar}, {Seifert}, {Sherbert}, {Sherwood-Taylor}, {Shih}, {Sick}, {Silbiger}, {Singanamalla}, {Singer}, {Sladen}, {Sooley}, {Sornarajah}, {Streicher}, {Teuben}, {Thomas}, {Tremblay}, {Turner}, {Terr{\'o}n}, {van Kerkwijk}, {de la Vega}, {Watkins}, {Weaver}, {Whitmore}, {Woillez}, {Zabalza}, \& {Astropy Contributors}}]{astropy:2018}
{Astropy Collaboration}, {Price-Whelan}, A.~M., {Sip{\H{o}}cz}, B.~M., {et~al.} 2018, \aj, 156, 123, \dodoi{10.3847/1538-3881/aabc4f}

\bibitem[{{Astropy Collaboration} {et~al.}(2022){Astropy Collaboration}, {Price-Whelan}, {Lim}, {Earl}, {Starkman}, {Bradley}, {Shupe}, {Patil}, {Corrales}, {Brasseur}, {N{"o}the}, {Donath}, {Tollerud}, {Morris}, {Ginsburg}, {Vaher}, {Weaver}, {Tocknell}, {Jamieson}, {van Kerkwijk}, {Robitaille}, {Merry}, {Bachetti}, {G{"u}nther}, {Aldcroft}, {Alvarado-Montes}, {Archibald}, {B{'o}di}, {Bapat}, {Barentsen}, {Baz{'a}n}, {Biswas}, {Boquien}, {Burke}, {Cara}, {Cara}, {Conroy}, {Conseil}, {Craig}, {Cross}, {Cruz}, {D'Eugenio}, {Dencheva}, {Devillepoix}, {Dietrich}, {Eigenbrot}, {Erben}, {Ferreira}, {Foreman-Mackey}, {Fox}, {Freij}, {Garg}, {Geda}, {Glattly}, {Gondhalekar}, {Gordon}, {Grant}, {Greenfield}, {Groener}, {Guest}, {Gurovich}, {Handberg}, {Hart}, {Hatfield-Dodds}, {Homeier}, {Hosseinzadeh}, {Jenness}, {Jones}, {Joseph}, {Kalmbach}, {Karamehmetoglu}, {Ka{l}uszy{'n}ski}, {Kelley}, {Kern}, {Kerzendorf}, {Koch}, {Kulumani}, {Lee}, {Ly}, {Ma}, {MacBride}, {Maljaars}, {Muna}, {Murphy}, {Norman}, {O'Steen},
  {Oman}, {Pacifici}, {Pascual}, {Pascual-Granado}, {Patil}, {Perren}, {Pickering}, {Rastogi}, {Roulston}, {Ryan}, {Rykoff}, {Sabater}, {Sakurikar}, {Salgado}, {Sanghi}, {Saunders}, {Savchenko}, {Schwardt}, {Seifert-Eckert}, {Shih}, {Jain}, {Shukla}, {Sick}, {Simpson}, {Singanamalla}, {Singer}, {Singhal}, {Sinha}, {Sip{H{o}}cz}, {Spitler}, {Stansby}, {Streicher}, {{{S}}umak}, {Swinbank}, {Taranu}, {Tewary}, {Tremblay}, {Val-Borro}, {Van Kooten}, {Vasovi{'c}}, {Verma}, {de Miranda Cardoso}, {Williams}, {Wilson}, {Winkel}, {Wood-Vasey}, {Xue}, {Yoachim}, {Zhang}, {Zonca}, \& {Astropy Project Contributors}}]{astropy:2022}
{Astropy Collaboration}, {Price-Whelan}, A.~M., {Lim}, P.~L., {et~al.} 2022, \apj, 935, 167, \dodoi{10.3847/1538-4357/ac7c74}

\bibitem[{Bacon {et~al.}(2023)Bacon, Brinchmann, Conseil, Maseda, Nanayakkara, Wendt, Bacher, Mary, Weilbacher, Krajnović, Boogaard, Bouché, Contini, Epinat, Feltre, Guo, Herenz, Kollatschny, Kusakabe, Leclercq, Michel-Dansac, Pello, Richard, Roth, Salvignol, Schaye, Steinmetz, Tresse, Urrutia, Verhamme, Vitte, Wisotzki, \& Zoutendijk}]{Bacon_2023}
Bacon, R., Brinchmann, J., Conseil, S., {et~al.} 2023, Astronomy \& Astrophysics, 670, A4, \dodoi{10.1051/0004-6361/202244187}

\bibitem[{Banik {et~al.}(2018)Banik, Tan, \& Monaco}]{Banik_2018}
Banik, N., Tan, J.~C., \& Monaco, P. 2018, Monthly Notices of the Royal Astronomical Society, 483, 3592–3606, \dodoi{10.1093/mnras/sty3298}

\bibitem[{{Banik} {et~al.}(2019){Banik}, {Tan}, \& {Monaco}}]{banik_2019popIII.1}
{Banik}, N., {Tan}, J.~C., \& {Monaco}, P. 2019, \mnras, 483, 3592, \dodoi{10.1093/mnras/sty3298}

\bibitem[{Barro {et~al.}(2023)Barro, Perez-Gonzalez, Kocevski, McGrath, Trump, Simons, Somerville, Yung, Haro, Bagley, Cleri, Costantin, Davis, Dickinson, Finkelstein, Giavalisco, Gomez-Guijarro, Hathi, Hirschmann, Akins, Holwerda, Huertas-Company, Lucas, Papovich, Seille, Tacchella, Wilkins, de~la Vega, Yang, \& Zavala}]{barro2023LRD}
Barro, G., Perez-Gonzalez, P.~G., Kocevski, D.~D., {et~al.} 2023, Extremely red galaxies at $z=5-9$ with MIRI and NIRSpec: dusty galaxies or obscured AGNs?
\newblock \doarXiv{2305.14418}

\bibitem[{Bañados {et~al.}(2017)Bañados, Venemans, Mazzucchelli, Farina, Walter, Wang, Decarli, Stern, Fan, Davies, Hennawi, Simcoe, Turner, Rix, Yang, Kelson, Rudie, \& Winters}]{Banados_2017_highzqso}
Bañados, E., Venemans, B.~P., Mazzucchelli, C., {et~al.} 2017, Nature, 553, 473–476, \dodoi{10.1038/nature25180}

\bibitem[{{Begelman} {et~al.}(2006){Begelman}, {Volonteri}, \& {Rees}}]{begelman_DCBH_2006}
{Begelman}, M.~C., {Volonteri}, M., \& {Rees}, M.~J. 2006, \mnras, 370, 289, \dodoi{10.1111/j.1365-2966.2006.10467.x}

\bibitem[{Bennett {et~al.}(2023)Bennett, Sijacki, Costa, Laporte, \& Witten}]{bennett2023growthgargantuanblackholes}
Bennett, J.~S., Sijacki, D., Costa, T., Laporte, N., \& Witten, C. 2023, The growth of the gargantuan black holes powering high-redshift quasars and their impact on the formation of early galaxies and protoclusters.
\newblock \doarXiv{2305.11932}

\bibitem[{{Bertin} {et~al.}(2002){Bertin}, {Mellier}, {Radovich}, {Missonnier}, {Didelon}, \& {Morin}}]{SWarp_Bertin_2002}
{Bertin}, E., {Mellier}, Y., {Radovich}, M., {et~al.} 2002, in Astronomical Society of the Pacific Conference Series, Vol. 281, Astronomical Data Analysis Software and Systems XI, ed. D.~A. {Bohlender}, D.~{Durand}, \& T.~H. {Handley}, 228

\bibitem[{{Bertin, E.} \& {Arnouts, S.}(1996)}]{SExtractor}
{Bertin, E.}, \& {Arnouts, S.} 1996, Astron. Astrophys. Suppl. Ser., 117, 393, \dodoi{10.1051/aas:1996164}

\bibitem[{Boquien {et~al.}(2019)Boquien, Burgarella, Roehlly, Buat, Ciesla, Corre, Inoue, \& Salas}]{Boquien_cigale_2019}
Boquien, M., Burgarella, D., Roehlly, Y., {et~al.} 2019, Astronomy \&; Astrophysics, 622, A103, \dodoi{10.1051/0004-6361/201834156}

\bibitem[{{Boucaud, A.} {et~al.}(2016){Boucaud, A.}, {Dole, H.}, {Abergel, A.}, {Ayasso, H.}, \& {Orieux, F.}}]{Boucaud_pypher_2016}
{Boucaud, A.}, {Dole, H.}, {Abergel, A.}, {Ayasso, H.}, \& {Orieux, F.} 2016, EAS Publications Series, 78, 275, \dodoi{10.1051/eas/1678013}

\bibitem[{{Brammer}(2016)}]{Brammer_TVB_2016}
{Brammer}, G. 2016, {Reprocessing WFC3/IR Exposures Affected by Time-Variable Backgrounds}, Instrument Science Report WFC3 2016-16, 12 pages

\bibitem[{{Bressan} {et~al.}(2012){Bressan}, {Marigo}, {Girardi}, {Salasnich}, {Dal Cero}, {Rubele}, \& {Nanni}}]{Bressan2012}
{Bressan}, A., {Marigo}, P., {Girardi}, L., {et~al.} 2012, \mnras, 427, 127, \dodoi{10.1111/j.1365-2966.2012.21948.x}

\bibitem[{Bromm \& Loeb(2003)}]{Bromm_2003dcbh}
Bromm, V., \& Loeb, A. 2003, The Astrophysical Journal, 596, 34–46, \dodoi{10.1086/377529}

\bibitem[{{Bruzual} \& {Charlot}(2003)}]{BruzualCharlot_2003}
{Bruzual}, G., \& {Charlot}, S. 2003, \mnras, 344, 1000, \dodoi{10.1046/j.1365-8711.2003.06897.x}

\bibitem[{{Bunker} {et~al.}(2024){Bunker}, {Cameron}, {Curtis-Lake}, {Jakobsen}, {Carniani}, {Curti}, {Witstok}, {Maiolino}, {D'Eugenio}, {Looser}, {Willott}, {Bonaventura}, {Hainline}, {{\"U}bler}, {Willmer}, {Saxena}, {Smit}, {Alberts}, {Arribas}, {Baker}, {Baum}, {Bhatawdekar}, {Bowler}, {Boyett}, {Charlot}, {Chen}, {Chevallard}, {Circosta}, {DeCoursey}, {de Graaff}, {Egami}, {Eisenstein}, {Endsley}, {Ferruit}, {Giardino}, {Hausen}, {Helton}, {Hviding}, {Ji}, {Johnson}, {Jones}, {Kumari}, {Laseter}, {L{\"u}tzgendorf}, {Maseda}, {Nelson}, {Parlanti}, {Perna}, {Rauscher}, {Rawle}, {Rix}, {Rieke}, {Robertson}, {Rodr{\'\i}guez Del Pino}, {Sandles}, {Scholtz}, {Sharpe}, {Skarbinski}, {Stark}, {Sun}, {Tacchella}, {Topping}, {Villanueva}, {Wallace}, {Williams}, \& {Woodrum}}]{bunkerspecz2023}
{Bunker}, A.~J., {Cameron}, A.~J., {Curtis-Lake}, E., {et~al.} 2024, \aap, 690, A288, \dodoi{10.1051/0004-6361/202347094}

\bibitem[{{Burke} {et~al.}(2022){Burke}, {Liu}, {Shen}, {Phadke}, {Yang}, {Hartley}, {Harrison}, {Palmese}, {Guo}, {Zhang}, {Kron}, {Turner}, {Giles}, {Lidman}, {Chen}, {Gruendl}, {Choi}, {Amon}, {Sheldon}, {Aguena}, {Allam}, {Andrade-Oliveira}, {Bacon}, {Bertin}, {Brooks}, {Rosell}, {Kind}, {Carretero}, {Conselice}, {Costanzi}, {da Costa}, {Pereira}, {Davis}, {De Vicente}, {Desai}, {Diehl}, {Everett}, {Ferrero}, {Flaugher}, {Garc{\'\i}a-Bellido}, {Gaztanaga}, {Gruen}, {Gschwend}, {Gutierrez}, {Hinton}, {Hollowood}, {Honscheid}, {Hoyle}, {James}, {Kuehn}, {Maia}, {Marshall}, {Menanteau}, {Miquel}, {Morgan}, {Paz-Chinch{\'o}n}, {Pieres}, {Malag{\'o}n}, {Reil}, {Romer}, {Sanchez}, {Schubnell}, {Serrano}, {Sevilla-Noarbe}, {Smith}, {Suchyta}, {Tarle}, {Thomas}, {To}, {Varga}, {Wilkinson}, \& {DES Collaboration}}]{Burke_dwarfgals_2022}
{Burke}, C.~J., {Liu}, X., {Shen}, Y., {et~al.} 2022, \mnras, 516, 2736, \dodoi{10.1093/mnras/stac2262}

\bibitem[{Burnham \& Anderson(2002)}]{statsAICbook}
Burnham, K.~P., \& Anderson, D.~R. 2002, Model Selection and Multimodel Inference: \textit{A Practical Information-Theoretic Approach}, 2nd edn. (Springer)

\bibitem[{{Calzetti} {et~al.}(2000){Calzetti}, {Armus}, {Bohlin}, {Kinney}, {Koornneef}, \& {Storchi-Bergmann}}]{calzetti_2000}
{Calzetti}, D., {Armus}, L., {Bohlin}, R.~C., {et~al.} 2000, \apj, 533, 682, \dodoi{10.1086/308692}

\bibitem[{Cammelli {et~al.}(2024)Cammelli, Monaco, Tan, Singh, Fontanot, Lucia, Hirschmann, \& Xie}]{cammelli2024popiii.1}
Cammelli, V., Monaco, P., Tan, J.~C., {et~al.} 2024, The formation of supermassive black holes from Population III.1 seeds. III. Galaxy evolution and black hole growth from semi-analytic modelling.
\newblock \doarXiv{2407.09949}

\bibitem[{Cammelli {et~al.}(2025)Cammelli, Tan, Young, Hayes, Singh, Ellis, Saxena, Laporte, Monaco, \& Keller}]{cammelli_glimmers25}
Cammelli, V., Tan, J.~C., Young, A.~R., {et~al.} 2025, Glimmers in the Cosmic Dawn. II. A variability census of supermassive black holes across the Universe.
\newblock \doarXiv{2501.17675}

\bibitem[{Carnall {et~al.}(2018)Carnall, McLure, Dunlop, \& Davé}]{BagpipesCarnall_2018}
Carnall, A.~C., McLure, R.~J., Dunlop, J.~S., \& Davé, R. 2018, Monthly Notices of the Royal Astronomical Society, 480, 4379–4401, \dodoi{10.1093/mnras/sty2169}

\bibitem[{Chon {et~al.}(2016)Chon, Hirano, Hosokawa, \& Yoshida}]{Chon_2016DCBH}
Chon, S., Hirano, S., Hosokawa, T., \& Yoshida, N. 2016, The Astrophysical Journal, 832, 134, \dodoi{10.3847/0004-637x/832/2/134}

\bibitem[{{Cohen} {et~al.}(2006){Cohen}, {Ryan}, {Straughn}, {Hathi}, {Windhorst}, {Koekemoer}, {Pirzkal}, {Xu}, {Mobasher}, {Malhotra}, {Strolger}, \& {Rhoads}}]{cohen_2006}
{Cohen}, S.~H., {Ryan}, Jr., R.~E., {Straughn}, A.~N., {et~al.} 2006, \apj, 639, 731, \dodoi{10.1086/499577}

\bibitem[{Dalla~Bontà {et~al.}(2025)Dalla~Bontà, Peterson, Grier, Berton, Brandt, Ciroi, Corsini, Dalla~Barba, Davies, Dehghanian, Edelson, Foschini, Gasparri, Ho, Horne, Iodice, Morelli, Pizzella, Portaluri, Shen, Schneider, \& Vestergaard}]{Dalla_Bont_2025compGRAVtoSE}
Dalla~Bontà, E., Peterson, B.~M., Grier, C.~J., {et~al.} 2025, Astronomy \& Astrophysics, 696, A48, \dodoi{10.1051/0004-6361/202452746}

\bibitem[{DeCoursey {et~al.}(2025)DeCoursey, Egami, Pierel, Sun, Rest, Coulter, Engesser, Siebert, Hainline, Johnson, Bunker, Cargile, Charlot, Chen, Curti, DeFour-Remy, Eisenstein, Fox, Gezari, Gomez, Jencson, Joshi, Khairnar, Lyu, Maiolino, Moriya, Quimby, Rieke, Rieke, Robertson, Shahbandeh, Strolger, Tacchella, Wang, Williams, Willmer, Willott, \& Zenati}]{decoursey2025_1511stellar}
DeCoursey, C., Egami, E., Pierel, J. D.~R., {et~al.} 2025, The JADES Transient Survey: Discovery and Classification of Supernovae in the JADES Deep Field.
\newblock \doarXiv{2406.05060}

\bibitem[{{Eisenstein} {et~al.}(2023){Eisenstein}, {Johnson}, {Robertson}, {Tacchella}, {Hainline}, {Jakobsen}, {Maiolino}, {Bonaventura}, {Bunker}, {Cameron}, {Cargile}, {Curtis-Lake}, {Hausen}, {Pusk{\'a}s}, {Rieke}, {Sun}, {Willmer}, {Willott}, {Alberts}, {Arribas}, {Baker}, {Baum}, {Bhatawdekar}, {Carniani}, {Charlot}, {Chen}, {Chevallard}, {Curti}, {DeCoursey}, {D'Eugenio}, {de Graaff}, {Egami}, {Helton}, {Ji}, {Jones}, {Kumari}, {L{\"u}tzgendorf}, {Laseter}, {Looser}, {Lyu}, {Maseda}, {Nelson}, {Parlanti}, {Rauscher}, {Rawle}, {Rieke}, {Rix}, {Rujopakarn}, {Sandles}, {Saxena}, {Scholtz}, {Sharpe}, {Shivaei}, {Simmonds}, {Smit}, {Topping}, {{\"U}bler}, {Venturi}, {Williams}, {Witstok}, \& {Woodrum}}]{JADES_DR2}
{Eisenstein}, D.~J., {Johnson}, B.~D., {Robertson}, B., {et~al.} 2023, arXiv e-prints, arXiv:2310.12340, \dodoi{10.48550/arXiv.2310.12340}

\bibitem[{{Falc{\'o}n-Barroso} {et~al.}(2011){Falc{\'o}n-Barroso}, {S{\'a}nchez-Bl{\'a}zquez}, {Vazdekis}, {Ricciardelli}, {Cardiel}, {Cenarro}, {Gorgas}, \& {Peletier}}]{MILES2011}
{Falc{\'o}n-Barroso}, J., {S{\'a}nchez-Bl{\'a}zquez}, P., {Vazdekis}, A., {et~al.} 2011, \aap, 532, A95, \dodoi{10.1051/0004-6361/201116842}

\bibitem[{Fan {et~al.}(2022)Fan, Banados, \& Simcoe}]{fan2022}
Fan, X., Banados, E., \& Simcoe, R.~A. 2022, Quasars and the Intergalactic Medium at Cosmic Dawn.
\newblock \doarXiv{2212.06907}

\bibitem[{Fan {et~al.}(1999)Fan, Strauss, Schneider, Gunn, Lupton, Yanny, Anderson, Anderson, Annis, Bahcall, Bakken, Bastian, Berman, Boroski, Briegel, Briggs, Brinkmann, Carr, Colestock, Connolly, Crocker, Csabai, Czarapata, Davis, Doi, Elms, Evans, Federwitz, Frieman, Fukugita, Gurbani, Harris, Heckman, Hennessy, Hindsley, Holmgren, Hull, Ichikawa, Ichikawa, Ivezić, Kent, Knapp, Kron, Lamb, Leger, Limmongkol, Lindenmeyer, Long, Loveday, MacKinnon, Mannery, Mantsch, Margon, McKay, Munn, Nash, Newberg, Nichol, Nicinski, Okamura, Ostriker, Owen, Pauls, Peoples, Petravick, Pier, Pordes, Prosapio, Rechenmacher, Richards, Richmond, Rivetta, Rockosi, Sandford, Sergey, Sekiguchi, Shimasaku, Siegmund, Smith, Stoughton, Szalay, Szokoly, Tucker, Vogeley, Waddell, Wang, Weinberg, Yasuda, \& York}]{Fan_1999}
Fan, X., Strauss, M.~A., Schneider, D.~P., {et~al.} 1999, The Astronomical Journal, 118, 1–13, \dodoi{10.1086/300944}

\bibitem[{Fan {et~al.}(2001)Fan, Narayanan, Lupton, Strauss, Knapp, Becker, White, Pentericci, Leggett, Haiman, Gunn, Ivezić, Schneider, Anderson, Brinkmann, Bahcall, Connolly, Csabai, Doi, Fukugita, Geballe, Grebel, Harbeck, Hennessy, Lamb, Miknaitis, Munn, Nichol, Okamura, Pier, Prada, Richards, Szalay, \& York}]{Fan_2001}
Fan, X., Narayanan, V.~K., Lupton, R.~H., {et~al.} 2001, The Astronomical Journal, 122, 2833–2849, \dodoi{10.1086/324111}

\bibitem[{{Ferland} {et~al.}(2017){Ferland}, {Chatzikos}, {Guzm{\'a}n}, {Lykins}, {van Hoof}, {Williams}, {Abel}, {Badnell}, {Keenan}, {Porter}, \& {Stancil}}]{Ferland2017}
{Ferland}, G.~J., {Chatzikos}, M., {Guzm{\'a}n}, F., {et~al.} 2017, \rmxaa, 53, 385.
\newblock \doarXiv{1705.10877}

\bibitem[{{Fonseca Alvarez} {et~al.}(2020){Fonseca Alvarez}, {Trump}, {Homayouni}, {Grier}, {Shen}, {Horne}, {Li}, {Brandt}, {Ho}, {Peterson}, \& {Schneider}}]{SDSS_signleepoch_2020}
{Fonseca Alvarez}, G., {Trump}, J.~R., {Homayouni}, Y., {et~al.} 2020, \apj, 899, 73, \dodoi{10.3847/1538-4357/aba001}

\bibitem[{Foreman-Mackey {et~al.}(2013)Foreman-Mackey, Hogg, Lang, \& Goodman}]{emcee_Foreman_Mackey_2013}
Foreman-Mackey, D., Hogg, D.~W., Lang, D., \& Goodman, J. 2013, Publications of the Astronomical Society of the Pacific, 125, 306–312, \dodoi{10.1086/670067}

\bibitem[{Freitag {et~al.}(2006)Freitag, Rasio, \& Baumgardt}]{Freitag_2006mergsc}
Freitag, M., Rasio, F.~A., \& Baumgardt, H. 2006, Monthly Notices of the Royal Astronomical Society, 368, 121–140, \dodoi{10.1111/j.1365-2966.2006.10095.x}

\bibitem[{{Furtak} {et~al.}(2023){Furtak}, {Zitrin}, {Plat}, {Fujimoto}, {Wang}, {Nelson}, {Labb{\'e}}, {Bezanson}, {Brammer}, {van Dokkum}, {Endsley}, {Glazebrook}, {Greene}, {Leja}, {Price}, {Smit}, {Stark}, {Weaver}, {Whitaker}, {Atek}, {Chevallard}, {Curtis-Lake}, {Dayal}, {Feltre}, {Franx}, {Fudamoto}, {Marchesini}, {Mowla}, {Pan}, {Suess}, {Vidal-Garc{\'\i}a}, \& {Williams}}]{LRDFurtak_2023}
{Furtak}, L.~J., {Zitrin}, A., {Plat}, A., {et~al.} 2023, \apj, 952, 142, \dodoi{10.3847/1538-4357/acdc9d}

\bibitem[{Furtak {et~al.}(2024)Furtak, Labbé, Zitrin, Greene, Dayal, Chemerynska, Kokorev, Miller, Goulding, de~Graaff, Bezanson, Brammer, Cutler, Leja, Pan, Price, Wang, Weaver, Whitaker, Atek, Bogdán, Charlot, Curtis-Lake, van Dokkum, Endsley, Feldmann, Fudamoto, Fujimoto, Glazebrook, Juneau, Marchesini, Maseda, Nelson, Oesch, Plat, Setton, Stark, \& Williams}]{furtak_high_2024}
Furtak, L.~J., Labbé, I., Zitrin, A., {et~al.} 2024, Nature, 628, 57, \dodoi{10.1038/s41586-024-07184-8}

\bibitem[{{Granato} {et~al.}(2004){Granato}, {De Zotti}, {Silva}, {Bressan}, \& {Danese}}]{Granato_2004}
{Granato}, G.~L., {De Zotti}, G., {Silva}, L., {Bressan}, A., \& {Danese}, L. 2004, \apj, 600, 580, \dodoi{10.1086/379875}

\bibitem[{{Gravity Collaboration} {et~al.}(2018){Gravity Collaboration}, {Sturm}, {Dexter}, {Pfuhl}, {Stock}, {Davies}, {Lutz}, {Cl{\'e}net}, {Eckart}, {Eisenhauer}, {Genzel}, {Gratadour}, {H{\"o}nig}, {Kishimoto}, {Lacour}, {Millour}, {Netzer}, {Perrin}, {Peterson}, {Petrucci}, {Rouan}, {Waisberg}, {Woillez}, {Amorim}, {Brandner}, {F{\"o}rster Schreiber}, {Garcia}, {Gillessen}, {Ott}, {Paumard}, {Perraut}, {Scheithauer}, {Straubmeier}, {Tacconi}, \& {Widmann}}]{gravityquasarmass_2018}
{Gravity Collaboration}, {Sturm}, E., {Dexter}, J., {et~al.} 2018, \nat, 563, 657, \dodoi{10.1038/s41586-018-0731-9}

\bibitem[{Greene \& Ho(2005)}]{Greene_2005}
Greene, J.~E., \& Ho, L.~C. 2005, The Astrophysical Journal, 630, 122–129, \dodoi{10.1086/431897}

\bibitem[{Greene {et~al.}(2020)Greene, Strader, \& Ho}]{Greene_IMBHreview_2020}
Greene, J.~E., Strader, J., \& Ho, L.~C. 2020, Annual Review of Astronomy and Astrophysics, 58, 257–312, \dodoi{10.1146/annurev-astro-032620-021835}

\bibitem[{{Greene} {et~al.}(2024){Greene}, {Labbe}, {Goulding}, {Furtak}, {Chemerynska}, {Kokorev}, {Dayal}, {Volonteri}, {Williams}, {Wang}, {Setton}, {Burgasser}, {Bezanson}, {Atek}, {Brammer}, {Cutler}, {Feldmann}, {Fujimoto}, {Glazebrook}, {de Graaff}, {Khullar}, {Leja}, {Marchesini}, {Maseda}, {Matthee}, {Miller}, {Naidu}, {Nanayakkara}, {Oesch}, {Pan}, {Papovich}, {Price}, {van Dokkum}, {Weaver}, {Whitaker}, \& {Zitrin}}]{Greene_2024LRDs}
{Greene}, J.~E., {Labbe}, I., {Goulding}, A.~D., {et~al.} 2024, \apj, 964, 39, \dodoi{10.3847/1538-4357/ad1e5f}

\bibitem[{{Haehnelt} \& {Rees}(1993)}]{haehneltrees_1993DCBH}
{Haehnelt}, M.~G., \& {Rees}, M.~J. 1993, \mnras, 263, 168, \dodoi{10.1093/mnras/263.1.168}

\bibitem[{{Haiman} \& {Loeb}(2001)}]{haiman_loeb_2001}
{Haiman}, Z., \& {Loeb}, A. 2001, \apj, 552, 459, \dodoi{10.1086/320586}

\bibitem[{{Harikane} {et~al.}(2023){Harikane}, {Zhang}, {Nakajima}, {Ouchi}, {Isobe}, {Ono}, {Hatano}, {Xu}, \& {Umeda}}]{Harikane_23MBH}
{Harikane}, Y., {Zhang}, Y., {Nakajima}, K., {et~al.} 2023, \apj, 959, 39, \dodoi{10.3847/1538-4357/ad029e}

\bibitem[{Hayes {et~al.}(2024)Hayes, Tan, Ellis, Young, Cammelli, Singh, Runnholm, Saxena, Lunnan, Keller, Monaco, Laporte, \& Melinder}]{Hayes_Glimmers_2024}
Hayes, M.~J., Tan, J.~C., Ellis, R.~S., {et~al.} 2024, The Astrophysical Journal Letters, 971, L16, \dodoi{10.3847/2041-8213/ad63a7}

\bibitem[{Ho \& Kim(2016)}]{Ho_2016}
Ho, L.~C., \& Kim, M. 2016, The Astrophysical Journal, 821, 48, \dodoi{10.3847/0004-637x/821/1/48}

\bibitem[{{Hogg} \& {Foreman-Mackey}(2018)}]{Hogg_formanMackey_2018MCMCrecipes}
{Hogg}, D.~W., \& {Foreman-Mackey}, D. 2018, \apjs, 236, 11, \dodoi{10.3847/1538-4365/aab76e}

\bibitem[{{Hook} \& {McMahon}(1998)}]{hook_1998}
{Hook}, I.~M., \& {McMahon}, R.~G. 1998, \mnras, 294, L7, \dodoi{10.1046/j.1365-8711.1998.01368.x10.1111/j.1365-8711.1998.01368.x}

\bibitem[{{Hopkins} {et~al.}(2006){Hopkins}, {Hernquist}, {Cox}, {Di Matteo}, {Robertson}, \& {Springel}}]{hopkins_2006}
{Hopkins}, P.~F., {Hernquist}, L., {Cox}, T.~J., {et~al.} 2006, \apjs, 163, 1, \dodoi{10.1086/499298}

\bibitem[{{Illingworth} {et~al.}(2013){Illingworth}, {Magee}, {Oesch}, {Bouwens}, {Labb{\'e}}, {Stiavelli}, {van Dokkum}, {Franx}, {Trenti}, {Carollo}, \& {Gonzalez}}]{Illingworth_XDF_2013}
{Illingworth}, G.~D., {Magee}, D., {Oesch}, P.~A., {et~al.} 2013, \apjs, 209, 6, \dodoi{10.1088/0067-0049/209/1/6}

\bibitem[{Inayoshi {et~al.}(2020)Inayoshi, Visbal, \& Haiman}]{InayoshiReview_2020}
Inayoshi, K., Visbal, E., \& Haiman, Z. 2020, Annual Review of Astronomy and Astrophysics, 58, 27–97, \dodoi{10.1146/annurev-astro-120419-014455}

\bibitem[{{Johnson} \& {Bromm}(2007)}]{johnsonbromm_100Myr_2007}
{Johnson}, J.~L., \& {Bromm}, V. 2007, \mnras, 374, 1557, \dodoi{10.1111/j.1365-2966.2006.11275.x}

\bibitem[{{Kaspi} {et~al.}(2000){Kaspi}, {Smith}, {Netzer}, {Maoz}, {Jannuzi}, \& {Giveon}}]{kaspi_2000_MBH}
{Kaspi}, S., {Smith}, P.~S., {Netzer}, H., {et~al.} 2000, \apj, 533, 631, \dodoi{10.1086/308704}

\bibitem[{Kimbro {et~al.}(2025)Kimbro, Baldassare, Worthey, Geha, \& Greene}]{Kimbro_dwarf_2025}
Kimbro, E., Baldassare, V., Worthey, G., Geha, M., \& Greene, J. 2025, The Astrophysical Journal, 980, 215, \dodoi{10.3847/1538-4357/adaebf}

\bibitem[{{Kocevski} {et~al.}(2023){Kocevski}, {Onoue}, {Inayoshi}, {Trump}, {Arrabal Haro}, {Grazian}, {Dickinson}, {Finkelstein}, {Kartaltepe}, {Hirschmann}, {Aird}, {Holwerda}, {Fujimoto}, {Juneau}, {Amor{\'\i}n}, {Backhaus}, {Bagley}, {Barro}, {Bell}, {Bisigello}, {Calabr{\`o}}, {Cleri}, {Cooper}, {Ding}, {Grogin}, {Ho}, {Hutchison}, {Inoue}, {Jiang}, {Jones}, {Koekemoer}, {Li}, {Li}, {McGrath}, {Molina}, {Papovich}, {P{\'e}rez-Gonz{\'a}lez}, {Pirzkal}, {Wilkins}, {Yang}, \& {Yung}}]{Kocevski_monsters_23MBH}
{Kocevski}, D.~D., {Onoue}, M., {Inayoshi}, K., {et~al.} 2023, \apjl, 954, L4, \dodoi{10.3847/2041-8213/ace5a0}

\bibitem[{{Koekemoer} {et~al.}(2013){Koekemoer}, {Ellis}, {McLure}, {Dunlop}, {Robertson}, {Ono}, {Schenker}, {Ouchi}, {Bowler}, {Rogers}, {Curtis-Lake}, {Schneider}, {Charlot}, {Stark}, {Furlanetto}, {Cirasuolo}, {Wild}, \& {Targett}}]{Koekemoer_HUDF_2012}
{Koekemoer}, A.~M., {Ellis}, R.~S., {McLure}, R.~J., {et~al.} 2013, \apjs, 209, 3, \dodoi{10.1088/0067-0049/209/1/3}

\bibitem[{Kokorev {et~al.}(2023)Kokorev, Fujimoto, Labbe, Greene, Bezanson, Dayal, Nelson, Atek, Brammer, Caputi, Chemerynska, Cutler, Feldmann, Fudamoto, Furtak, Goulding, de~Graaff, Leja, Marchesini, Miller, Nanayakkara, Oesch, Pan, Price, Setton, Smit, Stefanon, Wang, Weaver, Whitaker, Williams, \& Zitrin}]{kokorev_uncover_2023}
Kokorev, V., Fujimoto, S., Labbe, I., {et~al.} 2023, The Astrophysical Journal Letters, 957, L7, \dodoi{10.3847/2041-8213/ad037a}

\bibitem[{{Kokorev} {et~al.}(2024){Kokorev}, {Caputi}, {Greene}, {Dayal}, {Trebitsch}, {Cutler}, {Fujimoto}, {Labb{\'e}}, {Miller}, {Iani}, {Navarro-Carrera}, \& {Rinaldi}}]{LRDKokorev_2024}
{Kokorev}, V., {Caputi}, K.~I., {Greene}, J.~E., {et~al.} 2024, \apj, 968, 38, \dodoi{10.3847/1538-4357/ad4265}

\bibitem[{Korista \& Goad(2004)}]{Korista_2004}
Korista, K.~T., \& Goad, M.~R. 2004, The Astrophysical Journal, 606, 749–762, \dodoi{10.1086/383193}

\bibitem[{Kormendy \& Ho(2013)}]{KormendyHo_2013}
Kormendy, J., \& Ho, L.~C. 2013, Annual Review of Astronomy and Astrophysics, 51, 511–653, \dodoi{10.1146/annurev-astro-082708-101811}

\bibitem[{Koudmani {et~al.}(2022)Koudmani, Sijacki, \& Smith}]{Koudmani_2022}
Koudmani, S., Sijacki, D., \& Smith, M.~C. 2022, Monthly Notices of the Royal Astronomical Society, 516, 2112–2141, \dodoi{10.1093/mnras/stac2252}

\bibitem[{{Kroupa}(2001)}]{Kroupa2001}
{Kroupa}, P. 2001, \mnras, 322, 231, \dodoi{10.1046/j.1365-8711.2001.04022.x}

\bibitem[{Labbe {et~al.}(2024)Labbe, Greene, Matthee, Treiber, Kokorev, Miller, Kramarenko, Setton, Ma, Goulding, Bezanson, Naidu, Williams, Atek, Brammer, Cutler, Chemerynska, Cloonan, Dayal, de~Graaff, Fudamoto, Fujimoto, Furtak, Glazebrook, Heintz, Leja, Marchesini, Nanayakkara, Nelson, Oesch, Pan, Price, Shivaei, Sobral, Suess, van Dokkum, Wang, Weaver, Whitaker, \& Zitrin}]{labbe2024LRD}
Labbe, I., Greene, J.~E., Matthee, J., {et~al.} 2024, An unambiguous AGN and a Balmer break in an Ultraluminous Little Red Dot at z=4.47 from Ultradeep UNCOVER and All the Little Things Spectroscopy.
\newblock \doarXiv{2412.04557}

\bibitem[{Larson {et~al.}(2023)Larson, Finkelstein, Kocevski, Hutchison, Trump, Arrabal~Haro, Bromm, Cleri, Dickinson, Fujimoto, Kartaltepe, Koekemoer, Papovich, Pirzkal, Tacchella, Zavala, Bagley, Behroozi, Champagne, Cole, Jung, Morales, Yang, Zhang, Zitrin, Amorín, Burgarella, Casey, Chávez~Ortiz, Cox, Chworowsky, Fontana, Gawiser, Grazian, Grogin, Harish, Hathi, Hirschmann, Holwerda, Juneau, Leung, Lucas, McGrath, Pérez-González, Rigby, Seillé, Simons, de~la Vega, Weiner, Wilkins, Yung, \& Team}]{larson_ceers_2023}
Larson, R.~L., Finkelstein, S.~L., Kocevski, D.~D., {et~al.} 2023, The Astrophysical Journal Letters, 953, L29, \dodoi{10.3847/2041-8213/ace619}

\bibitem[{Leung {et~al.}(2024)Leung, Finkelstein, Pérez-González, Morales, Taylor, Barro, Kocevski, Akins, Carnall, Óscar A.~Chávez~Ortiz, Cleri, Cullen, Donnan, Dunlop, Ellis, Grogin, Hirschmann, Koekemoer, Kokorev, Lucas, McLeod, Papovich, \& Yung}]{leung2024_LRDs}
Leung, G. C.~K., Finkelstein, S.~L., Pérez-González, P.~G., {et~al.} 2024, Exploring the Nature of Little Red Dots: Constraints on AGN and Stellar Contributions from PRIMER MIRI Imaging.
\newblock \doarXiv{2411.12005}

\bibitem[{{Li} {et~al.}(2007){Li}, {Hernquist}, {Robertson}, {Cox}, {Hopkins}, {Springel}, {Gao}, {Di Matteo}, {Zentner}, {Jenkins}, \& {Yoshida}}]{Li_2007}
{Li}, Y., {Hernquist}, L., {Robertson}, B., {et~al.} 2007, \apj, 665, 187, \dodoi{10.1086/519297}

\bibitem[{{Lim} {et~al.}(2020){Lim}, {Davis}, {Hack}, {Grogin}, {Ogaz}, {Ubeda}, {Cara}, {Borncamp}, \& {Miles}}]{acstools}
{Lim}, P.~L., {Davis}, M., {Hack}, W., {et~al.} 2020, {ACStools: Python tools for Hubble Space Telescope Advanced Camera for Surveys data}, Astrophysics Source Code Library, record ascl:2011.024

\bibitem[{Lin {et~al.}(2025)Lin, Fan, Wang, Sun, Champagne, Egami, Kakiichi, Lyu, Tee, Yang, Bian, Bosman, Cai, Casey, Decarli, Faisst, Fujimoto, Harish, Ilbert, Inoue, Jin, Kartaltepe, Kocevski, Li, Liu, Liu, Schindler, Shuntov, Tanaka, Vestergaard, Wu, Zhang, \& Zhang}]{lin_2025}
Lin, X., Fan, X., Wang, F., {et~al.} 2025, Bridging {Quasars} and {Little} {Red} {Dots}: {Insights} into {Broad}-{Line} {AGNs} at \$z=5-8\$ from the {First} {JWST} {COSMOS}-{3D} {Dataset},  arXiv, \dodoi{10.48550/arXiv.2504.08039}

\bibitem[{Lyu {et~al.}(2022)Lyu, Alberts, Rieke, \& Rujopakarn}]{Lyu_2022goods-sAGN}
Lyu, J., Alberts, S., Rieke, G.~H., \& Rujopakarn, W. 2022, The Astrophysical Journal, 941, 191, \dodoi{10.3847/1538-4357/ac9e5d}

\bibitem[{Ma {et~al.}(2025)Ma, Greene, Setton, Volonteri, Leja, Wang, Bezanson, Brammer, Cutler, Dayal, van Dokkum, Furtak, Glazebrook, Goulding, de~Graaff, Kokorev, Labbe, Pan, Price, Weaver, Williams, Whitaker, \& Zitrin}]{Ma_2025_LRDs}
Ma, Y., Greene, J.~E., Setton, D.~J., {et~al.} 2025, The Astrophysical Journal, 981, 191, \dodoi{10.3847/1538-4357/ada613}

\bibitem[{{Madau}(1995)}]{Madau_IGM_1995}
{Madau}, P. 1995, \apj, 441, 18, \dodoi{10.1086/175332}

\bibitem[{{Madau} \& {Rees}(2001)}]{Madau_Rees_2001}
{Madau}, P., \& {Rees}, M.~J. 2001, \apjl, 551, L27, \dodoi{10.1086/319848}

\bibitem[{Maiolino {et~al.}(2024{\natexlab{a}})Maiolino, Scholtz, Curtis-Lake, Carniani, Baker, de~Graaff, Tacchella, Übler, D’Eugenio, Witstok, Curti, Arribas, Bunker, Charlot, Chevallard, Eisenstein, Egami, Ji, Jones, Lyu, Rawle, Robertson, Rujopakarn, Perna, Sun, Venturi, Williams, \& Willott}]{Maiolino_2024}
Maiolino, R., Scholtz, J., Curtis-Lake, E., {et~al.} 2024{\natexlab{a}}, Astronomy \& Astrophysics, 691, A145, \dodoi{10.1051/0004-6361/202347640}

\bibitem[{Maiolino {et~al.}(2024{\natexlab{b}})Maiolino, {Scholtz}, {Witstok}, {Carniani}, {D'Eugenio}, {de Graaff}, {{\"U}bler}, {Tacchella}, {Curtis-Lake}, {Arribas}, {Bunker}, {Charlot}, {Chevallard}, {Curti}, {Looser}, {Maseda}, {Rawle}, {Rodr{\'\i}guez del Pino}, {Willott}, {Egami}, {Eisenstein}, {Hainline}, {Robertson}, {Williams}, {Willmer}, {Baker}, {Boyett}, {DeCoursey}, {Fabian}, {Helton}, {Ji}, {Jones}, {Kumari}, {Laporte}, {Nelson}, {Perna}, {Sandles}, {Shivaei}, \& {Sun}}]{maiolino_gnz112024a}
Maiolino, R., {Scholtz}, J., {Witstok}, J., {et~al.} 2024{\natexlab{b}}, \nat, 627, 59, \dodoi{10.1038/s41586-024-07052-5}

\bibitem[{{Marigo} {et~al.}(2013){Marigo}, {Bressan}, {Nanni}, {Girardi}, \& {Pumo}}]{Marigo2013}
{Marigo}, P., {Bressan}, A., {Nanni}, A., {Girardi}, L., \& {Pumo}, M.~L. 2013, \mnras, 434, 488, \dodoi{10.1093/mnras/stt1034}

\bibitem[{Marshall {et~al.}(2022)Marshall, Auger-Williams, Banerji, Maiolino, \& Bowler}]{Marshall_2022}
Marshall, A., Auger-Williams, M.~W., Banerji, M., Maiolino, R., \& Bowler, R. 2022, Monthly Notices of the Royal Astronomical Society, 515, 5617–5628, \dodoi{10.1093/mnras/stac1619}

\bibitem[{{Matsuoka} {et~al.}(2018){Matsuoka}, {Strauss}, {Kashikawa}, {Onoue}, {Iwasawa}, {Tang}, {Lee}, {Imanishi}, {Nagao}, {Akiyama}, {Asami}, {Bosch}, {Furusawa}, {Goto}, {Gunn}, {Harikane}, {Ikeda}, {Izumi}, {Kawaguchi}, {Kato}, {Kikuta}, {Kohno}, {Komiyama}, {Lupton}, {Minezaki}, {Miyazaki}, {Murayama}, {Niida}, {Nishizawa}, {Noboriguchi}, {Oguri}, {Ono}, {Ouchi}, {Price}, {Sameshima}, {Schulze}, {Shirakata}, {Silverman}, {Sugiyama}, {Tait}, {Takada}, {Takata}, {Tanaka}, {Toba}, {Utsumi}, {Wang}, \& {Yamashita}}]{Matsuoka_2018a_highzqso}
{Matsuoka}, Y., {Strauss}, M.~A., {Kashikawa}, N., {et~al.} 2018, \apj, 869, 150, \dodoi{10.3847/1538-4357/aaee7a}

\bibitem[{{Matsuoka} {et~al.}(2019){Matsuoka}, {Onoue}, {Kashikawa}, {Strauss}, {Iwasawa}, {Lee}, {Imanishi}, {Nagao}, {Akiyama}, {Asami}, {Bosch}, {Furusawa}, {Goto}, {Gunn}, {Harikane}, {Ikeda}, {Izumi}, {Kawaguchi}, {Kato}, {Kikuta}, {Kohno}, {Komiyama}, {Koyama}, {Lupton}, {Minezaki}, {Miyazaki}, {Murayama}, {Niida}, {Nishizawa}, {Noboriguchi}, {Oguri}, {Ono}, {Ouchi}, {Price}, {Sameshima}, {Schulze}, {Shirakata}, {Silverman}, {Sugiyama}, {Tait}, {Takada}, {Takata}, {Tanaka}, {Tang}, {Toba}, {Utsumi}, {Wang}, \& {Yamashita}}]{Matsuoka_2019b_highzqso}
{Matsuoka}, Y., {Onoue}, M., {Kashikawa}, N., {et~al.} 2019, \apjl, 872, L2, \dodoi{10.3847/2041-8213/ab0216}

\bibitem[{{Matthee} {et~al.}(2024){Matthee}, {Naidu}, {Brammer}, {Chisholm}, {Eilers}, {Goulding}, {Greene}, {Kashino}, {Labbe}, {Lilly}, {Mackenzie}, {Oesch}, {Weibel}, {Wuyts}, {Xiao}, {Bordoloi}, {Bouwens}, {van Dokkum}, {Illingworth}, {Kramarenko}, {Maseda}, {Mason}, {Meyer}, {Nelson}, {Reddy}, {Shivaei}, {Simcoe}, \& {Yue}}]{LRDMatthee_2024}
{Matthee}, J., {Naidu}, R.~P., {Brammer}, G., {et~al.} 2024, \apj, 963, 129, \dodoi{10.3847/1538-4357/ad2345}

\bibitem[{McCaffrey {et~al.}(2025)McCaffrey, Regan, Smith, Wise, O’Shea, \& Norman}]{McCaffrey_2025_heavyseeds}
McCaffrey, J., Regan, J., Smith, B., {et~al.} 2025, The Open Journal of Astrophysics, 8, \dodoi{10.33232/001c.129138}

\bibitem[{McKee \& Tan(2008)}]{McKeetan_2008popiii.1}
McKee, C.~F., \& Tan, J.~C. 2008, The Astrophysical Journal, 681, 771–797, \dodoi{10.1086/587434}

\bibitem[{{Mezcua} {et~al.}(2023){Mezcua}, {Siudek}, {Suh}, {Valiante}, {Spinoso}, \& {Bonoli}}]{Mezcua_dwarf_2023}
{Mezcua}, M., {Siudek}, M., {Suh}, H., {et~al.} 2023, \apjl, 943, L5, \dodoi{10.3847/2041-8213/acae25}

\bibitem[{{Mortlock} {et~al.}(2011){Mortlock}, {Warren}, {Venemans}, {Patel}, {Hewett}, {McMahon}, {Simpson}, {Theuns}, {Gonz{\'a}les-Solares}, {Adamson}, {Dye}, {Hambly}, {Hirst}, {Irwin}, {Kuiper}, {Lawrence}, \& {R{\"o}ttgering}}]{mortlock_2011_highzqso}
{Mortlock}, D.~J., {Warren}, S.~J., {Venemans}, B.~P., {et~al.} 2011, \nat, 474, 616, \dodoi{10.1038/nature10159}

\bibitem[{Naidu {et~al.}(2025)Naidu, Matthee, Katz, Graaff, Oesch, Smith, Greene, Brammer, Weibel, Hviding, Chisholm, Labbé, Simcoe, Witten, Atek, Baggen, Belli, Bezanson, Boogaard, Bose, Covelo-Paz, Dayal, Fudamoto, Furtak, Giovinazzo, Goulding, Gronke, Heintz, Hirschmann, Illingworth, Inoue, Johnson, Leja, Leonova, McConachie, Maseda, Natarajan, Nelson, Setton, Shivaei, Sobral, Stefanon, Tacchella, Toft, Torralba, Dokkum, Wel, Volonteri, Walter, Wang, \& Watson}]{naidu_2025}
Naidu, R.~P., Matthee, J., Katz, H., {et~al.} 2025, A "{Black} {Hole} {Star}" {Reveals} the {Remarkable} {Gas}-{Enshrouded} {Hearts} of the {Little} {Red} {Dots},  arXiv, \dodoi{10.48550/arXiv.2503.16596}

\bibitem[{{Newville} {et~al.}(2016){Newville}, {Stensitzki}, {Allen}, {Rawlik}, {Ingargiola}, \& {Nelson}}]{lmfit_2016}
{Newville}, M., {Stensitzki}, T., {Allen}, D.~B., {et~al.} 2016, {Lmfit: Non-Linear Least-Square Minimization and Curve-Fitting for Python}, Astrophysics Source Code Library, record ascl:1606.014

\bibitem[{O'Brien {et~al.}(2024)O'Brien, Jansen, Grogin, Cohen, Smith, Silver, III, Windhorst, Carleton, Koekemoer, Hathi, Willmer, Frye, Alpaslan, Ashby, Ashcraft, Bonoli, Brisken, Cappelluti, Civano, Conselice, Dhillon, Driver, Duncan, Dupke, Elvis, Fazio, Finkelstein, Gim, Griffiths, Hammel, Hyun, Im, Jones, Kim, Ladjelate, Larson, Malhotra, Marshall, Milam, Pierel, Rhoads, Rodney, Röttgering, Rutkowski, R.~E.~Ryan, Ward, White, van Weeren, Zhao, Summers, D'Silva, III, Robotham, Coe, Nonino, Pirzkal, Yan, \& Acharya}]{obrien2024}
O'Brien, R., Jansen, R.~A., Grogin, N.~A., {et~al.} 2024, TREASUREHUNT: Transients and Variability Discovered with HST in the JWST North Ecliptic Pole Time Domain Field.
\newblock \doarXiv{2401.04944}

\bibitem[{{Oke} \& {Gunn}(1983)}]{ABmag_OkeandGunn_1983}
{Oke}, J.~B., \& {Gunn}, J.~E. 1983, \apj, 266, 713, \dodoi{10.1086/160817}

\bibitem[{{Perrin} {et~al.}(2012){Perrin}, {Soummer}, {Elliott}, {Lallo}, \& {Sivaramakrishnan}}]{STPSF}
{Perrin}, M.~D., {Soummer}, R., {Elliott}, E.~M., {Lallo}, M.~D., \& {Sivaramakrishnan}, A. 2012, in Society of Photo-Optical Instrumentation Engineers (SPIE) Conference Series, Vol. 8442, Space Telescopes and Instrumentation 2012: Optical, Infrared, and Millimeter Wave, ed. M.~C. {Clampin}, G.~G. {Fazio}, H.~A. {MacEwen}, \& J.~M. {Oschmann}, Jr., 84423D, \dodoi{10.1117/12.925230}

\bibitem[{{Pirzkal} {et~al.}(2024){Pirzkal}, {Rothberg}, {Papovich}, {Shen}, {Leung}, {Bagley}, {Finkelstein}, {Vanderhoof}, {Lotz}, {Koekemoer}, {Hathi}, {Cheng}, {Cleri}, {Grogin}, {Yung}, {Dickinson}, {Ferguson}, {Gardner}, {Jung}, {Kartaltepe}, {Ryan}, {Simons}, {Ravindranath}, {Berg}, {Backhaus}, {Casey}, {Castellano}, {Ch{\'a}vez Ortiz}, {Chworowsky}, {Cox}, {Dav{\'e}}, {Davis}, {Estrada-Carpenter}, {Fontana}, {Fujimoto}, {Giavalisco}, {Grazian}, {Hutchison}, {Jaskot}, {Kewley}, {Kirkpatrick}, {Kocevski}, {Larson}, {Matharu}, {Natarajan}, {Pentericci}, {P{\'e}rez-Gonz{\'a}lez}, {Snyder}, {Somerville}, {Trump}, \& {Wilkins}}]{pirzkal+2023_NGDEEP_paper}
{Pirzkal}, N., {Rothberg}, B., {Papovich}, C., {et~al.} 2024, \apj, 969, 90, \dodoi{10.3847/1538-4357/ad429c}

\bibitem[{Pouliasis {et~al.}(2019)Pouliasis, Georgantopoulos, Bonanos, Yang, Sokolovsky, Hatzidimitriou, Mountrichas, Gavras, Charmandaris, Bellas-Velidis, Spetsieri, \& Tsinganos}]{Pouliasis_2019}
Pouliasis, E., Georgantopoulos, I., Bonanos, A.~Z., {et~al.} 2019, Monthly Notices of the Royal Astronomical Society, 487, 4285–4304, \dodoi{10.1093/mnras/stz1483}

\bibitem[{{Prevot} {et~al.}(1984){Prevot}, {Lequeux}, {Maurice}, {Prevot}, \& {Rocca-Volmerange}}]{Prevot_SMC_1984}
{Prevot}, M.~L., {Lequeux}, J., {Maurice}, E., {Prevot}, L., \& {Rocca-Volmerange}, B. 1984, \aap, 132, 389

\bibitem[{Rafelski {et~al.}(2015)Rafelski, Teplitz, Gardner, Coe, Bond, Koekemoer, Grogin, Kurczynski, McGrath, Bourque, Atek, Brown, Colbert, Codoreanu, Ferguson, Finkelstein, Gawiser, Giavalisco, Gronwall, Hanish, Lee, Mehta, de~Mello, Ravindranath, Ryan, Scarlata, Siana, Soto, \& Voyer}]{Rafelski_2015_UVUDF}
Rafelski, M., Teplitz, H.~I., Gardner, J.~P., {et~al.} 2015, The Astronomical Journal, 150, 31, \dodoi{10.1088/0004-6256/150/1/31}

\bibitem[{{Rees}(1978)}]{Rees_1978}
{Rees}, M.~J. 1978, in IAU Symposium, Vol.~77, Structure and Properties of Nearby Galaxies, ed. E.~M. {Berkhuijsen} \& R.~{Wielebinski}, 237

\bibitem[{Regan \& Volonteri(2024)}]{Regan_2024}
Regan, J., \& Volonteri, M. 2024, The Open Journal of Astrophysics, 7, \dodoi{10.33232/001c.123239}

\bibitem[{{Reines} {et~al.}(2013){Reines}, {Greene}, \& {Geha}}]{reines_dwarfgal_2013}
{Reines}, A.~E., {Greene}, J.~E., \& {Geha}, M. 2013, \apj, 775, 116, \dodoi{10.1088/0004-637X/775/2/116}

\bibitem[{Reines \& Volonteri(2015)}]{Reines_Volonteri_2015}
Reines, A.~E., \& Volonteri, M. 2015, The Astrophysical Journal, 813, 82, \dodoi{10.1088/0004-637x/813/2/82}

\bibitem[{{Rieke} {et~al.}(2023){Rieke}, {Robertson}, {Tacchella}, {Hainline}, {Johnson}, {Hausen}, {Ji}, {Willmer}, {Eisenstein}, {Pusk{\'a}s}, {Alberts}, {Arribas}, {Baker}, {Baum}, {Bhatawdekar}, {Bonaventura}, {Boyett}, {Bunker}, {Cameron}, {Carniani}, {Charlot}, {Chevallard}, {Chen}, {Curti}, {Curtis-Lake}, {Danhaive}, {DeCoursey}, {Dressler}, {Egami}, {Endsley}, {Helton}, {Hviding}, {Kumari}, {Looser}, {Lyu}, {Maiolino}, {Maseda}, {Nelson}, {Rieke}, {Rix}, {Sandles}, {Saxena}, {Sharpe}, {Shivaei}, {Skarbinski}, {Smit}, {Stark}, {Stone}, {Suess}, {Sun}, {Topping}, {{\"U}bler}, {Villanueva}, {Wallace}, {Williams}, {Willott}, {Whitler}, {Witstok}, \& {Woodrum}}]{JADES_DR1}
{Rieke}, M.~J., {Robertson}, B., {Tacchella}, S., {et~al.} 2023, \apjs, 269, 16, \dodoi{10.3847/1538-4365/acf44d}

\bibitem[{{Robitaille} {et~al.}(2020){Robitaille}, {Deil}, \& {Ginsburg}}]{astropyreproject}
{Robitaille}, T., {Deil}, C., \& {Ginsburg}, A. 2020, {reproject: Python-based astronomical image reprojection}, Astrophysics Source Code Library, record ascl:2011.023

\bibitem[{Sanati {et~al.}(2025)Sanati, Devriendt, SergioMartin-Alvarez, Slyz, \& Tan}]{sanati2025}
Sanati, M., Devriendt, J., SergioMartin-Alvarez, Slyz, A., \& Tan, J.~C. 2025, On the rapid growth of SMBHs in high-z galaxies: the aftermath of Population III.1 stars.
\newblock \doarXiv{2507.02058}

\bibitem[{{Schleicher} {et~al.}(2023){Schleicher}, {Reinoso}, \& {Klessen}}]{Schleicher_2023mergsc}
{Schleicher}, D. R.~G., {Reinoso}, B., \& {Klessen}, R.~S. 2023, \mnras, 521, 3972, \dodoi{10.1093/mnras/stad807}

\bibitem[{{Schneider} {et~al.}(1994){Schneider}, {Schmidt}, \& {Gunn}}]{Schneider_qsoz>41994}
{Schneider}, D.~P., {Schmidt}, M., \& {Gunn}, J.~E. 1994, \aj, 107, 880, \dodoi{10.1086/116901}

\bibitem[{{Schneider} {et~al.}(2000){Schneider}, {Fan}, {Strauss}, {Gunn}, {Richards}, {Knapp}, {Lupton}, {Saxe}, {Anderson}, {Bahcall}, {Brinkmann}, {Brunner}, {Csabai}, {Fukugita}, {Hennessy}, {Hindsley}, {Ivezi{\'c}}, {Nichol}, {Pier}, \& {York}}]{schneider2000}
{Schneider}, D.~P., {Fan}, X., {Strauss}, M.~A., {et~al.} 2000, \aj, 120, 2183, \dodoi{10.1086/316834}

\bibitem[{Setton {et~al.}(2024)Setton, Greene, de~Graaff, Ma, Leja, Matthee, Bezanson, Boogaard, Cleri, Katz, Labbe, Maseda, McConachie, Miller, Price, Suess, van Dokkum, Wang, Weibel, Whitaker, \& Williams}]{setton2024_LRDs}
Setton, D.~J., Greene, J.~E., de~Graaff, A., {et~al.} 2024, Little Red Dots at an Inflection Point: Ubiquitous "V-Shaped" Turnover Consistently Occurs at the Balmer Limit.
\newblock \doarXiv{2411.03424}

\bibitem[{Sijacki {et~al.}(2009)Sijacki, Springel, \& Haehnelt}]{Sijacki_2009}
Sijacki, D., Springel, V., \& Haehnelt, M.~G. 2009, Monthly Notices of the Royal Astronomical Society, 400, 100–122, \dodoi{10.1111/j.1365-2966.2009.15452.x}

\bibitem[{Singh {et~al.}(2023)Singh, Monaco, \& Tan}]{Singh_2023}
Singh, J., Monaco, P., \& Tan, J.~C. 2023, Monthly Notices of the Royal Astronomical Society, 525, 969–982, \dodoi{10.1093/mnras/stad2346}

\bibitem[{{Siudek} {et~al.}(2023){Siudek}, {Mezcua}, \& {Krywult}}]{Siudek_dwarf_2023}
{Siudek}, M., {Mezcua}, M., \& {Krywult}, J. 2023, \mnras, 518, 724, \dodoi{10.1093/mnras/stac3092}

\bibitem[{{STSCI Development Team}(2012)}]{DrizzlePac}
{STSCI Development Team}. 2012, {DrizzlePac: HST image software}, Astrophysics Source Code Library, record ascl:1212.011

\bibitem[{{Tacchella} {et~al.}(2023){Tacchella}, {Eisenstein}, {Hainline}, {Johnson}, {Baker}, {Helton}, {Robertson}, {Suess}, {Chen}, {Nelson}, {Pusk{\'a}s}, {Sun}, {Alberts}, {Egami}, {Hausen}, {Rieke}, {Rieke}, {Shivaei}, {Williams}, {Willmer}, {Bunker}, {Cameron}, {Carniani}, {Charlot}, {Curti}, {Curtis-Lake}, {Looser}, {Maiolino}, {Maseda}, {Rawle}, {Rix}, {Smit}, {{\"U}bler}, {Willott}, {Witstok}, {Baum}, {Bhatawdekar}, {Boyett}, {Danhaive}, {de Graaff}, {Endsley}, {Ji}, {Lyu}, {Sandles}, {Saxena}, {Scholtz}, {Topping}, \& {Whitler}}]{tacchella_2023gnz11}
{Tacchella}, S., {Eisenstein}, D.~J., {Hainline}, K., {et~al.} 2023, \apj, 952, 74, \dodoi{10.3847/1538-4357/acdbc6}

\bibitem[{Taylor {et~al.}(2025)Taylor, Kokorev, Kocevski, Akins, Cullen, Dickinson, Finkelstein, Haro, Bromm, Giavalisco, Inayoshi, Juneau, Leung, Perez-Gonzalez, Somerville, Trump, Amorin, Barro, Burgarella, Brooks, Carnall, Casey, Cheng, Chisholm, Chworowsky, Davis, Donnan, Dunlop, Ellis, Fernandez, Fujimoto, Grogin, Gupta, Hathi, Jung, Hirschmann, Kartaltepe, Koekemoer, Larson, Leung, Llerena, Lucas, McLeod, McLure, Napolitano, Papovich, Stanton, Tripodi, Wang, Wilkins, Yung, \& Zavala}]{taylor2025_capersz9LRD}
Taylor, A.~J., Kokorev, V., Kocevski, D.~D., {et~al.} 2025, CAPERS-LRD-z9: A Gas Enshrouded Little Red Dot Hosting a Broad-line AGN at z=9.288.
\newblock \doarXiv{2505.04609}

\bibitem[{Temple {et~al.}(2021)Temple, Hewett, \& Banerji}]{Temple_2021}
Temple, M.~J., Hewett, P.~C., \& Banerji, M. 2021, Monthly Notices of the Royal Astronomical Society, 508, 737–754, \dodoi{10.1093/mnras/stab2586}

\bibitem[{Teplitz {et~al.}(2013)Teplitz, Rafelski, Kurczynski, Bond, Grogin, Koekemoer, Atek, Brown, Coe, Colbert, Ferguson, Finkelstein, Gardner, Gawiser, Giavalisco, Gronwall, Hanish, Lee, de~Mello, Ravindranath, Ryan, Siana, Scarlata, Soto, Voyer, \& Wolfe}]{Teplitz_2013_UVUDF}
Teplitz, H.~I., Rafelski, M., Kurczynski, P., {et~al.} 2013, The Astronomical Journal, 146, 159, \dodoi{10.1088/0004-6256/146/6/159}

\bibitem[{Trinca {et~al.}(2022)Trinca, Schneider, Valiante, Graziani, Zappacosta, \& Shankar}]{Trinca_2022}
Trinca, A., Schneider, R., Valiante, R., {et~al.} 2022, Monthly Notices of the Royal Astronomical Society, 511, 616–640, \dodoi{10.1093/mnras/stac062}

\bibitem[{Tripodi {et~al.}(2024)Tripodi, Martis, Markov, Bradač, Mascia, Cammelli, D'Eugenio, Willott, Curti, Bhatt, Gallerani, Rihtaršič, Singh, Gaspar, Harshan, Judež, Merida, Desprez, Sawicki, Goovaerts, Muzzin, Noirot, Sarrouh, Abraham, Asada, Brammer, Carpenter, Felicioni, Fujimoto, Iyer, Mowla, \& Strait}]{tripodi_2024}
Tripodi, R., Martis, N., Markov, V., {et~al.} 2024, Red, hot, and very metal poor: extreme properties of a massive accreting black hole in the first 500 {Myr},  arXiv, \dodoi{10.48550/arXiv.2412.04983}

\bibitem[{{{\"U}bler} {et~al.}(2023){{\"U}bler}, {Maiolino}, {Curtis-Lake}, {P{\'e}rez-Gonz{\'a}lez}, {Curti}, {Perna}, {Arribas}, {Charlot}, {Marshall}, {D'Eugenio}, {Scholtz}, {Bunker}, {Carniani}, {Ferruit}, {Jakobsen}, {Rix}, {Rodr{\'\i}guez Del Pino}, {Willott}, {Boeker}, {Cresci}, {Jones}, {Kumari}, \& {Rawle}}]{Ubler_23MBH}
{{\"U}bler}, H., {Maiolino}, R., {Curtis-Lake}, E., {et~al.} 2023, \aap, 677, A145, \dodoi{10.1051/0004-6361/202346137}

\bibitem[{Valiante {et~al.}(2016)Valiante, Schneider, Volonteri, \& Omukai}]{Valiante_2016}
Valiante, R., Schneider, R., Volonteri, M., \& Omukai, K. 2016, Monthly Notices of the Royal Astronomical Society, 457, 3356–3371, \dodoi{10.1093/mnras/stw225}

\bibitem[{Virtanen {et~al.}(2020)Virtanen, Gommers, Oliphant, Haberland, Reddy, Cournapeau, Burovski, Peterson, Weckesser, Bright, {van der Walt}, Brett, Wilson, Millman, Mayorov, Nelson, Jones, Kern, Larson, Carey, Polat, Feng, Moore, {VanderPlas}, Laxalde, Perktold, Cimrman, Henriksen, Quintero, Harris, Archibald, Ribeiro, Pedregosa, {van Mulbregt}, \& {SciPy 1.0 Contributors}}]{2020SciPy}
Virtanen, P., Gommers, R., Oliphant, T.~E., {et~al.} 2020, Nature Methods, 17, 261, \dodoi{10.1038/s41592-019-0686-2}

\bibitem[{{Volonteri}(2010)}]{volonteri_2010}
{Volonteri}, M. 2010, \aapr, 18, 279, \dodoi{10.1007/s00159-010-0029-x}

\bibitem[{Volonteri {et~al.}(2003)Volonteri, Haardt, \& Madau}]{Volonteri_2003}
Volonteri, M., Haardt, F., \& Madau, P. 2003, The Astrophysical Journal, 582, 559–573, \dodoi{10.1086/344675}

\bibitem[{Volonteri {et~al.}(2023)Volonteri, Habouzit, \& Colpi}]{Volonteri_2023}
Volonteri, M., Habouzit, M., \& Colpi, M. 2023, Monthly Notices of the Royal Astronomical Society, 521, 241–250, \dodoi{10.1093/mnras/stad499}

\bibitem[{Wang {et~al.}(2025)Wang, de~Graaff, Davies, Greene, Leja, Brammer, Goulding, Miller, Suess, Weibel, Williams, Bezanson, Boogaard, Cleri, Hirschmann, Katz, Labbé, Maseda, Matthee, McConachie, Naidu, Oesch, Rix, Setton, \& Whitaker}]{Wang_2025_LRDs}
Wang, B., de~Graaff, A., Davies, R.~L., {et~al.} 2025, The Astrophysical Journal, 984, 121, \dodoi{10.3847/1538-4357/adc1ca}

\bibitem[{{Wang} {et~al.}(2018){Wang}, {Yang}, {Fan}, {Yue}, {Wu}, {Schindler}, {Bian}, {Li}, {Farina}, {Ba{\~n}ados}, {Davies}, {Decarli}, {Green}, {Jiang}, {Hennawi}, {Huang}, {Mazzucchelli}, {McGreer}, {Venemans}, {Walter}, \& {Beletsky}}]{wang_2018_highzqso}
{Wang}, F., {Yang}, J., {Fan}, X., {et~al.} 2018, \apjl, 869, L9, \dodoi{10.3847/2041-8213/aaf1d2}

\bibitem[{{Warren} {et~al.}(1987){Warren}, {Hewett}, {Irwin}, {McMahon}, \& {Bridgeland}}]{Warren_1987qsoz>4}
{Warren}, S.~J., {Hewett}, P.~C., {Irwin}, M.~J., {McMahon}, R.~G., \& {Bridgeland}, M.~T. 1987, \nat, 325, 131, \dodoi{10.1038/325131a0}

\bibitem[{{Wasleske} \& {Baldassare}(2023)}]{Wasleske_Baldassare2024dwarfxray}
{Wasleske}, E.~J., \& {Baldassare}, V.~F. 2023, \aj, 166, 64, \dodoi{10.3847/1538-3881/ace16b}

\bibitem[{{Weaver} {et~al.}(2024){Weaver}, {Cutler}, {Pan}, {Whitaker}, {Labb{\'e}}, {Price}, {Bezanson}, {Brammer}, {Marchesini}, {Leja}, {Wang}, {Furtak}, {Zitrin}, {Atek}, {Chemerynska}, {Coe}, {Dayal}, {van Dokkum}, {Feldmann}, {F{\"o}rster Schreiber}, {Franx}, {Fujimoto}, {Fudamoto}, {Glazebrook}, {de Graaff}, {Greene}, {Juneau}, {Kassin}, {Kriek}, {Khullar}, {Maseda}, {Mowla}, {Muzzin}, {Nanayakkara}, {Nelson}, {Oesch}, {Pacifici}, {Papovich}, {Setton}, {Shapley}, {Shipley}, {Smit}, {Stefanon}, {Taylor}, {Weibel}, \& {Williams}}]{Weaver_UNCOVERphot_2024}
{Weaver}, J.~R., {Cutler}, S.~E., {Pan}, R., {et~al.} 2024, \apjs, 270, 7, \dodoi{10.3847/1538-4365/ad07e0}

\bibitem[{Wiener(1949)}]{wiener_1949}
Wiener, N. 1949, Extrapolation, Interpolation, and Smoothing of Stationary Time Series: With Engineering Applications (The MIT Press)

\bibitem[{{Williams} {et~al.}(2023){Williams}, {Tacchella}, {Maseda}, {Robertson}, {Johnson}, {Willott}, {Eisenstein}, {Willmer}, {Ji}, {Hainline}, {Helton}, {Alberts}, {Baum}, {Bhatawdekar}, {Boyett}, {Bunker}, {Carniani}, {Charlot}, {Chevallard}, {Curtis-Lake}, {de Graaff}, {Egami}, {Franx}, {Kumari}, {Maiolino}, {Nelson}, {Rieke}, {Sandles}, {Shivaei}, {Simmonds}, {Smit}, {Suess}, {Sun}, {{\"U}bler}, \& {Witstok}}]{williams_2023_JEMS}
{Williams}, C.~C., {Tacchella}, S., {Maseda}, M.~V., {et~al.} 2023, \apjs, 268, 64, \dodoi{10.3847/1538-4365/acf130}

\bibitem[{Wise {et~al.}(2019)Wise, Regan, O’Shea, Norman, Downes, \& Xu}]{Wise_2019DCBH}
Wise, J.~H., Regan, J.~A., O’Shea, B.~W., {et~al.} 2019, Nature, 566, 85–88, \dodoi{10.1038/s41586-019-0873-4}

\bibitem[{{Yang} {et~al.}(2019){Yang}, {Wang}, {Fan}, {Yue}, {Wu}, {Li}, {Bian}, {Jiang}, {Ba{\~n}ados}, \& {Beletsky}}]{Yang2019_highzqso}
{Yang}, J., {Wang}, F., {Fan}, X., {et~al.} 2019, \aj, 157, 236, \dodoi{10.3847/1538-3881/ab1be1}

\bibitem[{Yang {et~al.}(2020)Yang, Wang, Fan, Hennawi, Davies, Yue, Banados, Wu, Venemans, Barth, Bian, Boutsia, Decarli, Farina, Green, Jiang, Li, Mazzucchelli, \& Walter}]{Yang_2020b_highzqso}
Yang, J., Wang, F., Fan, X., {et~al.} 2020, The Astrophysical Journal Letters, 897, L14, \dodoi{10.3847/2041-8213/ab9c26}

\bibitem[{{Yoshida}(2006)}]{Yoshida_2006_100Myr}
{Yoshida}, N. 2006, \nar, 50, 19, \dodoi{10.1016/j.newar.2005.11.011}

\bibitem[{{Zaw} {et~al.}(2020){Zaw}, {Rosenthal}, {Katkov}, {Gelfand}, {Chen}, {Greenhill}, {Brisken}, \& {Noori}}]{Zaw_lowzMbhMstellar_catalogue_2020}
{Zaw}, I., {Rosenthal}, M.~J., {Katkov}, I.~Y., {et~al.} 2020, \apj, 897, 111, \dodoi{10.3847/1538-4357/ab9944}

\bibitem[{Zhang {et~al.}(2025)Zhang, Egami, Sun, Lin, Lyu, Zhu, Rinaldi, Sun, Bunker, Bhatawdekar, Helton, Maiolino, Ma, Robertson, Tacchella, Venturi, Williams, \& Willott}]{zhang2025_SCJCMBH}
Zhang, J., Egami, E., Sun, F., {et~al.} 2025, Abundant Population of Broad H$\alpha$ Emitters in the GOODS-N Field Revealed by CONGRESS, FRESCO, and JADES.
\newblock \doeprint{2505.02895}

\end{thebibliography}
\bibliographystyle{aasjournal}



\end{document}